\documentclass[amsmath,amssymb,11pt]{article}
\usepackage{jheppub2}
\pdfoutput=1
\usepackage{graphicx,epsfig}
\usepackage{float}
\usepackage{subfig}
\usepackage{subfloat}
\usepackage[utf8]{inputenc}
\usepackage{hyperref}
\usepackage{caption}
\usepackage{color}

\newcommand{\beq}{\begin{equation}}

\newcommand{\eeq}{\end{equation}}

 \newcommand{\be}{\begin{equation}}
 \newcommand{\ee}{\end{equation}}
 \newcommand{\bea}{\begin{eqnarray}}
 \newcommand{\eea}{\end{eqnarray}}

\usepackage{tikz}
\usetikzlibrary{positioning}
\usetikzlibrary{intersections}
\usetikzlibrary{fadings} 
\usetikzlibrary{arrows.meta} 
\usetikzlibrary{arrows}

\tikzfading[name=fade out,
inner color=transparent!0,
outer color=transparent!100]

\definecolor{cherryblossompink}{rgb}{1.0, 0.72, 0.77}
\definecolor{lightblue}{rgb}{0.68, 0.85, 0.9}

\usetikzlibrary{decorations.pathmorphing}
\usetikzlibrary{decorations.pathreplacing,decorations.markings}

\usetikzlibrary{backgrounds,automata}

\title{Quantum Kerr-de Sitter black holes in three dimensions}

\author{Emanuele Panella}
\emailAdd{emanuele.panella.21@ucl.ac.uk}
\author{and Andrew Svesko}
\emailAdd{a.svesko@ucl.ac.uk}

\affiliation{
Department of Physics and Astronomy, University College London,\\
Gower Street, London, WC1E 6BT, United Kingdom}

\abstract{We use braneworld holography to construct a three-dimensional quantum-corrected Kerr-de Sitter black hole, exactly accounting for semi-classical backreaction effects due to a holographic conformal field theory. By contrast, classically there are no de Sitter black holes in three-dimensions, only geometries with a single cosmological horizon. The quantum Kerr black hole shares many qualitative features with the classical four-dimensional Kerr-de Sitter solution. Of note, backreaction induces inner and outer black hole horizons which hide a ring singularity. Moreover, the quantum-corrected geometry has extremal, Nariai, and ultracold limits, which appear as fibered products of a circle and two-dimensional anti-de Sitter, de Sitter, and Minkowski space, respectively. The thermodynamics of the classical bulk black hole, described by the rotating four-dimensional anti-de Sitter C-metric, has an interpretation on the brane as thermodynamics of the quantum black hole, obeying a semi-classical first law where the Bekenstein-Hawking area entropy is replaced by the generalized entropy. For purposes of comparison, we derive the renormalized quantum stress-tensor due to a free conformally coupled scalar field in the classical Kerr-de Sitter conical geometry and perturbatively solve for its backreaction.

}

\begin{document}

\maketitle

\section{Introduction} \label{sec:intro}

Developing a consistent theory of quantum gravity remains a difficult open problem in theoretical physics. To make progress, it is standard practice to simplify the problem. One approach is to focus on spacetimes exhibiting features quantum gravity is expected to display, such as black holes, whose thermodynamics %\cite{Bekenstein:1972tm,Bekenstein:1973ur,Hawking:1974sw,Hawking:1976de}
lay the foundation for holography. Another approach is to study quantum gravity in spacetimes with fewer dimensions than our own, where we often have better analytic control. Combining both views has proven successful in the context of the anti-de Sitter/conformal field theory (AdS/CFT) correspondence, where, for example, physics of three-dimensional AdS black holes \cite{Banados:1992wn,Banados:1992gq}, including a statistical interpretation of their thermal entropy \cite{Strominger:1997eq,Birmingham:1998jt} and computation of the partition function \cite{Dijkgraaf:2000fq}, has a dual description in terms of a CFT living on the two-dimensional boundary of AdS.

%to Einstein's general relativity, namely, black holes. Indeed, black hole thermodynamics lays the foundation for holography, a defining feature which quantum gravity is expected to display. Black holes of particular interest are those in backgrounds approximating our universe in its phase of accelerated expansion. 

%Black hole physics continues to play a critical role in our understanding of the universe, both on a classical and quantum level. 

Classically, however, there are no black hole solutions to Einstein's field equations in three-dimensional de Sitter space ($\text{dS}_{3}$). Rather, the geometry of a point mass in $\text{dS}_{3}$ describes a conical defect
%\footnote{There are in fact two conical singularities, one at each pole of the 2-sphere arising from a constant timeslice of $\text{dS}_{3}$.}
with a single cosmological horizon and no black hole horizon \cite{Deser:1983nh}.
%\footnote{The reason there are no such classical black holes can be seen via dimensional analysis. In three-dimensional gravity with a positive cosmological constant $\Lambda_{3}$, a particle of mass $M$ does not introduce a length scale since $G_{3}M$ has a vanishing scaling dimension while the presence of $\Lambda_{3}$ only introduces a cosmological horizon.}
This is unfortunate since de Sitter space, having a positive cosmological constant, is a reasonable approximation of our universe during its inflationary past and current phase of accelerated expansion. Consequently, the uncharged Kerr-de Sitter spacetime is, arguably, the most astrophysically relevant black hole to study. 
However, it seems we cannot learn about the microphysics of higher-dimensional de Sitter black holes by appealing to lower dimensions. 

Despite the lack of classical $\text{dS}_{3}$ black holes, quantum effects dramatically alter the situation: de-Sitter black holes with horizons significantly larger than the Planck length arise due to semi-classical backreaction \cite{Emparan:2022ijy}.\footnote{The basic reasoning follows from dimensional analysis. The Planck length in three-dimensions is $L_{\text{P}}=\hbar G_{3}$, where we work in units with the speed of light set to unity. A collection of $c\gg1$ quantum fields will introduce a combined quantum effect of $c\hbar$ which may gravitate giving rise to a black hole with horizon radii $cL_{\text{P}}\gg L_{\text{P}}$. Crucially, even in the limit of vanishing quantum gravity effects, where $c\to\infty$ and $L_{\text{P}}\to0$ with $cL_{\text{P}}$ fixed, \emph{classical} backreaction due to the quantum fields remain finite.} Cursory evidence for this can be seen by analyzing the semi-classical Einstein equations
\beq G_{\mu\nu}+\Lambda_{3}g_{\mu\nu}=8\pi G_{3}\langle T_{\mu\nu}\rangle\;\label{eq:semiEin}\eeq
of a massless conformally coupled scalar field with stress-energy tensor $T_{\mu\nu}$ in a three-dimensional Schwarzschild-de Sitter  background. To leading order in backreaction, the geometry receives a correction, $\delta g_{tt}\sim L_{\text{P}}/r$ in static coordinates, such that a black hole horizon appears \cite{Emparan:2022ijy}. In fact, the $(t,r)$-sector of the resulting geometry looks like a classical four-dimensional Schwarzschild-dS black hole, heralding a holographic pedigree. 

The purpose of this article is to analyze semi-classical backreaction of a conformal field theory in a Kerr-$\text{dS}_{3}$ ($\text{KdS}_{3}$) background. Despite the inclusion of rotation, the $\text{KdS}_{3}$ metric also describes a conical defect with a single cosmological horizon (see, \emph{e.g.,} \cite{Klemm:2002ir}). As we will show, semi-classical backreaction leads to the development of inner and outer black hole horizons, as is standard for the classical higher-dimensional Kerr-dS geometry.
%and heralds a holographic pedigree. 
However, as in the static case, to confirm such a quantum-corrected black hole does arise, we must consistently solve the semi-classical Einstein equations for a large number of quantum fields perturbatively in $L_{\text{P}}$ beyond leading order, another challenging open problem. 

%To circumvent the challenge of consistently solving the semi-classical Einstein equations (\ref{eq:semiEin}), 

Fortunately, there is an alternative framework, dubbed `braneworld holography' \cite{deHaro:2000wj}, that may \emph{exactly} account for quantum backreaction, without explicitly solving the semi-classical Einstein equations (\ref{eq:semiEin}). In this setting, one couples Einstein gravity in a holographic asymptotically $(d+1)$-dimensional AdS background to a $d$-dimensional Randall-Sundrum brane \cite{Randall:1999vf}. Integrating out the bulk gravitational degrees of freedom up to the brane leads to a theory of gravity induced on the brane with, schematically, the following field equations
\beq G_{\mu\nu}+\Lambda_{d}g_{\mu\nu}+(\text{higher-curvature})=8\pi G_{d}\langle T_{\mu\nu}\rangle\;.\label{eq:semigrav}\eeq
This leads to two equivalent perspectives, bulk and brane. The bulk is described by classical general relativity in $\text{AdS}_{d+1}$, with a dual $\text{CFT}_{d}$ description, coupled to a codimension-1 brane, while an observer confined to the brane experiences a $d$-dimensional semi-classical theory of gravity coupled to the holographic $\text{CFT}_{d}$ with an ultraviolet cutoff and renormalized stress-tensor $\langle T_{\mu\nu}\rangle$. Consistency between these two pictures demands \emph{quantum} dynamics of the brane gravity theory be entirely encoded in the \emph{classical} dynamics of the bulk. Consequently, black holes localized on a brane in $\text{AdS}_{d+1}$ -- found by solving the classical bulk Einstein equations with brane boundary conditions -- are, from the brane point of view,  black holes corrected by the backreaction of the $d$-dimensional CFT \cite{Emparan:2002px}. Thus far, this procedure has been carried out analytically in $d=3$, uncovering a family of quantum $\text{AdS}_{3}$ black holes, collectively referred to as the quantum BTZ (qBTZ) solution \cite{Emparan:2020znc}, and the quantum Schwarzschild-dS (qSdS) black hole \cite{Emparan:2022ijy}.\footnote{Historically, exact three-dimensional black hole solutions on the brane were uncovered in \cite{Emparan:1999wa,Emparan:1999fd} and later interpreted as holographic quantum black holes in \cite{Emparan:2002px}, however, the higher curvature corrections of the induced gravity action on the brane were not explicitly accounted for until \cite{Emparan:2020znc}.}

Following suit, here we use braneworld holography to find an exact quantum-corrected rotating black hole in $\text{dS}_{3}$,  a non-trivial extension of \cite{Emparan:2022ijy}. Our starting point, as in \cite{Emparan:2020znc}, is the rotating $\text{AdS}_{4}$ C-metric, however, coupled to an asymptotically $\text{dS}_{3}$ Randall-Sundrum brane. As an exact solution to the bulk Einstein equations, we are guaranteed the brane geometry is an exact solution to the full semi-classical theory (\ref{eq:semigrav}), in the planar limit of the CFT, resulting in the quantum Kerr-dS (qKdS) black hole, presented in (\ref{eq:branegeomv2}). The essential new correction to the geometry is a term which goes like $1/r$ in appropriate static coordinates. This feature, combined with the $1/r^{2}$ behavior typical in a Kerr-black hole (including classical Kerr-$\text{dS}_{3}$), results in a three-dimensional geometry with three horizons, an inner and outer black hole horizon and a cosmological horizon, leading to three limiting behaviors:  (i) an extremal limit, when the inner and outer horizons coincide, (ii) the rotating quantum Nariai black hole, when the outer and cosmological horizons coincide, and (iii) an ultracold limit when all three horizons coincide. There is also a `lukewarm' quantum black hole, where the surface gravities of outer and cosmological horizons are equal. Further, unlike three-dimensional KdS, the quantum-corrected geometry has a ring singularity.

Quantum backreaction also enriches the thermodynamic behavior of the black hole. Via holography, thermodynamics of the bulk black hole is interpreted as thermodynamics of the semi-classical brane black hole. Crucially, the Bekenstein-Hawking entropy of the bulk black hole $S_{\text{BH}}$, proportional to the area $A_{4}$ of its event horizon, is identified with the three-dimensional generalized entropy $S_{\text{gen}}$,
\beq S_{\text{BH}}=\frac{A_{4}}{4G_{4}}\leftrightarrow S_{\text{gen}}=\frac{A_{3}}{4G_{3}}+S_{\text{Wald}}+S_{\text{CFT}}\;.\eeq
Here $A_{3}$ represents the area of the horizon of the brane black hole, $S_{\text{Wald}}$ is the Wald entropy \cite{Wald:1993nt} accounting for the higher-curvature corrections in the theory, and $S_{\text{CFT}}$ is the CFT entanglement entropy due to quantum fluctuations outside of each horizon. With this identification, we will uncover the first law of thermodynamics of quantum-corrected Kerr-de Sitter black holes, which, again, is guaranteed to hold to all orders in backreaction. 

An outline of the remainder of this article is as follows. In Section \ref{sec:Branesdsrev} we briefly review braneworld holography and the construction of the qSdS solution. Section \ref{sec:GeomKdS3} is primarily devoted to the geometric construction of the qKdS black hole, where include an analysis of each of its Nariai, extremal, and ultracold limits, and compute the renormalized stress-tensor of the holographic CFT. A detailed account of the horizon thermodynamics is given in Section \ref{sec:ThermoqKdS}. We conclude in Section \ref{sec:disc} where we describe multiple future research directions.
%including a brief discussion on a doubly-holographic scenario where the quantum Kerr-$\text{dS}_{3}$ system has a dual two-dimensional description. 
To keep this article complete and self-contained, we include a number of appendices. Appendix \ref{app:KerrdS3} summarizes the basic elements of the classical Kerr-$\text{dS}_{3}$ geometry. Since we have not seen the computation performed in the literature, in Appendix \ref{app:pertbackKdS} we provide an analysis of perturbative backreaction due to a massless conformally coupled scalar field in classical Kerr-$\text{dS}_{3}$. Appendix \ref{app:AdSCmetprops} reviews the geometry of the $\text{AdS}_{4}$ C-metric along with additional details of the braneworld construction. Appendix \ref{app:limitsqKdS} expounds on the extremal, Nariai, ultracold, and lukewarm limits of the quantum black hole.

%%%%%%%%%%%%%%%%%%%%%%%%%%%%%%%%%%%%%%%%%%%%%%%%%%%%%%%%%%%%%%%%%
\section{Braneworlds and quantum SdS black hole: review} \label{sec:Branesdsrev}

\subsection*{Braneworld holography and induced gravity}

Consider an asymptotically $\text{AdS}_{d+1}$ spacetime $\mathcal{M}$ of curvature scale $L_{d+1}$, with a dual description in terms of a $\text{CFT}_{d}$ on the asymptotic boundary $\partial\mathcal{M}$. The standard AdS/CFT dictionary of Gubser, Klebanov, Polyakov and Witten  (GKPW) \cite{Gubser:1998bc,Witten:1998qj} relates the CFT generating functional to the on-shell gravitational action. Even at tree level, the on-shell action will have long-distance infrared (IR) divergences since the metric will grow to infinity as the asymptotic AdS boundary is approached. These correspond to ultraviolet (UV) divergences due to quantum fluctuations of the dual CFT. Holographic renormalization \cite{deHaro:2000vlm,Skenderis:2002wp} is a prescription to remove the IR divergences by adding appropriate local counterterms \cite{Kraus:1999di,Emparan:1999pm,deHaro:2000vlm,Papadimitriou:2004ap} in a minimal subtraction scheme. The divergent contribution to the action may be cast in terms of the curvature invariants with respect to the induced metric $h_{ij}$ near $\partial\mathcal{M}$
\beq I_{\text{div}}\hspace{-1mm}=\hspace{-1mm}\frac{L_{d+1}}{16\pi G_{d+1}}\hspace{-1mm}\int_{\partial\mathcal{M}}\hspace{-4mm}d^{d}x\sqrt{-h}\biggr[\frac{2(d-1)}{L_{d+1}^{2}}+\frac{R}{(d-2)}+\frac{L^{2}_{d+1}}{(d-2)^{2}(d-4)}\hspace{-1mm}\left(R_{ij}^{2}-\frac{dR^{2}}{4(d-1)}\right)+...\biggr].\label{eq:Idivd} \eeq
Technically, this action arises by introducing an IR cutoff surface at some small finite distance away from $\partial\mathcal{M}$. Integrating out the bulk degrees freedom up to the cutoff surface, the regulated bulk gravity action (Einstein-Hilbert action plus a Gibbons-Hawking-York boundary term) is a sum of a divergent contribution (\ref{eq:Idivd}) and a finite action in the limit the bulk cutoff surface approaches the boundary. Holographic renormalization is complete by adding a counterterm $I_{\text{ct}}=-I_{\text{div}}$ to the regulated action and taking the boundary limit.

In braneworld holography \cite{deHaro:2000wj}, the bulk IR regulator surface is replaced by a $d$-dimensional end-of-the-world brane~$\mathcal{B}$ (typically taken to be near the boundary), as in the Randall-Sundrum braneworld construction \cite{Randall:1999vf}, such that the limiting procedure is not taken. Consequently, the physical space is cutoff, however, there are no longer any divergences to remove and $I_{\text{div}}$ is finite.  Additionally, the metric on the brane is dynamical, characterized by a holographically induced higher curvature theory of gravity coupled to a CFT with a UV cutoff. More precisely, the induced theory on the brane is found by adding to the bulk Einstein theory the brane action
\beq 
I_{\text{brane}}=-\tau \int_{\mathcal{B}}d^{d}x\sqrt{-h}\,,
\label{eq:braneaction}\eeq
where $\tau$ is a parameter controlling the tension of the brane. Integrating out the bulk between $\partial\mathcal{M}$ up to $\mathcal{B}$ as in holographic regularization, the induced theory on the brane is
\be
I=I_{\text{Bgrav}}[\mathcal{B}]+I_{\text{CFT}}[\mathcal{B}]\,.
\label{eq:inductheorygen}\ee
The induced gravity theory on the brane is (see, \emph{e.g.}, \cite{Chen:2020uac,Bueno:2022log})\footnote{The gravitational sector of the induced theory here is technically given by the sum $I_{\text{Bgrav}}=2I_{\text{div}}+I_{\text{brane}}$. The factor of two depends on the specific braneworld construction. Namely, whether or not we consider a $\mathbb{Z}_{2}$ construction by gluing a second copy of the spacetime along the cutoff region such that one integrates out the bulk geometry on both sides of the brane.}
\beq
\begin{split}
I_{\text{Bgrav}}&=\frac{1}{16\pi G_{d}}\int_{\mathcal{B}} \hspace{-1mm}d^{d}x\sqrt{-h}\biggr[R+\frac{2(d-1)(d-2)}{L_{d+1}^{2}}\left(1-\frac{4\pi G_{d+1}L_{d+1}}{(d-1)}\tau\right)\\
&+\frac{L_{d+1}^{2}}{(d-4)(d-2)}\left(R_{ij}^{2}-\frac{dR^{2}}{4(d-1)}\right)+\cdots\biggr]\,,
\end{split}
\eeq
where the ellipsis corresponds to higher curvature densities, which have thus far been efficiently computed up to quintic order in curvature for arbitrary $d$ and sextic order for $d=3$ \cite{Bueno:2022log}. Here $G_{d}$ represents the effective Newton's constant endowed from the bulk 
\beq G_{d}=\frac{d-2}{2L_{d+1}}G_{d+1} \,.\label{eq:effGd}\eeq
It is also natural to introduce another induced length scale $L_{d}$ on the brane, expressed in terms of $L_{d+1},G_{d+1}$ and $\tau$, which would represent the induced dS radius on the brane. We will do this explicitly momentarily. 
%\beq 
%\frac{1}{L_{d}^2}=\frac{2}{L_{d+1}^2}\left(1-\frac{4\pi G_{d+1}L_{d+1}}{d-1}\tau\right)\,. \label{eq:effLd}
%\eeq
 The second action $I_{\text{CFT}}$ corresponds to the finite contribution to the regulated bulk action upon integrating out the bulk. This contribution is not determined by the boundary metric and thus corresponds to the state of the $\text{CFT}_{d}$.

There are two ways to interpret the theory $I$ (\ref{eq:inductheorygen}). From the bulk perspective, $I$ characterizes a theory of a finite $(d+1)$-dimensional system with dynamics ruled by general relativity and a brane. Alternatively, from the brane perspective, $I$ represents a specific holographically induced higher-curvature theory in $d$ dimensions coupled to a CFT which backreacts on the brane metric $h_{ij}$. Notably, the CFT has a UV cutoff corresponding to the IR cutoff surface introduced in holographic regularization. Additionally, the induced theory of gravity is said to be `massive' since a massive graviton bound state will localize on the brane \cite{Karch:2000ct}, however, this mass will become negligible for a brane very near the boundary.\footnote{A construction with two branes would also have a remaining scalar mode, the radion \cite{Garriga:1999yh}, representing the displacement between the branes \cite{Garriga:1999yh,Charmousis:1999rg}. Here we consider a bulk with a single brane and thence no radion.}

By consistency, there is an equivalence between the bulk and brane viewpoints, leading to a powerful computational device: classical solutions to the bulk Einstein equations (with appropriate brane boundary conditions) exactly correspond to solutions to the semi-classical equations of motion on the brane.\footnote{It is worth emphasizing that exact bulk solutions lead to exact solutions to the entire gravity theory on the brane, including the whole tower of higher-derivative terms. Importantly, while general higher derivative theories of gravity are pathological since they are typically accompanied with ghosts, one does not expect the brane theory to be pathological (assuming one does not truncate the series of terms) since the bulk theory and the procedure of integrating out the bulk are not pathological.} Specifically, classical black holes map to quantum-corrected black holes, accounting for all orders of backreaction \cite{Emparan:2002px}. To illustrate this point, we briefly summarize a relevant construction below, the quantum $\text{SdS}_{3}$ black hole \cite{Emparan:2022ijy}.

\subsection*{The qSdS black hole}

Consider the four-dimensional $\text{AdS}_{4}$ C-metric, a solution to Einstein's equation with a negative cosmological and belongs to the general Plebanski-Demianski type-D class \cite{Plebanski:1976gy}
\beq ds^{2}=\frac{\ell^{2}}{(\ell+xr)^{2}}\left[-H(r)dt^{2}+\frac{dr^{2}}{H(r)}+r^{2}\left(\frac{dx^{2}}{G(x)}+G(x)d\phi^{2}\right)\right]\;,\label{eq:AdS4Ccoord}\eeq
with metric functions $H(r)$ and $G(x)$
\beq H(r)= 1-\frac{r^{2}}{R_{3}^{2}}-\frac{\mu\ell}{r} \;,\qquad G(x)=1-x^{2}-\mu x^{3}\;.\label{eq:Hrfunc}\eeq
Our conventions primarily follow \cite{Emparan:2020znc}, however, with $\kappa=+1$ and set $\ell_{3}^{2}=-R_{3}^{2}$ such that the brane we eventually introduce is a $\text{dS}_{3}$ brane of radius $R_{3}$.\footnote{The cases $\kappa=0$ or $\kappa=-1$ exclude the possibility of a $\text{dS}_{3}$ brane, since the roots of $H(r)$ do not represent a cosmological horizon in those cases.} The C-metric is known to describe accelerating black holes, where the real, positive  parameter $\ell$ is equal to the (inverse) acceleration. Meanwhile, $\mu>0$ is interpreted to be a mass parameter of the four-dimensional black hole. The $\text{AdS}_{4}$ length scale $L_{4}$ is related to the parameters $R_{3}$ and $\ell$ via 
\beq L_{4}^{-2}=\ell^{-2}\left[1-\left(\frac{\ell}{R_{3}}\right)^{2}\right]\;.\label{eq:L4bulk}\eeq
For $L_{4}^{2}>0$ such that the bulk cosmological constant is negative, we require $R_{3}^{2}>\ell^{2}$.

\begin{figure}[t!]
\begin{center}
\includegraphics[width=.2\textwidth, height=.32\textwidth]{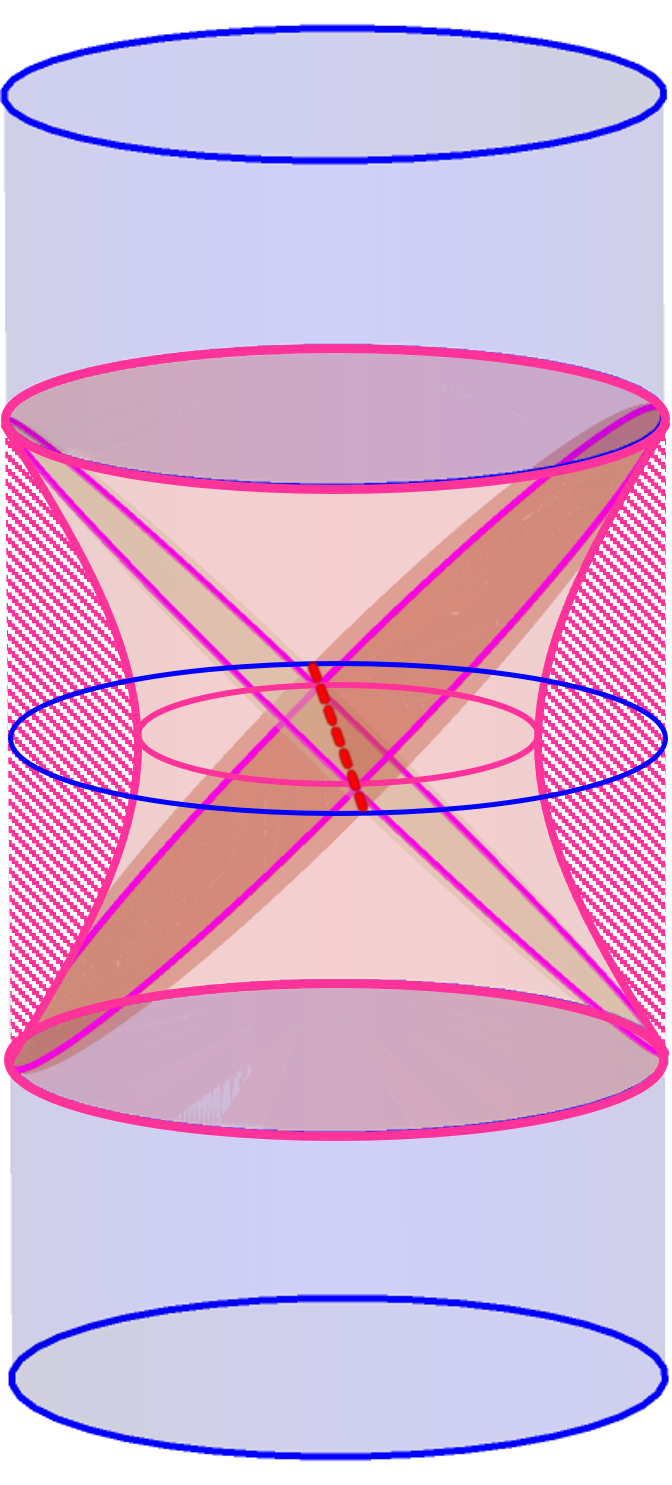} $\quad\quad\quad\quad$ \includegraphics[width=.3\textwidth]{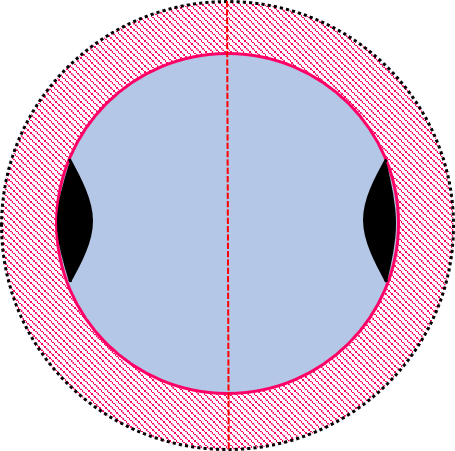}
\end{center}
\caption{\small \textbf{Left:} Bulk $\text{AdS}_{4}$ with a de Sitter$_3$ brane. The brane is represented as a (magenta) hyperboloid. The bulk region up to the brane ($x<0$, dashed magenta region) is excluded. To complete the construction, we glue a second copy along the two-sided brane. Cosmological horizons on the dS brane corresponds to the bulk acceleration horizons intersecting the brane (red dashed line). \textbf{Right:} Constant $t$-time slice of a single $\text{AdS}_{4}$ cylinder with a de Sitter brane (thick red circle) containing black holes. The coordinates cover only half of the disk, containing only a single black hole and cosmological horizon, where the other half is obtained via an appropriate analytic continuation.}
\label{fig:dSbranebhs} 
\end{figure}

Following the construction of \cite{Emparan:1999wa,Emparan:1999fd}, a Randall-Sundrum brane with tension $\tau$ and action (\ref{eq:braneaction}) is placed at the umbilic $x=0$ surface, resulting in a tension
\beq \tau=\frac{1}{2\pi G_{4}\ell}\;.\label{eq:branetens}\eeq
The tension may be read off from the Israel junction conditions which determine the location of the brane, such that tuning the tension corresponds to changing the position of the brane. We review this construction more carefully in Appendix \ref{app:AdSCmetprops}. Further, recall that the brane effectively cuts off the bulk space. For a dS braneworld, we keep only the $x>0$ portion of the bulk, eliminating all but one of the roots of $G(x)$, which we denote as $x_{1}$. This root corresponds to an axis for the rotational Killing symmetry $\partial_{\phi}$ resulting in a conical singularity at $x=x_{1}$,\footnote{In fact, the vector $\partial_{\phi}$ will have vanishing norm on the locus $x=x_{i}$ for each zero $x_{i}$ of $G(x)$.} and is removed via the identification
\beq \phi\sim \phi+\Delta\phi \;,\quad \Delta\phi=\frac{4\pi}{|G'(x_{1})|}=\frac{4\pi x_{1}}{3-x_{1}^{2}}\;.\label{eq:conedefidsds}\eeq
To complete the space, we perform surgery by cutting the bulk at $x=0$, keeping only the range $0\leq x\leq x_{1}$, where there are no conical singularities, and glue a second copy along $\mathcal{B}$ to complete the space. See Figure \ref{fig:dSbranebhs} for an illustration.

The geometry induced on the brane at $x=0$ will result in a metric in $(t,r,\phi)$-coordinates which has a conical deficit angle due to the identification (\ref{eq:conedefidsds}). To respect regularity in the bulk, one thus rescales coordinates $(t,r,\phi)\to(\bar{t},\bar{r},\bar{\phi})$, where $t=\eta\bar{t}$, $r=\eta^{-1}\bar{r}$, and $\phi=\eta\bar{\phi}$, and $2\pi\eta\equiv\Delta\phi$, such that $\bar{\phi}$ is periodic in $2\pi$, and results in the geometry \cite{Emparan:2022ijy}
\beq ds^{2}_{\text{qSdS}}=-H(\bar{r})d\bar{t}^{2}+H^{-1}(\bar{r})d\bar{r}^{2}+\bar{r}^{2}d\bar{t}^{2}\;,\quad H(\bar{r})=1-8\mathcal{G}_{3}M-\frac{\ell F(M)}{\bar{r}}-\frac{\bar{r}^{2}}{R_{3}^{2}}\;.\label{eq:metqsds}\eeq
This is the three-dimensional quantum Schwarzschild-de Sitter black hole. Depending on the size of $\ell F(M)$, there exist two positive roots to $H(\bar{r})=0$, denoted $\bar{r}_{+}$ and $\bar{r}_{c}$, the black hole and cosmological horizon, respectively. From the bulk perspective, the cosmological horizon on the brane arises due to the brane intersecting the acceleration horizon of the bulk black hole. Moreover, here the mass $M$ of the brane black hole and the form factor $F(M)$ are
\beq 8\mathcal{G}_{3}M\equiv 1-\frac{4x_{1}^{2}}{(3-x_{1}^{2})^{2}}\;,\quad F(M)\equiv\frac{8(1-x_{1}^{2})}{(3-x_{1}^{2})^{3}}\;,\label{eq:MformMsds}\eeq
with renormalized Newton's constant $\mathcal{G}_{3}\equiv L_{4} G_{3}/\ell$. To arrive at the expression for $F(M)$, the parameter $\mu$ is treated as being ``derived'' from $G(x_{1})=0$, such that $x_{1}\in(0,1]$, where $x_{1}=1$ coincides with $\mu=0$ \cite{Emparan:2020znc}.

We can think of the metric (\ref{eq:metqsds}) as a semi-classical black hole because it is an exact solution to the holographically induced theory of gravity
\beq I=\frac{1}{16\pi G_{3}}\int_{\mathcal{B}}\hspace{-1mm}d^{3}x\sqrt{-h}\left[R-\frac{2}{L_{3}^{2}}-L_{4}^{2}\left(R_{ij}^{2}-\frac{3}{8}R^{2}\right)+...\right]+I_{\text{CFT}}\;,\label{eq:indthebrane3D}\eeq
with semi-classical equations of motion
\begin{align} \label{eq:semiclasseom}
&8\pi G_{3}\langle T^{\text{CFT}}_{ij}\rangle=G_{ij}+\frac{h_{ij}}{L_{3}^{2}}\\
&+\ell^{2}\biggr[4R_{i}^{\;k}\tilde{R}_{jk}-\frac{9}{4}RR_{ij}-\Box R_{ij}+\frac{1}{4}\nabla_{i}\nabla_{j}R+\frac{1}{2}h_{ij}\left(\frac{13}{8}R^{2}-3R_{kl}^{2}+\frac{1}{2}\Box R\right)\biggr]+ ...\;. \nonumber
\end{align}
The CFT stress-energy tensor sources the effective three-dimensional gravity theory such that backreaction is accounted for by $\langle T_{ij}^{\text{CFT}}\rangle$. Here we work in the limit where the effective three-dimensional theory obeys $L_{4}\ll L_{3}$, or, equivalently, $\ell\sim L_{4}$ such that $\ell/R_{3}$ is taken to be a small expansion parameter. Thus, the higher curvature terms in the action are multiplied by higher powers of $\ell$, where, from the brane perspective, $\ell$ captures the strength of backreaction. Hence, the higher-derivative corrections can be understood as a series of corrections due to semi-classical backreaction. In the limit of small backreaction, moreover, $L_{3}^{2}\approx R_{3}^{2}$ while the central charge $c\equiv L_{4}^{2}/G_{4}$ of the $\text{CFT}_{3}$ satisfies $2cG_{3}=L_{4}\approx\ell$. Therefore, for fixed $c$, gravity becomes weak on the brane as $\ell\to0$ such that there is no backreaction due to the CFT.\footnote{The limit of vanishing backreaction looks singular from the bulk perspective, since keeping $R_3$ finite would then require $L_4\to 0$. Instead take the limit $L_4\to 0$ while rescaling the bulk metric by a factor $L_4^2$,  then the brane is pushed to the boundary and gravitational dynamics on the brane is turned off, while still keeping a non-trivial state of the non-backreacting CFT$_3$ \cite{Emparan:2022ijy}.} Lastly,  $\ell\approx 2cL_{\text{P}}$, where $L_{\text{P}}=G_{3}$ is the Planck length (since we set $\hbar=1$).

Returning to the qSdS geometry (\ref{eq:metqsds}), we see that the $\ell F(M)/\bar{r}$ contribution characterizes quantum-corrections to the classical $\text{SdS}_{3}$ solution. Since $\ell\approx 2cL_{\text{P}}$ and $c\gg1$, as required by holography, the qSdS is not a Planck-sized black hole, but rather has a horizon much larger than the Planck length. Further, the renormalized Newton's constant $\mathcal{G}_{3}$ (\ref{eq:MformMsds}) takes into account the modification of the definition of mass due to the higher curvature corrections \cite{Emparan:2020znc}. Finally, we emphasize that analyzing the semi-classical Einstein's equations for a free conformally coupled scalar results in the metric (\ref{eq:metqsds}) to leading order in $L_{\text{P}}$ \cite{Emparan:2022ijy}.

%%%%%%%%%%%%%%%%%%%%%%%%%%%%%%%%%%%%%%%%%%%%%%%%%%%%%%%%%%%%%%%%%
\section{Geometry of quantum Kerr-$\text{dS}_{3}$ black holes} \label{sec:GeomKdS3}

As reviewed above, braneworld holography grants us the ability to study the problem of semi-classical backreaction without having to explicitly solve semi-classical equations of motion. By a judicious choice of a bulk spacetime, the bulk black hole localizes on the brane and leads to the qSdS on the brane -- the first known example of an exact quantum de Sitter black hole in the sense that the solution encodes all orders of semi-classical backreaction. Here we use braneworld holography to uncover the quantum Kerr-de Sitter black hole, focusing primarily on the geometry, leaving the thermodynamic analysis for the subsequent section.

\subsection{Bulk and brane geometry}

Analogous to the rotating quantum BTZ black hole \cite{Emparan:2020znc}, our starting point is the rotating $\text{AdS}_{4}$ C-metric, describing accelerating Kerr-$\text{AdS}_{4}$ black holes and has the line element
\beq
\begin{split}
\hspace{-1.8mm}ds^{2}&=\omega^{2}\biggr(\hspace{-1mm}-\frac{H(r)}{\Sigma}(dt-ax^{2}d\phi)^{2}+\frac{\Sigma}{H(r)}dr^{2}+r^{2}\left[\frac{\Sigma}{G(x)}dx^{2}+\frac{G(x)}{\Sigma}\left(d\phi+\frac{a}{r^{2}}dt\right)^{\hspace{-1mm}2}\right]\biggr)
\end{split}
\label{eq:rotCmet}\eeq
with metric functions
\beq H(r)=1-\frac{r^{2}}{R_{3}^{2}}-\frac{\mu\ell}{r}+\frac{a^{2}}{r^{2}}\;,\quad G(x)=1-x^{2}-\mu x^{3}-\frac{a^{2}}{R_{3}^{2}}x^{4}\;,\eeq
\beq \omega^{2}=\frac{\ell^{2}}{(\ell+xr)^{2}}\;,\quad \Sigma=1+\frac{a^{2}x^{2}}{r^{2}}\;.\eeq
Our conventions largely follow \cite{Emparan:2020znc}, apart from the substitutions $R_{3}^{2}=-\ell_{3}^{2}$ (or $\ell_{3}=iR_{3}$) and $a\to-a$. Here $a$ is a parameter encoding the rotation of the bulk black hole (the angular momentum per unit mass), and in the limit $a=0$ we recover the static C-metric (\ref{eq:AdS4Ccoord}). Evaluating the bulk Kretschmann scalar invariant $\hat{R}^{abcd}\hat{R}_{abcd}$, there is a curvature singularity when $r^{2}\Sigma=r^{2}+a^{2}x^{2}=0$, i.e., when both $r=0$ and $x=0$. This is the familiar ring singularity in Kerr black holes.\footnote{This is clarified when one moves to coordinates where $x=\cos\theta$, such that the singularity lies at the $\theta=\pi/2$ edge of the $r=0$ disk.}

Despite rotation, the $x=0$ hypersurface remains umbilic, obeying $K_{ij}=-\ell^{-1}h_{ij}$, and is thus a natural location to place the de Sitter brane. The geometry on the brane is\footnote{These coordinates are Boyer-Lindquist-like. To see this, perform the successive coordinate transformations $t\to\hat{t}+a\phi$ and $\phi\to-\hat{\phi}/(1+a^{2}/R_{3}^{2})$ on (\ref{eq:KdSnaivemet}). The resulting geometry is the $(\hat{t},\hat{\phi},r)$ submanifold of the four-dimensional Kerr-dS metric in Boyer-Lindquist coordinates; \emph{e.g.}, Eq. (2.2) of \cite{Anninos:2009yc} with $\theta=\pi/2$, $\mu\ell=2M$, and where our blackening factor $H(r)$ is different by a shift in $a^{2}/R_{3}^{2}$.}
\beq ds^{2}|_{x=0}=-H(r)dt^{2}+H^{-1}(r)dr^{2}+r^{2}\left(d\phi+\frac{a}{r^{2}}dt\right)^{\hspace{-1mm}2}\;.\label{eq:KdSnaivemet}\eeq
Since the rotating C-metric (\ref{eq:rotCmet}) is a solution to the bulk Einstein equations, we are guaranteed the brane geometry is a solution to the induced theory of gravity (\ref{eq:indthebrane3D}). However, at this stage it would be naive to interpret this solution as the quantum Kerr-$\text{dS}_{3}$ black hole. This is because we have not yet accounted for bulk regularity conditions, which will affect more than just the periodicity of the angular variable $\phi$. In fact, we know the `naive metric' (\ref{eq:KdSnaivemet}) does not capture all of the correct features because the ring singularity lives on the brane, yet the above metric does not have a ring singularity at $r=0$ but rather a standard curvature singularity. We will see momentarily how the ring singularity makes an appearance. 

\subsubsection*{Bulk regularity}

Notice that the Killing vector $\partial_{\phi}$ no longer has vanishing norm at a zero $x_{i}$ of $G(x)$. Rather, the Killing vector 
\beq \xi^{b}=\partial^{b}_{\phi}+ax_{i}^{2}\partial^{b}_{t}\;,\label{eq:phitvec}\eeq
obeys $\xi^{2}|_{x_{i}}=0$. Avoiding conical defects at $x=x_{i}$ requires us to identify points along the integral curves of the vector (\ref{eq:phitvec}) with an appropriate period. To determine the correct periodicity, consider the rotating C-metric (\ref{eq:rotCmet}) near a zero $x=x_{i}$ such that $G(x)\sim G'(x_{i})(x-x_{i})$ (see Appendix \ref{app:AdSCmetprops}). Removal of a conical singularity at $x=x_{i}$ requires one simultaneously perform the coordinate transformation $\tilde{t}=t-ax_{i}^{2}\phi$ together with the same periodicity condition on $\phi$ (\ref{eq:conedefidsds}). Specifically, singling out the smallest positive root $x=x_{1}$, then 
\beq \phi\sim\phi+\Delta\phi\;,\quad \Delta\phi=\frac{4\pi}{|G'(x_{1})|}=\frac{4\pi x_{1}}{3-x_{1}^{2}+\tilde{a}^{2}}\;,\label{eq:periodx1KdS}\eeq
where to arrive to the second equality we recast the parameter $\mu$ in terms of $x_{1}$
\beq \mu=\frac{1-x_{1}^{2}-\tilde{a}^{2}}{x_{1}^{3}}\;,\quad \tilde{a}\equiv\frac{ax_{1}^{2}}{R_{3}}\;.\label{eq:muatdef}\eeq
Thus, identifying points along the orbits of $\xi^{b}$ are made on surfaces of constant 
\beq \tilde{t}\equiv t-ax_{1}^{2}\phi\;.\label{eq:tildet}\eeq
The remaining zeros $x_{i}\neq x_{1}$ are dealt with by cutting off the bulk spacetime at $x=0$, and gluing to a second region such that the complete space is comprised of a bulk region with $x\in[0,x_{1}]$, leaving a space which is free of conical singularities at $x=x_{i}$. 

%Similarly, in the static case roots of $H(r)$ correspond to the Killing horizons of the Killing vector $\partial_{t}$, whereas with rotation,  the Killing vector 
%\beq \zeta^{b}=\partial_{t}-\frac{a}{r_{i}^{2}}\partial_{\phi}\;\label{eq:zetavecv1}\eeq
%becomes null at roots $r_{i}$ of $H(r)$. Thus, the roots $r_{i}$ of $H(r)$ correspond to rotating horizons with angular velocity $\Omega=-ar_{i}^{2}$. Working in coordinate frame $(\tilde{t},\phi)$, as required for bulk regularity, the angular velocity is $\tilde{\Omega}=-a/(r_{i}^{2}+a^{2}x_{1}^{2})$ \cite{Emparan:1999fd}, such that non-zero $a$ leads to an additional root corresponding to an inner horizon. 
%We will analyze these horizons in detail momentarily.

Returning to the naive geometry at $x=0$ (\ref{eq:KdSnaivemet}), consider the asymptotic limit $r\to\infty$. The metric is asymptotic to `rotating $\text{dS}_{3}$', where the $dtd\phi$ component grows like a constant. Unfortunately, the coordinates are not canonically normalized due to the periodicity in $\phi$ (\ref{eq:periodx1KdS}). In fact, since points along orbits of (\ref{eq:phitvec}) are identified, the periodicity in $\phi$ (\ref{eq:periodx1KdS}) returns one to a different point in time $t$: from (\ref{eq:tildet}), we see that with $\tilde{t}\sim\tilde{t}$ then $t\sim t+2\pi\eta ax_{1}^{2}$, where $\eta\equiv \Delta\phi/2\pi$. This means we cannot just rescale coordinates $(t,r,\phi)\to(\bar{t},\bar{r},\bar{\phi})$  as done in the static case (\ref{eq:metqsds}). Additionally, the periodicity alters the asymptotic form of the metric such that the $dtd\phi$ grows as $r^{2}$, which would seem to imply a diverging angular momentum.\footnote{To see this, perform the following coordinate transformation in the brane geometry (\ref{eq:KdSnaivemet}) $t\to \tilde{t}+ax_{1}^{2}\tilde{\phi}$ and $\phi\to\tilde{\phi}$. Then, it is straightforward to show for large $r$ the $h_{\tilde{t}\tilde{\phi}}$ component of the geometry diverges as $r^{2}$.}

We can remedy the situation by changing coordinates $(t,\phi)$ to $(\tilde{t},\tilde{\phi})$ where $t=\tilde{t}+ax_{1}^{2}\tilde{\phi}$ and $\phi=\tilde{\phi}+C\tilde{t}$ for some constant $C$. In the asymptotic limit, the $\tilde{t}-\tilde{\phi}$ component of the naive brane metric (\ref{eq:KdSnaivemet}) will have go as
\beq h_{\tilde{t}\tilde{\phi}}=\left(C+\frac{ax_{1}^{2}}{R_{3}^{2}}\right)r^{2}+(a-ax_{1}^{2}+Ca^{2}x_{1}^{2})+\mathcal{O}(1/r)\;.\eeq
Judiciously, we choose $C\equiv -ax_{1}^{2}/R_{3}^{2}=-\tilde{a}/R_{3}$ to eliminate the $r^{2}$ divergence. Making this choice deals with the undesired asymptotic growth, however, $\tilde{\phi}$ is still not periodic in $2\pi$. This is now easily resolved by a simple rescaling, $\tilde{t}=\eta\bar{t}$ and $\tilde{\phi}=\eta\bar{\phi}$, such that the transformation
\beq t=\eta(\bar{t}+\tilde{a}R_{3}\bar{\phi})\;,\quad \phi=\eta\left(\bar{\phi}-\frac{\tilde{a}}{R_{3}}\bar{t}\right)\;,\label{eq:coordtbphib}\eeq
puts the brane geometry in a more canonical form.\footnote{We can recover the canonically normalized coordinates to describe rotating $\text{AdS}_{3}$ black holes \cite{Emparan:2020znc} upon the double replacement $\ell_{3}\to iR_{3}$ and $a_{\text{AdS}_{3}}\to-a_{\text{dS}_{3}}$, such that $\tilde{a}_{\text{AdS}_{3}}\to i\tilde{a}_{\text{dS}_{3}}$.}  Inverting the transformation (\ref{eq:coordtbphib}),
\beq \bar{t}=\frac{1}{\eta(1+\tilde{a}^{2})}(t-\tilde{a}R_{3}\phi)\;,\quad \bar{\phi}=\frac{1}{\eta(1+\tilde{a}^{2})}\left(\phi+\frac{\tilde{a}}{R_{3}}t\right)\;,\eeq
we see the Killing vectors $\partial_{t}$ and $\partial_{\phi}$ transform as
\beq \partial_{t}=\frac{1}{\eta(1+\tilde{a}^{2})}\left(\partial_{\bar{t}}+\frac{\tilde{a}}{R_{3}}\partial_{\bar{\phi}}\right)\;,\quad \partial_{\phi}=\frac{1}{\eta(1+\tilde{a}^{2})}\left(\partial_{\bar{\phi}}-\tilde{a}R_{3}\partial_{\bar{t}}\right)\;.\label{eq:Killvecstrans}\eeq
Consequently, now the rotational Killing vector (\ref{eq:phitvec}) is $\xi^{b}=\eta^{-1}\partial_{\bar{\phi}}$. 

With the coordinate change (\ref{eq:coordtbphib}), the brane metric does not quite have the canonical asymptotic form of a rotating de Sitter black hole. We still need to redefine the radial coordinate $r$. Following \cite{Emparan:2020znc}, let
\beq r^{2}\equiv\frac{\bar{r}^{2}-r_{s}^{2}}{(1+\tilde{a}^{2})\eta^{2}}\;,\quad  r_{s}=-\frac{R_{3}\tilde{a}\eta}{x_{1}}\sqrt{2-x_{1}^{2}}=-\frac{2\tilde{a}R_{3}\sqrt{2-x_{1}^{2}}}{3-x_{1}^{2}+\tilde{a}^{2}}\;.\label{eq:rsdef}\eeq
Altogether, the geometry on the brane in the canonically normalized coordinates $(\bar{t},\bar{r},\bar{\phi})$ is
\beq
\begin{split}
ds^{2}|_{x=0}&=-\left(\eta^{2}\left(1-\tilde{a}^{2}+\frac{4\tilde{a}^{2}}{x_{1}^{2}}\right)-\frac{\bar{r}^{2}}{R_{3}^{2}}-\frac{\mu\ell\eta^{2}}{r}\right)d\bar{t}^{2}\\
&+\left(\eta^{2}\left(1-\tilde{a}^{2}+\frac{4\tilde{a}^{2}}{x_{1}^{2}}\right)-\frac{\bar{r}^{2}}{R_{3}^{2}}-\frac{\mu\ell(1+\tilde{a}^{2})^{2}\eta^{4}r}{\bar{r}^{2}}+\frac{R_{3}^{2}\tilde{a}^{2}\mu^{2}x_{1}^{2}\eta^{4}}{\bar{r}^{2}}\right)^{-1}d\bar{r}^{2}\\
&+\left(\bar{r}^{2}+\frac{\mu\ell\tilde{a}^{2}R_{3}^{2}\eta^{2}}{r}\right)d\bar{\phi}^{2}+R_{3}\tilde{a}\mu x_{1}\eta^{2}\left(1+\frac{\ell}{x_{1}r}\right)(d\bar{\phi}d\bar{t}+d\bar{t}d\bar{\phi})\;,
\end{split}
\label{eq:branegeomv1}\eeq
where we have kept both $r$ and $\bar{r}$ when convenient.

\subsection{Black hole on the brane}

Let us now scrutinize the brane geometry (\ref{eq:branegeomv1}). First, we identify the mass $M$ as
\beq 8\mathcal{G}_{3}M\equiv 1-\eta^{2}\left(1-\tilde{a}^{2}+\frac{4\tilde{a}^{2}}{x_{1}^{2}}\right)=1-\frac{4[x_{1}^{2}-\tilde{a}^{2}(x_{1}^{2}-4)]}{(3-x_{1}^{2}+\tilde{a}^{2})^{2}}\;,\label{eq:Massid}\eeq
where $\mathcal{G}_{3}\equiv L_{4}G_{3}/\ell$ is the `renormalized' three-dimensional Newton's constant\footnote{Here we operate under the assumption that $\mathcal{G}_{3}\equiv L_{4}G_{3}/\ell$ holds to all orders in $\ell$.} \cite{Emparan:2020znc}. Since the brane theory is generally three-dimensional Einstein-de Sitter gravity plus higher-derivative corrections, we do not have a generic Komar-like mass integral in which we compute $M$. Rather, here we have identified the mass as the subleading constant term in $h_{\bar{t}\bar{t}}$, as done in Einstein-de Sitter gravity, and used $\mathcal{G}_{3}$ to encompass all of the higher-derivative corrections entering at order $\ell^{2}$ in the brane action (\ref{eq:indthebrane3D}) \cite{Cremonini:2009ih}. Similarly, we have identified the three-dimensional angular momentum $J$ to be
\beq 4\mathcal{G}_{3}J\equiv-R_{3}\tilde{a}\mu x_{1}\eta^{2}=\frac{4R_{3}\tilde{a}(x_{1}^{2}+\tilde{a}^{2}-1)}{(3-x_{1}^{2}+\tilde{a}^{2})^{2}}\;,\label{eq:angJ}\eeq
where again the renormalized Newton's constant plays the role of accounting for higher-derivative corrections to the angular momentum. 
%Additional justification for these identifications of mass (\ref{eq:Massid}) and angular momentum (\ref{eq:angJ}) will be given when we study the thermodynamics of the black hole in Section \ref{sec:ThermoqKdS}.
 Importantly, notice $M$ and $J$ depend on $\tilde{a}^{2}$ and $x_{1}^{2}$, and the parameter $\ell$ does not make an explicit appearance.

We emphasize, at this stage, the mass $M$ (\ref{eq:Massid}) and angular momentum $J$ (\ref{eq:angJ}) are identifications. Justification for this, in part, comes from the fact that these quantities satisfy a first law of thermodynamics, as we demonstrate in the next section.\footnote{An additional argument from thermodynamics, independent of the first law, is that one can directly compute the thermodynamics of the bulk black hole + braneworld system (with either dS or AdS slicing of the brane) via an on-shell Euclidean action approach. This was done in the case of non-rotating AdS C-metric with a brane of AdS or dS slicing \cite{Kudoh:2004ub} where one finds precisely the same formulae for the mass, given by a temperature derivative of the on-shell action.} Essentially, as argued in \cite{Emparan:2002px}
%in either braneworld construction -- the Karch-Randall braneworld in an AdS C-metric as in the quantum BTZ black hole \cite{Emparan:2020znc} or the Randall-Sundrum model of the  $\text{dS}_{3}$ black holes \cite{Emparan:2022ijy} and here -- 
the mass of the black hole on the brane is identified as the mass of the bulk black hole intersecting the brane. A feature distinguishing AdS and dS braneworld constructions is how the mass (\ref{eq:Massid}) coincides with a conserved charge. This is because asymptotically dS spacetimes do not have a boundary which makes providing an invariant notion of conserved charges more difficult. From the brane perspective,  one could compute conserved charges, for example, by calculating the Brown-York quasi-local stress tensor on slices at past and future infinity \cite{Balasubramanian:2001nb}. The mass found should then coincide with the mass of the bulk black hole intersecting the brane at $I^{\pm}$. In practice this is difficult, however,  because the theory on the brane is a complicated higher-order theory of gravity, a context in which defining conserved charges is also a subtle matter (AdS braneworld models encounter the same subtlety in this regard). Alternatively, one can use the method developed in \cite{Dolan:2018hpl}, which does not require entering an asymptotic region. It would be worthwhile to explore this question and verify the mass identified in the first law coincides with an invariant conserved charge.

%From the bulk perspective, the brane has removed a portion of the bulk AdS boundary between past and future infinity, $I{\pm}$. 

With the substitutions (\ref{eq:Massid}) and $J$ (\ref{eq:angJ})s, the brane geometry (\ref{eq:branegeomv1}) takes the form
%\footnote{The coordinate system $(\bar{t},\bar{r},\bar{\phi})$ closely resembles the coordinates employed by Henneaux and Teitelboim \cite{Henneaux:1985tv} to study Kerr-Newman-AdS black holes. Explicitly, consider, \emph{e.g.}, Eq. (2) of \cite{Winstanley:2001nx} when $\theta=\pi/2$ and perform make the parameter identifications: $\ell_{\text{there}}^{2}=-R_{3}^{2}$, $r_{+}\to r_{i}x_{1}^{2}$, $(1+a^{2}/R_{3}^{2})^{2}\to\eta^{2}$, along with coordinate shifts in $\bar{\phi}$ and $\bar{t}$.}
\beq
\begin{split}
ds^{2}_{\text{qKdS}}&=-\left(1-8\mathcal{G}_{3}M-\frac{\bar{r}^{2}}{R_{3}^{2}}-\frac{\mu\ell\eta^{2}}{r}\right)d\bar{t}^{2}\\
&+\left(1-8\mathcal{G}_{3}M-\frac{\bar{r}^{2}}{R_{3}^{2}}+\frac{(4\mathcal{G}_{3}J)^{2}}{\bar{r}^{2}}-\frac{\mu\ell(1+\tilde{a}^{2})^{2}\eta^{4}r}{\bar{r}^{2}}\right)^{-1}d\bar{r}^{2}\\
&+\left(\bar{r}^{2}+\frac{\mu\ell\tilde{a}^{2}R_{3}^{2}\eta^{2}}{r}\right)d\bar{\phi}^{2}-4\mathcal{G}_{3}J\left(1+\frac{\ell}{x_{1}r}\right)(d\bar{\phi}d\bar{t}+d\bar{t}d\bar{\phi})\;.
\end{split}
\label{eq:branegeomv2}\eeq
Since the metric (\ref{eq:branegeomv2}) is an \emph{exact} solution to the full semi-classical theory of gravity on the brane (\ref{eq:semiclasseom}), we refer to the three-dimensional spacetime as the quantum Kerr-$\text{dS}_{3}$ black hole (qKdS). We say `black hole' because, as we describe below, this geometry possesses both an inner and outer black hole horizon, shrouding a ring singularity, and a cosmological horizon. We say `quantum' because it includes all orders of semi-classical backreaction due to the CFT, where terms in the metric proportional to $\mu \ell$ are understood to be quantum corrections to the classical Kerr-$\text{dS}_{3}$ conical defect. Justification of this terminology will be given when we compute the renormalized CFT stress-tensor $\langle T^{\text{CFT}}_{ij}\rangle$.

Before we analyze the brane geometry (\ref{eq:branegeomv2}) in more detail, there are a few special limits to consider. First, clearly, when the rotation $a\to0$, then $J=0$ and the geometry reduces to the static metric (\ref{eq:metqsds}), the quantum Schwarzschild-de Sitter black hole \cite{Emparan:2022ijy}. Next, in the limit of vanishing backreaction $\ell\to0$, in which the gravitational effects of the cutoff CFT are suppressed (where $\mathcal{G}_{3}\to G_{3}$), the metric (\ref{eq:branegeomv2}) takes the form of the classical Kerr-$\text{dS}_{3}$ conical defect spacetime (see Appendix \ref{app:KerrdS3}). Thirdly, when the parameter $\mu$ (\ref{eq:muatdef}) vanishes, i.e., $x_{1}=\sqrt{1-\tilde{a}^{2}}$, then both $M=J=0$, resulting in the empty $\text{dS}_{3}$ geometry. The mass $M$ will also be zero when $x_{1}=\sqrt{9-\tilde{a}^{2}}$. When this is the case, $J\neq0$ and $\mu\neq0$, 
\beq 4\mathcal{G}_{3}J=\frac{32\tilde{a}R_{3}}{(6-2\tilde{a}^{2})^{2}}\;,\quad \mu=\frac{512 \tilde{a}R_{3}}{(\tilde{a}^{2}-9)^{3}(\tilde{a}^{2}-3)^{2}}\;,\eeq
and we can think of the brane geometry as quantum rotating $\text{dS}_{3}$.

\subsubsection*{Horizons and closed timelike curves}

While the metric (\ref{eq:branegeomv2}) is in the correct canonically normalized coordinates $(\bar{t},\bar{r},\bar{\phi})$, in what follows we will perform calculations in the naive background $(t,r,\phi)$ (\ref{eq:KdSnaivemet}), and perform the appropriate coordinate transformation. This is largely done for convenience, but also because both metrics share nearly all of the same qualitative features. 

In the static case, roots of $H(r)$ correspond to the Killing horizons of the Killing vector $\partial_{t}$. With rotation,  the Killing vector 
\beq \zeta^{b}=\partial_{t}-\frac{a}{r_{i}^{2}}\partial_{\phi}\;\label{eq:zetavecv1}\eeq
becomes null at roots $r_{i}$ of $H(r)$.
%\footnote{This can be explicitly verified in using either the bulk metric or naive brane geometry.} 
Define the function $Q(r)\equiv r^{2}H(r)$. Since $Q(r)$ is a quartic polynomial in $r$, it will generally have either four, two, or zero real roots. Here we focus on the case when there are four real roots, which we will see later enforces conditions on the physical parameters $a$ and $\mu$. The three positive roots to $Q(r)$ are  the cosmological horizon $r_{c}$, the outer black hole horizon $r_{+}$ and inner black hole horizon $r_{-}$, obeying $r_{-}\leq r_{+}\leq r_{c}$. The fourth root, $r_{n}$, is negative and resides behind the singularity at $r=0$. Using $H(r_{c})=0$, and $H(r_{\pm})=0$, we can express
\beq
\begin{split}
&R_{3}^{2}=r_{c}^{2}+r_{+}^{2}+r_{c}r_{+}+r_{-}(r_{c}+r_{+}+r_{-})\;,\\
&\mu\ell=\frac{(r_{c}+r_{+})(r_{c}+r_{-})(r_{+}+r_{-})}{r_{c}^{2}+r_{+}^{2}+r_{c}r_{+}+r_{-}(r_{c}+r_{+}+r_{-})}\;,\\
&a^{2}=\frac{r_{c}r_{+}r_{-}(r_{c}+r_{+}+r_{-})}{r_{c}^{2}+r_{+}^{2}+r_{c}r_{+}+r_{-}(r_{c}+r_{+}+r_{-})}\;.
\end{split}
\label{eq:paramsintermsofri}\eeq
The blackening factor $H(r)$ factorizes as 
\beq 
\begin{split}
H(r)&=
%=\frac{1}{R_{3}^{2}r^{2}}(R_{3}^{2}r^{2}-r^{4}+a^{2}R_{3}^{2}-r\mu\ell R_{3}^{2})\\
\frac{1}{R_{3}^{2}r^{2}}(r_{c}-r)(r-r_{+})(r-r_{-})(r+r_{c}+r_{+}+r_{-})\;.
\end{split}
\eeq
The limit $r_{-}\to0$ coincides with $a=0$, while $r_{+}=r_{-}=0$ corresponds to $\mu\to0$, resulting in the Kerr-$\text{dS}_{3}$ geometry with a single cosmological horizon.

Since the black hole is stationary, the positive roots $r_{i}$ to $H(r)$ correspond to rotating horizons with rotation $\Omega_{i}$, 
\beq \Omega_{i}\equiv \frac{a}{R_{3}^{2}}\frac{(x_{1}^{2}r_{i}^{2}-R_{3}^{2})}{(r_{i}^{2}+a^{2}x_{1}^{2})}\;,\label{eq:Omi}\eeq
where we used the transformations (\ref{eq:Killvecstrans}), to express $\zeta^{b}$ and define $\bar{\zeta}^{b}$
%\beq \zeta^{b}=\frac{\left(1+\frac{a^{2}x_{1}^{2}}{r_{i}^{2}}\right)}{\eta(1+\tilde{a}^{2})}\left(\partial^{b}_{\bar{t}}+\frac{a}{R_{3}^{2}}\frac{(x_{1}^{2}r_{i}^{2}-R_{3}^{2})}{(r_{i}^{2}+a^{2}x_{1}^{2})}\partial^{b}_{\bar{\phi}}\right)\;.\eeq
\beq \bar{\zeta}^{b}\equiv\frac{\eta(1+\tilde{a}^{2})}{\left(1+\frac{a^{2}x_{1}^{2}}{r_{i}^{2}}\right)}\zeta^{b}=\partial_{\bar{t}}^{b}+\Omega_{i}\partial_{\bar{\phi}}^{b}\;.\label{eq:zetakillvec}\eeq
Further, relative to $\bar{\zeta}^{b}$, the surface gravity $\kappa_{i}$ associated with each horizon $r_{i}$ is given by 
\beq \kappa_{i}=\frac{\eta(1+\tilde{a}^{2})}{\left(r_{i}^{2}+a^{2}x_{1}^{2}\right)}\frac{r_{i}^{2}}{2}|H'(r_{i})|=\frac{\eta(1+\tilde{a}^{2})}{\left(r_{i}^{2}+a^{2}x_{1}^{2}\right)}\frac{1}{2R_{3}^{2}r_{i}}|R_{3}^{2}\mu\ell r_{i}-2r_{i}^{4}-2a^{2}R_{3}^{2}|
%=\frac{\eta(1+\tilde{a}^{2})}{2\left(r_{i}^{2}+a^{2}x_{1}^{2}\right)}\biggr|\mu\ell-\frac{2r_{i}^{3}}{R_{3}^{2}}-\frac{2a^{2}}{r_{i}}\biggr|
\;,\label{eq:surfacegravs}\eeq
where we used the definition $\bar{\zeta}^{b}\nabla_{b}\bar{\zeta}^{c}=\kappa\bar{\zeta}^{c}$. Explicitly, 
\beq 
\begin{split}
&\kappa_{c}=-\frac{\eta(1+\tilde{a}^{2})}{2R_{3}^{2}\left(r_{c}^{2}+a^{2}x_{1}^{2}\right)}(r_{c}-r_{+})(r_{c}-r_{-})(r_{+}+r_{-}+2r_{c})\;,\\
&\kappa_{+}=\frac{\eta(1+\tilde{a}^{2})}{2R_{3}^{2}\left(r_{+}^{2}+a^{2}x_{1}^{2}\right)}(r_{c}-r_{+})(r_{+}-r_{-})(r_{c}+r_{-}+2r_{+})\;,\\
&\kappa_{-}=-\frac{\eta(1+\tilde{a}^{2})}{2R_{3}^{2}\left(r_{-}^{2}+a^{2}x_{1}^{2}\right)}(r_{c}-r_{-})(r_{+}-r_{-})(r_{c}+r_{+}+2r_{-})\;.
\end{split}
\label{eq:surfacegravsv2}\eeq
Notice the cosmological horizon surface gravity $\kappa_{c}$ vanishes when $r_{c}=r_{+}$ or $r_{c}=r_{-}$, and similarly for the other surface gravities. We explore these extremal limits momentarily. When $r_{-}\to0$, i.e., vanishing rotation, we recover the surface gravities of the cosmological horizon and black hole horizon of the qSdS black hole \cite{Emparan:2022ijy}. Additionally, in the limit of vanishing backreaction, then $r_{\pm}\to0$ such that $\kappa_{\pm}\to0$. 

As mentioned previously, in the naive coordinates (\ref{eq:KdSnaivemet}), a computation of the Kretschmann scalar reveals a curvature singularity at $r=0$. In the canonically normalized coordinates (\ref{eq:branegeomv2}), $r=0$ corresponds $\bar{r}=r_{s}$, corresponding to a ring singularity, and is endowed from the bulk black hole solution. Moreover, near the ring singularity there exists the possibility of closed timelike curves. Relative to the canonically normalized metric (\ref{eq:branegeomv2}), the norm of the axial Killing vector $\partial_{\bar{\phi}}$ is 
\beq \partial_{\bar{\phi}}^{2}=h_{\bar{\phi}{\phi}}=\bar{r}^{2}+\frac{\mu\ell\tilde{a}^{2}R_{3}^{2}\eta^{2}}{r}\;.\eeq
Thus, for sufficiently small and negative $r$, the vector $\partial_{\bar{\phi}}$ becomes timelike, the orbits of which are closed curves around the rotation axis. However, unlike the rotating qBTZ black hole, these closed timelike curves do not become naked. 

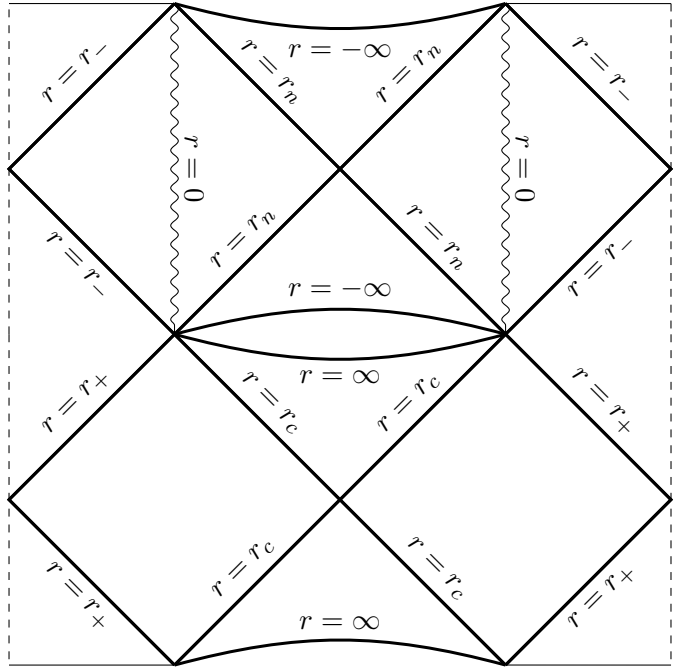
\begin{figure}[t!]
\centering
\begin{tikzpicture}[scale=1.1]
	\pgfmathsetmacro\myunit{4} 
           \draw [dashed, white]	(0,0)			coordinate (a)
		--++(90:\myunit)	coordinate (b);
	\draw [white] (b) --++(0:\myunit)		coordinate (c);
							
	\draw[dashed, white] (c) --++(-90:\myunit)	coordinate (d);
          \draw [line width = .4mm] (b)  --  node[pos=.5, above, sloped] {${\color{black} r=r_{-}}$} (-2,2) -- (a);
           \draw [line width = .4mm] (a)  --  node[pos=.5, below, sloped] {${\color{black} r=r_{-}}$} (-2,2) -- (b);
            \draw [line width = .4mm] (a)  --  node[pos=.5, above, sloped] {${\color{black} r=r_{+}}$} (-2,-2) -- (0,-4) coordinate (n3);
             \draw [line width = .4mm] (n3)  --  node[pos=.5, below, sloped] {${\color{black} r=r_{+}}$} (-2,-2) -- (a);
	\draw [line width = .4mm] (b)  --  node[pos=.5, above, sloped] {${\color{black} r=r_{n}}$} (2,2) -- (d);
 \draw [line width = .4mm] (d)  --  node[pos=.5, above, sloped] {${\color{black} r=r_{n}}$} (2,2) -- (b);
           \draw [line width = .4mm] (c) --  node[pos=.5, above, sloped] {${\color{black} r=r_{n}}$} (2,2) -- (a);
            \draw [line width = .4mm] (a) --  node[pos=.5, above, sloped] {${\color{black} r=r_{n}}$} (2,2) -- (c);
            \draw [line width = .4mm] (c) -- node[pos=.5, above, sloped] {${\color{black} r=r_{-}}$} (6,2) -- (d);
              \draw [line width = .4mm] (d) -- node[pos=.5, below, sloped] {${\color{black} r=r_{-}}$} (6,2) -- (c);
    %\draw (a) -- (-2,0) coordinate (e)  node[pos=.5, below] {$r=1\infty$};
    \draw[dashed] (-2,0) coordinate (e) -- (-2,-4) coordinate (n1);
    \draw [line width = .4mm] (a) to [out=15, in=165] node[pos=.5, above] {$r=-\infty$} (d);
    \draw [line width = .4mm] (a) to [out=-15, in=-165] node[pos=.5, below] {$r=\infty$} (d);
    \draw [dashed]  (e) -- (-2,4) coordinate (f);
    \draw (f) -- (b);   %node[pos=.5, above] {$r=4\infty$}; 
    \draw [line width = .4mm]  (b) to [out=-15, in=-165] node[pos=.5, below] {$r=-\infty$} (c);
    \draw  (c) -- (6,4) coordinate (g);   %node[pos=.5, above] {$r=3\infty$}; 
    \draw [dashed] (g) -- (6,0) coordinate (h);
     \draw[dashed] (h) -- (6,-4) coordinate (n2);
    %\draw (h) -- (d)  node[pos=.5, below] {$r=2\infty$}; 
    \draw [decorate, decoration={snake, amplitude=0.5mm, segment length=2.5mm}] (c) -- (d)   node[pos=.5, above, sloped] {$r=0$};
     \draw [decorate, decoration={snake, amplitude=0.5mm, segment length=2.5mm}] (b) -- (a)   node[pos=.5, above, sloped] {$r=0$};
     \draw (n1) -- (n3);
     \draw (n2) -- (4,-4) coordinate (n4);
      \draw [line width = .4mm] (n4)  --  node[pos=.5, below, sloped] {${\color{black} r=r_{+}}$} (6,-2) -- (d);
             \draw [line width = .4mm] (d)  --  node[pos=.5, above, sloped] {${\color{black} r=r_{+}}$} (6,-2) -- (n4);
              \draw [line width = .4mm] (a) --  node[pos=.5, above, sloped] {${\color{black} r=r_{c}}$} (2,-2) -- (n4);
              \draw [line width = .4mm] (n4) --  node[pos=.5, above, sloped] {${\color{black} r=r_{c}}$} (2,-2) -- (a);
               \draw [line width = .4mm] (n3)  --  node[pos=.5, above, sloped] {${\color{black} r=r_{c}}$} (2,-2) -- (d);
                \draw [line width = .4mm] (d)  --  node[pos=.5, above, sloped] {${\color{black} r=r_{c}}$} (2,-2) -- (n3);
                \draw [line width = .4mm] (n3) to [out=15, in=165] node[pos=.5, above] {$r=\infty$} (n4);
\end{tikzpicture}
\caption{\small Penrose diagram of a neutral quantum Kerr black hole in dS$_3$. Shown here is the global structure with periodic identifications made along constant $\bar{t}$ hypersurfaces. The diagram has infinite extent in the vertical directions while the dashed edges are identified.}
\label{fig:penKdS}
\end{figure}

When all of the roots to $Q(r)$ are distinct, then standard methods \cite{Gibbons:1977mu} lead to a maximal extension of the quantum Kerr-$\text{dS}_{3}$ black hole. Generally, the resulting conformal diagram is infinite in extent and is nearly identical to the Kruskal extension of the classical four-dimensional Kerr-dS black hole. The aforementioned closed timelike curves may be eliminated by an appropriate periodic identification \cite{Booth:1998gf}, such that constant $\bar{t}$ hypersurfaces are closed  and span two black hole regions with opposite spin, cutting through intersections of $r=r_{c}$ and $r=r_{+}$ (see Figure \ref{fig:penKdS} for a diagram).

\subsubsection*{Ergoregions}

As with classical Kerr-de Sitter spacetimes, the qKdS black hole has a stationary limit surface and two ergoregions associated with the outer black hole and cosmological horizons. Explicitly, the time-translation Killing vector $\partial_{t}$ in the naive metric has the norm $\mathcal{N}$
\beq \mathcal{N}=-H(r)+\frac{a^{2}}{r^{2}}\;.\eeq
Clearly, at the outer and cosmological horizons, where $H=0$, then $\partial_{t}$ is spacelike. The locus of points where $\mathcal{N}=0$ yields a stationary limit surface, satisfying $r(R_{3}^{2}-r^{2})=R_{3}^{2}\mu\ell$. Since there exist regions in between the outer and cosmological horizons where $\partial_{t}$ is timelike, there are two ergoregions, where an observer is forced to move in the direction of rotation of the outer black hole horizon or cosmological horizon (the black hole and cosmological ergoregions, respectively). With the appearance of ergoregions, one can in principle examine the Penrose process of energy extraction in the qKdS solution in morally the same way as a classical four-dimensional Kerr-de Sitter black hole (see, \emph{e.g.}, \cite{Bhattacharya:2017scw}). At least for small backreaction, it is expected the Penrose process in the cosmological ergoregion is not possible.

\subsection{Extremal, Nariai, ultracold, and lukewarm limits}

As with the four-dimensional Kerr-de Sitter black hole, the quantum Kerr-$\text{dS}_{3}$ has a number of limiting geometries. Specifically, (i) extremal or `cold' limit, where $r_{+}=r_{-}$; (ii) rotating Nariai limit, where $r_{c}=r_{+}$; (iii) the `ultacold' limit where $r_{c}=r_{+}=r_{-}$, and (iv) the `lukewarm' limit, where the surface gravities $\kappa_{c}=\kappa_{+}$. Below we summarize each of these limiting geometries and briefly explore their features, leaving the details to Appendix \ref{app:limitsqKdS}. Our analysis primarily follows \cite{Booth:1998gf}, and for simplicity, we work with the naive metric $(t,r,\phi)$ (\ref{eq:KdSnaivemet}) except when stated otherwise.

\subsubsection*{Extremal black hole: $r_{+}=r_{-}$}

The extremal black hole corresponds to when the outer and inner black hole horizons coincide. In this limit the surface gravity of the outer horizon $\kappa_{+}=0$, and, correspondingly the Hawking temperature $T_{+}$ of the black hole vanishes, i.e., the black hole is `cold'. Moreover, parameters $a^{2}$ and $\mu\ell$ may be cast as
\beq a^{2}=r_{+}^{2}-\frac{3r_{+}^{4}}{R_{3}^{2}}\;,\quad \mu\ell=2r_{+}-\frac{4r_{+}^{3}}{R_{3}^{2}}\;.\label{eq:amuextlim}\eeq
In the extremal limit the global structure of the spacetime changes because now the (double) black hole horizon moves to an infinite proper distance away from all other portions of the geometry, such that the black hole interior is inaccessible from the rest of the spacetime. 

In the near horizon limit of extremal qKdS, we can no longer express the metric in coordinates $(t,r,\phi)$ as they become singular. Rather, we perform a coordinate transformation analogous to \cite{Bardeen:1999px,Hartman:2008pb}
\beq r=r_{+}+\lambda\rho\;,\quad t=\frac{\tau}{\lambda}\;,\quad \phi=\varphi-\frac{a\tau}{r_{+}^{2}\lambda}\;,\eeq
where upon taking $\lambda\to0$ we find 
\beq ds^{2}_{\text{ex}}=\Gamma\left(-\hat{\rho}^{2}d\hat{\tau}^{2}+\frac{d\hat{\rho}^{2}}{\hat{\rho}^{2}}\right)+r_{+}^{2}\left(d\varphi+k\hat{\rho}d\hat{\tau}\right)^{2}\;,\eeq
with
 \beq \Gamma=\frac{r_{+}^{2}}{1-6r_{+}^{2}/R_{3}^{2}}\;,\quad k=-\frac{2aR_{3}^{2}}{r_{+}(R_{3}^{2}-6r_{+}^{2})}\;.\eeq
 This is the near horizon extremal Kerr (NHEK) geometry for the quantum-corrected Kerr-$\text{dS}_{3}$. Formally it has the same structure as the NHEK region of four-dimensional Kerr-(A)dS spacetimes, and has the form of a fibered product of $\text{AdS}_{2}$ and the circle.\footnote{Upon direct comparison, the functions $\Gamma$ and $k$ do not match those presented in \cite{Hartman:2008pb} in the $\theta=\pi/2$ limit. This is because our form of the metric (\ref{eq:KdSnaivemet}) is not exactly of Boyer-Lindquist form. Putting the naive metric in such a form would lead to the same form functions. Likewise, had we started with the metric in $(\bar{t},\bar{r},\bar{\phi})$ coordinates (\ref{eq:branegeomv2}), the near horizon extremal geometry would have the same structure with appropriately modified form functions.} As such, following \cite{Hartman:2008pb}, the isometry group is $\text{SL}(2,\mathbb{R})\times U(1)$. 

 Notice from (\ref{eq:amuextlim}) that $a=0$ when $r_{+}=0$ or $r_{+}=R_{3}/\sqrt{3}$, which, respectively, corresponds to $\mu\ell=0$ or $\mu\ell=2R_{3}/3\sqrt{3}$. The latter is simply the Nariai limit of the quantum Schwarzschild-de Sitter black hole \cite{Emparan:2022ijy}, which we explore in more detail below.

\subsubsection*{Rotating Nariai black hole: $r_{c}=r_{+}$}

The Nariai solution occurs when the cosmological and outer black hole horizons coincide $r_{c}=r_{+}\equiv r_{\text{N}}$. Then
\beq a^{2}=\frac{r_{\text{N}}^{2}}{R_{3}^{2}}(R_{3}^{2}-3r_{\text{N}}^{2})\;,\quad (\mu\ell)_{\text{N}}=\frac{2r_{\text{N}}}{R_{3}^{2}}(R_{3}^{2}-2r_{\text{N}}^{2})\;.\eeq
Notice when $a=0$ we recover $r_{\text{N}}=R_{3}/\sqrt{3}$ and $(\mu\ell)_{\text{N}}=2R_{3}/3\sqrt{3}$, the Nariai limit of the static Schwarzschild-de Sitter black hole. Physically, the Nariai black hole is the largest black hole which may fit inside the cosmological horizon, saturating at $\mu_{\text{N}}$. Moreover, the rotating Nariai black hole is generally larger than the static Nariai solution, analogous to how the charged Nariai black hole is larger than the neutral geometry. 

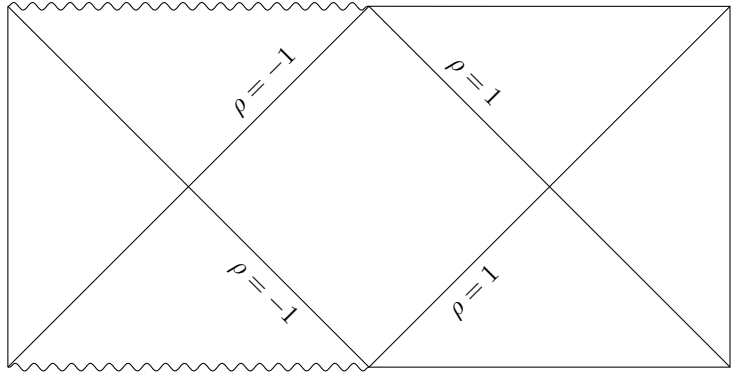
\begin{figure}[t!]
\centering
\begin{tikzpicture}[scale=1.2]
    \draw (0,0) coordinate (a) -- (0,4) coordinate (b); 
    \draw[decorate, decoration={snake, amplitude=0.5mm, segment length=2.5mm}] (b) -- (4,4) coordinate (c) ;
    \draw (c) -- (8,4) coordinate (d) -- (8,0) coordinate (e) -- (4,0) coordinate (f);
    \draw[decorate, decoration={snake, amplitude=0.5mm, segment length=2.5mm}] (f) -- (a); 
    \draw (a) -- (c) node[pos=.75, above, sloped] {{\small $\rho = -1$}} -- (e) node[pos=.25, above, sloped] {{\small $\rho = 1$}}; 
    \draw (b) -- (f) node[pos=.75, below, sloped] {{\small $\rho = -1$}} -- (d) node[pos=.25, below, sloped] {{\small $\rho = 1$}};
    %\draw[dashed] (c) -- (f) node[pos=.5, below, sloped] {{\small $\rho = \beta/2$}};
\end{tikzpicture}
\caption{Penrose diagram of the Nariai qKdS black hole. The black hole and cosmological horizons are located at $\rho=-1$ and $\rho=1$, respectively, and are in thermal equilibrium at a non-zero temperature. Clearly there is a finite proper distance between the two horizons. Future and past infinity $\mathcal{I}^{\pm}$ are located at $\rho=\infty$, while the past and future black hole singularities correspond to $\rho=-\infty$. The left and right sides of the diagram are identified.}
\label{fig:nariaiPen} 
\end{figure}

The blackening factor $H(r)$ vanishes when $r=r_{\text{N}}$ making the $(t,r,\phi)$ coordinate system incompatible in describing the Nariai geometry. Thus, introduce coordinates \cite{Booth:1998gf}
\beq r=r_{\text{N}}+\epsilon\rho\;,\quad t=\frac{\Gamma\hat{\tau}}{\epsilon}\;,\quad \phi=\varphi-\frac{a}{r_{\text{N}}^{2}\epsilon}\tau\;.\eeq
and send $\epsilon\to0$ such that the naive geometry (\ref{eq:KdSnaivemet}) becomes
\beq ds^{2}_{\text{N}}=\Gamma\left(-(1-\rho^{2})d\hat{\tau}^{2}+\frac{d\rho^{2}}{(1-\rho^{2})}\right)+r_{\text{N}}^{2}\left(d\varphi+k\rho d\hat{\tau}\right)^{2}\;,\label{eq:Nariaimet}\eeq
where
\beq \Gamma=\frac{R_{3}^{2}r_{\text{N}}^{2}}{6r_{\text{N}}^{2}-R_{3}^{2}}\;,\quad k=-\frac{2aR_{3}^{2}}{r_{\text{N}}(6r_{\text{N}}^{2}-R_{3}^{2})}\;.\eeq
Hence, the Nariai limit of the qKdS black hole has the product structure of $\text{dS}_{2}$ fibered over a circle, written here in static patch coordinates, and has the isometry group $U(1)\times \text{SL}(2,\mathbb{R})$. When $a=0$, then $\Gamma=R_{3}^{2}/3$, leading to the non-rotating Nariai metric \cite{Nariai99,Ginsparg:1982rs,Cardoso:2004uz} with product geometry $\text{dS}_{2}\times S_{2}$. A static patch observer is restricted to the region $\rho\in(-1,1)$, where $\rho=-1$ corresponds to the black hole horizon and $\rho=+1$ the cosmological horizon, at a finite proper distance apart. To draw the Penrose diagram (see Figure \ref{fig:nariaiPen}) it is useful to switch to global coordinates \cite{Anninos:2009yc}

\beq
\begin{split}
&\tan(\eta/2)=\tanh\left(\frac{1}{2}\sinh^{-1}(\sqrt{1-\rho^{2}}\sinh\hat{\tau})\right)\\
&\cos\psi=\rho(\cosh^{2}\hat{\tau}-\rho^{2}\sinh^{2}\hat{\tau})^{-1/2}\\
&\chi=\varphi+\frac{k}{2}\log\left(\frac{\sin(\eta+\psi)}{\sin(\eta-\psi)}\right)\;,
\end{split}
\eeq
such that 
\beq ds^{2}_{\text{N}}=\Gamma\left(\frac{-d\eta^{2}+d\psi^{2}}{\cos^{2}\eta}\right)+r_{\text{N}}^{2}(d\chi+k\tan\eta d\psi)^{2}\;,\eeq
where $\psi\sim\psi+2\pi$ and $\eta\in(-\pi/2,\pi/2)$ cover the all of the $\text{dS}_{2}$ portion. 

Naively, when $r_{c}=r_{+}$ the surface gravities (\ref{eq:surfacegravs}) of the cosmological  and black hole horizons vanish, $\kappa_{c}=\kappa_{+}=0$. However, in the Nariai geometry (\ref{eq:Nariaimet}), the two horizons are in thermal equilibrium at a non-zero temperature $T_{\text{N}}$.  We will return to this in Section \ref{sec:ThermoqKdS}.

\subsubsection*{Ultracold black hole: $r_{c}=r_{+}=r_{-}$}

The ultracold black hole is the limit when all of the horizons coincide, namely, $r_{c}=r_{+}=r_{-}\equiv r_{uc}$. The form of the metric can be found directly from the Nariai geometry (\ref{eq:Nariaimet}). Since the Nariai geometry becomes singular when $r_{\text{N}}=r_{-}$, the coordinates $(\rho,\hat{\tau})$ require an appropriate rescaling 
\beq \rho=\sqrt{\frac{2r_{\text{uc}}-\delta}{R_{3}}}X\;,\quad \tau=\sqrt{\frac{R_{3}}{2r_{\text{uc}}-\delta}}\frac{R_{3}r_{\text{uc}}}{4}T\;,\eeq
where $\tau=\Gamma\hat{\tau}$, and subsequently take the limit $\delta\to 2r_{\text{uc}}$. The resulting geometry is
\beq ds^{2}_{\text{uc}}=\frac{R_{3}r_{\text{uc}}}{4}(-dT^{2}+dX^{2})+r_{\text{uc}}^{2}\left(d\varphi-\frac{2aX}{r_{\text{uc}}^{3}}dT\right)^{2}\;.\eeq
This geometry is of the form of a fibered product of two-dimensional Minkowski space over a circle. Via an appropriate coordinate transformation (see \cite{Booth:1998gf}), the ultracold solution can also be expressed as a fibered product of two-dimensional Rindler space and a circle. In the limit of vanishing rotation there is no ultracold solution, but rather a static Nariai black hole.

\subsubsection*{Lukewarm black hole: $\kappa_{c}=\kappa_{+}$}

As with all Kerr-de Sitter black holes, the quantum Kerr-$\text{dS}_{3}$ has a lukewarm limit. This occurs when the surface gravities of the cosmological and outer black hole horizons coincide at a value different from the surface gravity of the Nariai black hole. Notably, the geometry is non-singular in $(t,r,\phi)$ coordinates.  Thermodynamically speaking, this spacetime is another example of where the black hole and cosmological horizon are in thermal equilibrium. We will return to this limit in Section \ref{sec:ThermoqKdS}, however, in Appendix \ref{app:limitsqKdS} we find its limiting form in the naive brane geometry.

\vspace{4mm}

Before moving on, we point out that the special limits of the quantum Kerr-$\text{dS}_{3}$ black hole  has qualitatively similar features to $\text{dS}_{3}$ black hole solutions to topologically massive gravity, cf. \cite{Nutku:1993eb,Anninos:2009jt,Anninos:2011vd}. Indeed, the asymptotically warped $\text{dS}_{3}$ black hole (obtained from discrete global identifications of warped $\text{dS}_{3}$) has a Nariai limit whose $U(1)\times U(1)$ isometry is enhanced to a $\text{SL}(2,\mathbb{R})\times U(1)$ isometry group. It would be interesting to understand the relation between quantum $\text{dS}_{3}$ black holes and warped $\text{dS}_{3}$ black holes in more detail.

\subsection{Holographic conformal matter stress-tensor}

We have been referring to the geometry on the brane (\ref{eq:branegeomv2}) as a quantum black hole since, via the holographic dictionary, it is a solution to the semi-classical equations of motion (\ref{eq:semiclasseom}) to all orders in backreaction. Let us now justify this claim and solve for the expectation value of the CFT stress-energy tensor $\langle T_{ij}\rangle$ sourcing the black hole. 

Following \cite{Emparan:2020znc}, we decompose $\langle T^{i}_{\;j}\rangle=\langle T^{i}_{\;j}\rangle_{0}+\ell^{2}\langle T^{i}_{\;j}\rangle_{2}+...$ in increasing powers of $\ell^{2}$. Specifically, the leading order contribution is 
\beq 8\pi G_{3}\langle T^{i}_{\;j}\rangle_{0}=R^{i}_{\;j}-\frac{1}{2}\delta^{i}_{\;j}\left(R-\frac{2}{R_{3}^{2}}\right)\;,\label{eq:Tab0gen}\eeq
while the $\mathcal{O}(\ell^{2})$ contribution is
\beq
\begin{split}
\hspace{-5mm} 8\pi G_{3}\langle T^{i}_{\;j}\rangle_{2}&=4R^{ik}R_{jk}-\Box R^{i}_{\;j}
-\frac{9}{4}RR^{i}_{\;j}+\frac{1}{4}\nabla^{i}\nabla_{j}R\\
&+\frac{1}{2}\delta^{i}_{\;j}\left(\frac{13}{8}R^{2}-3R_{kl}^{2}+\frac{1}{2}\Box R-\frac{1}{2R_{3}^{4}}\right)\;.
\end{split}
\label{eq:Tab2gen}\eeq
It proves is more computationally convenient to determine $\langle T_{ij}\rangle$ in the naive metric (\ref{eq:KdSnaivemet}) and then transform into the $(\bar{t},\bar{r},\bar{\phi})$ than working directly with the metric (\ref{eq:branegeomv2}). Thus, in the naive background we find the only non-vanishing components of the stress-tensor are
\beq
\begin{split}
&\langle T^{t}_{\;t}\rangle_{0}=\langle T^{r}_{\;r}\rangle_{0}=-\frac{1}{2}\langle T^{\phi}_{\;\phi}\rangle_{0}=\frac{1}{16\pi G_{3}}\frac{\mu\ell}{r^{3}}\;,\\
&\langle T^{\phi}_{\;t}\rangle_{0}=-\frac{1}{16\pi G_{3}}\frac{3\mu\ell a}{r^{5}}\;,
\end{split}
\eeq
and, for completeness,
\beq
\begin{split}
&\langle T^{t}_{\;t}\rangle_{2}=-\frac{\mu\ell}{32\pi G_{3}R_{3}^{2}r^{7}}\left[90a^{2}R_{3}^{2}-11r^{4}+R_{3}^{2}r(18r-19\mu\ell)\right]\;,\\
&\langle T^{r}_{\;r}\rangle_{2}=-\frac{\mu\ell}{32\pi G_{3}R_{3}^{2}r^{7}}\left[r^{4}+30a^{2}R_{3}^{2}+R_{3}^{2}r(6r-7\mu\ell)\right]\;,\\
&\langle T^{\phi}_{\;\phi}\rangle_{2}=-\frac{\mu\ell}{32\pi G_{3}R_{3}^{2}r^{7}}\left[10r^{4}-120a^{2}R_{3}^{2}-R_{3}^{2}r(24r-29\mu\ell)\right]\;,\\
&\langle T^{t}_{\;\phi}\rangle_{2}=-\frac{3a\mu\ell}{2\pi G_{3}r^{5}}\;,\\
&\langle T^{\phi}_{\;t}\rangle_{2}=-\frac{3a\mu\ell}{32\pi G_{3}R_{3}^{2}r^{9}}\left[23r^{4}-70a^{2}R_{3}^{2}-R_{3}^{2}r(30r-32\mu\ell)\right]\;
\end{split}
\label{eq:Tij2naive}\eeq
Notice that while $\langle T^{i}_{\;i}\rangle_{0}=0$, as one would expect for a CFT stress-tensor, we see $\langle T^{i}_{i}\rangle_{2}=-3(\mu\ell)^{2}/32\pi G_{3}r^{6}$. A non-termminating trace at higher order powers is a consequence of the fact that the CFT on the brane has an ultraviolet cutoff. 

Transforming to the $(\bar{t},\bar{r},\bar{\phi})$ coordinates
\beq \langle T^{\bar{i}}_{\;\bar{j}}\rangle=\Lambda^{\bar{i}}_{\;i}\Lambda^{j}_{\;\bar{j}}\langle T^{i}_{\;j}\rangle\;,\quad \Lambda^{\bar{i}}_{\;j}\equiv\frac{\partial\bar{x}^{i}}{\partial x^{j}}=\frac{1}{\eta(1+\tilde{a}^{2})}\begin{pmatrix}1&0&-\tilde{a}R_{3}\\ 0&r\sqrt{\frac{1+\tilde{a}^{2}}{r^{2}-r_{s}^{2}}}&0\\ -\frac{\tilde{a}}{R_{3}}&0&1\end{pmatrix}\;,\eeq
we find the leading order contribution to the stress-tensor is
\beq
\begin{split}
&\langle T^{\bar{t}}_{\;\bar{t}}\rangle_{0}=\frac{\mu\ell}{16\pi G_{3}(1+\tilde{a}^{2})r^{3}}\left(1-2\tilde{a}^{2}+\frac{3\tilde{a}^{2}R_{3}^{2}}{x_{1}^{2}r^{2}}\right)\;,\\
&\langle T^{\bar{r}}_{\;\bar{r}}\rangle_{0}=\frac{\mu\ell}{16\pi G_{3}r^{3}}\;,\\
&\langle T^{\bar{\phi}}_{\;\bar{\phi}}\rangle_{0}=-\frac{\mu\ell}{16\pi G_{3}(1+\tilde{a}^{2})r^{3}}\left(2-\tilde{a}^{2}+\frac{3\tilde{a}^{2}R_{3}^{2}}{x_{1}^{2}r^{2}}\right)\;,\\
&\langle T^{\bar{t}}_{\;\bar{\phi}}\rangle_{0}=\frac{3\mu\ell\tilde{a}R_{3}}{16\pi G_{3}(1+\tilde{a}^{2})r^{3}}\left(1+\frac{\tilde{a}^{2}R_{3}^{2}}{x_{1}^{2}r^{2}}\right)\;,\\
&\langle T^{\bar{\phi}}_{\;\bar{t}}\rangle_{0}=\frac{3\mu\ell\tilde{a}}{16\pi G_{3}(1+\tilde{a}^{2})R_{3}r^{3}}\left(1-\frac{R_{3}^{2}}{x_{1}^{2}r^{2}}\right)\;,
\end{split}
\label{eq:TijKdS0}\eeq
where recall $r$ is given in (\ref{eq:rsdef}). In what follows it suffices to only study the stress-tensor to this order and therefore we do not include the cumbersome expressions of $\langle T^{\bar{i}}_{\;\bar{j}}\rangle$ at higher orders in $\ell$. Notice these components are equivalent to the stress-tensor of the CFT in the rotating quantum BTZ black hole upon the simultaneous Wick rotations $\ell_{3}\to iR_{3}$ and $\tilde{a}\to i\tilde{a}$. 

For practical purposes, we can view the black hole as being characterized by $R_{3},x_{1}^{2}, \tilde{a}$ and $\ell$. Notably, the mass $M$ (\ref{eq:Massid}) and angular momentum $J$ (\ref{eq:angJ}) do not explicitly depend on $\ell$, they only depend on $\ell$ through the renormalized Newton's constant $\mathcal{G}_{3}$. Moreover, at least with respect to the leading order components of the stress-tensor (\ref{eq:TijKdS0}), the parameter $\ell$ only appears in the overall prefactor. Combining these two observations indicates $\langle T^{i}_{\;j}\rangle_{0}$ depends on backreaction only through $\mathcal{G}_{3}$. This is no longer the case at higher orders, however, as can be gleaned from the $\mathcal{O}(\ell^{2})$ contributions (\ref{eq:Tij2naive}). 

In the static case, the quantum SdS black hole (\ref{eq:metqsds}), the dependence of the stress-tensor on the mass was entirely captured by a single function $F(M)$ (\ref{eq:MformMsds}) \cite{Emparan:2022ijy},
\beq \langle T^{\bar{i}}_{\;\bar{j}}\rangle_{0}^{\text{qSdS}}=\frac{1}{16\pi G_{3}}\frac{\ell F(M)}{\bar{r}^{3}}\text{diag}(1,1,-2)\;.\label{eq:Tijstatic}\eeq
Unfortunately this is not possible when rotation is included: the dependence of the stress-tensor on $M$ and $J$ cannot be characterized solely by a single function $F(M,J)$. However, as in \cite{Emparan:2020znc}, we instead identify $F(M,J)$ with the leading contribution at large $\bar{r}$. Precisely, consider $\langle T^{\bar{t}}_{\;\bar{t}}\rangle_{0}$ at large $r$,
\beq \langle T^{\bar{t}}_{\bar{t}}\rangle_{0}= \frac{\mu\ell}{16\pi G_{3}\bar{r}^{3}}\sqrt{1+\tilde{a}^{2}}\eta^{3}(1-2\tilde{a}^{2})+\mathcal{O}(\bar{r}^{-5})\;,\eeq
where we used $r\approx\bar{r}/\sqrt{1+\tilde{a}^{2}}\eta$. We thus define
%\footnote{Our form factor $F(M,J)$ differs from the one presented in \cite{Emparan:2020znc} by an overall factor of $2$.}
\beq 
\begin{split}
  F(M,J)&\equiv\mu\eta^{3}\sqrt{1+\tilde{a}^{2}}(1-2\tilde{a}^{2})=\frac{8\sqrt{1+\tilde{a}^{2}}(1-2\tilde{a}^{2})}{(3-x_{1}^{2}+\tilde{a}^{2})^{3}}(1-x_{1}^{2}-\tilde{a}^{2})\;,
\end{split}
\eeq
such that for large $\bar{r}$
\beq \langle T^{\bar{t}}_{\;\bar{t}}\rangle_{0}\approx\frac{1}{16\pi G_{3}}\frac{\ell F(M,J)}{\bar{r}^{3}}\;,\eeq
and similarly for the other components of the stress-tensor (\ref{eq:TijKdS0}). Notice $F(M,J)$ will vanish when $\mu=0$ (i.e., $a^{2}=1-x_{1}^{2}$), the empty $\text{dS}_{3}$ solution, or when $\tilde{a}^{2}=1/2$, and it reduces to $F(M)$ (\ref{eq:MformMsds}) for the static solution when $\tilde{a}=0$.

It is worth repeating there are two perspectives to interpret the solution on the brane and the parameters defining the background. From the bulk viewpoint, the solution is naturally characterized by $L_{4},\ell,\mu$ and $a$. Meanwhile, from the point of view of the brane, the natural quantities parameterizing the solution include the radius $R_{3}$ fixing the scale of the brane geometry, $cG_{3}$, $\mathcal{G}_{3}M$, and $\mathcal{G}_{3}J$. The cutoff length of the three-dimensional effective theory is $L_{4}=cL_{\text{P}}$, such that for large $c$, this cutoff is much larger than the Planck length, where quantum gravity effects dominate. Thus, the `quantum' black holes constructed here, as described in the introduction, are much larger than the Planck length. Hence, our solution can be viewed as a valid solution to the problem of semi-classical backreaction.

%\subsubsection*{Stress-energy tensor in limiting geometries}

\subsubsection*{Comparison to perturbative backreaction}

It is illustrative to compare the holographic stress-tensor (\ref{eq:TijKdS0}) to the renormalized quantum stress-tensor due to perturbative backreaction of a free conformally coupled scalar field in conical Kerr-$\text{dS}_{3}$. We present the detailed computation in Appendix \ref{app:pertbackKdS}, summarizing the final result below:
\beq
\begin{split}
&\langle T^{t}_{\;t}\rangle=\frac{L_{\text{P}}}{8\pi G_{3}}\sum_{n=1}^{\infty}\frac{1}{r_{n}^{3}}\left(A_{n}+\frac{\tilde{A}_{n}}{r_{n}^{2}}\right)\;,\\
&\langle T^{r}_{\;r}\rangle=\frac{L_{\text{P}}}{16\pi G_{3}}\sum_{n=1}^{\infty}\frac{c_{n}}{r_{n}^{3}}\;,\\
&\langle T^{\phi}_{\;\phi}\rangle=-\frac{L_{\text{P}}}{8\pi G_{3}}\sum_{n=1}^{\infty}\frac{1}{r_{n}^{3}}\left(B_{n}+\frac{\tilde{A}_{n}}{r_{n}^{2}}\right)\;,\\
 &\langle T^{t}_{\;\phi}\rangle=-\frac{3R_{3}L_{\text{P}}}{8\pi G_{3}}\sum_{n=1}^{\infty}\frac{1}{r_{n}^{3}}\left(E_{n}+\frac{\tilde{E}_{n}}{r_{n}^{2}}\right)\;,\\
 &\langle T^{\phi}_{\;t}\rangle=-\frac{3 L_{\text{P}}}{8\pi G_{3}R_{3}}\sum_{n=1}^{\infty}\frac{1}{r_{n}^{3}}\left(E_{n}+\frac{F_{n}}{r_{n}^{2}}\right)\;.
\end{split}
\label{eq:compsstresstenscds3}\eeq
 Here the denominator $r_{n}\equiv \sqrt{r^{2}d_{n}^{(1)}+R_{3}^{2}d_{n}^{(2)}}$ with
\beq
\begin{split}
&d_{n}^{(1)}=\frac{16}{(\beta_{+}^{2}+\beta_{-}^{2})}\left[\sinh^{2}\left(\frac{n\pi\beta_{-}}{2}\right)+\sin^{2}\left(\frac{\pi n\beta_{+}}{2}\right)\right]\;,\\ 
&d^{(2)}_{n}=\frac{4}{(\beta_{+}^{2}+\beta_{-}^{2})}\biggr[\beta_{-}^{2}\sin^{2}\left(\frac{n\pi\beta_{+}}{2}\right)-\beta_{+}^{2}\sinh^{2}\left(\frac{\pi n\beta_{-}}{2}\right)\biggr]\;.
\end{split}
\eeq
The remaining coefficients $A_{n},\tilde{A}_{n}, c_{n}$, etc., are cumbersome to write here, but explicitly given in Appendix \ref{app:pertbackKdS} and satisfy $A_{n}+\frac{c_{n}}{2}-B_{n}=0$ . Moreover, the parameters $\beta_{+}\equiv 2r_{c}/R_{3}$ and $\beta_{-}=-8G_{3}J/r_{c}$ are related to the periodicity of coordinates in $t$ and $\phi$, respectively, where $r_{c}$ is the cosmological horizon radius.
% revealing Kerr-$\text{dS}_{3}$ is a conical defect geometry with cosmological horizon radius $r_{c}$. 
The infinite sum arises from using the method of images to determine the appropriate Green function solving the scalar equation of motion, where, unlike the Schwarzschild-$\text{dS}_{3}$ case \cite{Emparan:2022ijy}, there are a countably infinite number of distinct images. 

Comparing to the holographic stress-tensor (\ref{eq:TijKdS0}), we notice the tensor components share a similar structure. In particular, coefficients aside, the two sets of tensors have a comparable radial dependence, comparing the $\bar{r}$ dependence in (\ref{eq:TijKdS0}) and $r_{n}$ above. Of course, once the infinite sums are performed, the radial dependence in (\ref{eq:compsstresstenscds3}) is sufficiently more complicated than its holographic counterpart. Likewise, substituting in the explicit expressions of $\beta_{\pm}$ results in expressions with cumbersome dependence on $M$ and $J$. This is in contrast to the static case explored in \cite{Emparan:2022ijy}, where the radial dependence in either the holographic or perturbative methods was the same, going as $1/\bar{r}^{3}$ (\ref{eq:Tijstatic}). In summary, due to the complicated radial dependence, with non-zero rotation the result of a holographic CFT backreacting on the geometry is far simpler than that of a single conformally coupled scalar field. Indeed, the holographic stress-tensor (\ref{eq:TijKdS0}) is clearly non-singular everywhere outside of the ring singularity at $r=0$. This is far less obvious looking at the perturbative stress-tensor. 

Moreover, the complicated radial dependence in the perturbative backreaction (\ref{eq:compsstresstenscds3}) lead to far more complicated quantum corrections to the Kerr-$\text{dS}_{3}$ geometry, a result from solving the three-dimensional semi-classical Einstein equations 
\beq G_{\mu\nu}+\frac{1}{R_{3}^{2}}g_{\mu\nu}=8\pi G_{3}\langle T_{\mu\nu}\rangle\eeq
perturbatively in $L_{\text{P}}$. Leaving the details to Appendix \ref{app:pertbackKdS}, we expand the metric ansatz
\beq
ds^2 = N(r)^2 f(r) dt^2+ \frac{dr^2}{f(r)}+r^2(d\theta+k(r) dt)^2 \ 
\eeq
to linear order in $L_{\text{P}}$ such that
\beq
\begin{split}
N(r) = N_0(r) + L_{\text{P}} N_1(r)\;,\quad f(r) = f_0(r) + + L_{\text{P}} f_1(r)\;,\quad k(r) = k_0(r)+ L_{\text{P}} k_1(r)\;.
\end{split}
\eeq
At $\mathcal{O}(L^{0}_{\text{P}})$ we recover the classical Kerr-$\text{dS}_{3}$ geometry, while perturbatively solving the semi-classical Einstein equations yields
\beq
N_1(r) = \frac{R_3^2}{2(\beta_+^2 + \beta_-^2)}\sum^\infty_{n=1} \frac{a_n c_n-2\beta_+ \beta_- e_n}{b_n r_{n}^{3}} \ , 
\label{eq:N1func}\eeq
\beq
f_1(r) = \sum^\infty_{n=1} \frac{4h_{n}(r) (a_n c_n - 2\beta_+\beta_- e_n)-c_n r_{n}^{4} (\beta_+^2 + \beta_-^2)^{3}}{64r^2(\beta_+^2 + \beta_-^2)b_n^2 r_{n}^{3}} \ ,
\eeq
\beq
k_1(r) = -\frac{R_3}{8r^2}\sum^\infty_{n=1} \frac{(\beta_+^2-\beta_-^2)e_n+\beta_+ \beta_- c_n(c_n-4)}{b_n^2 r_{n}} \ .
\label{eq:k1func}\eeq
with coefficients $a_{n},b_{n}$ etc. are presented in Appendix \ref{app:pertbackKdS}. Clearly, the terms to linear order in $L_{\text{P}}$ are more cumbersome than the quantum corrected geometry due to the holographic stress-tensor. Since $f_1 \sim 1/r$ as $r\to \infty$, the correction to the blackening factor does resemble the 4D Schwarzschild-like contribution that emerges from the holographic calculations. However, as this derivation can only be accomplished to linear order in $L_{\text{P}}$, a limit of the perturbative approach, we cannot justify these quantum corrections induce a black hole horizon; one must consider higher order corrections. It is also  worth emphasizing that the backreacted geometry due to a single free field  would have lead to a black hole horizon on the scale of the Planck length, while the holographic quantum black hole horizon is of size $\ell\sim cL_{\text{P}}\gg L_{\text{P}}$.

%%%%%%%%%%%%%%%%%%%%%%%%%%%%%%%%%%%%%%%%%%%%%%%%%%%%%%%%%%%%%%%%%
\section{Thermodynamics of quantum Kerr-$\text{dS}_{3}$ black holes} \label{sec:ThermoqKdS}

Here we analyze the thermodynamics of the quantum Kerr-dS black hole. As with the geometry, there are two perspectives to view the thermodynamics of the system: the thermodynamics of the classical bulk black hole, and the thermodynamics of the quantum black hole on the brane. Due to the holographic construction, the formulae we derive in either perspective appear the same, however, with conceptually different interpretations. Since the parent solution is well understood, we begin with the thermodynamics of the bulk.

\subsection{Bulk thermodynamics}

The C-metric (\ref{eq:rotCmet}) is known to describe a uniformly accelerating black hole or a pair of such black holes, whose acceleration is mediated by a cosmic string. Since the bulk black hole is accelerating it is natural to wonder whether it is sensible to study the thermodynamics of accelerating black holes. It is worth emphasizing that while the black hole is accelerating, it is nonetheless stationary, having a time-translation Killing symmetry $\partial_{t}$.\footnote{We point out that $\partial_{t}$ is not globally timelike. Rather, it is timelike in the region between the acceleration horizon and the outer black hole horizon. The bulk thermodynamics we describe correspond to this region. Alternatively, $\partial_{t}$ becomes spacelike in the regions between the inner and outer black hole horizons and acceleration horizon and null infinity. In those regions the roles of coordinates $(t,r)$ are switched.} Moreover, the black hole(s) are held fixed at a proper distance away from the acceleration horizon. Consequently, the black hole has a sensible thermodynamic interpretation (see, e.g., \cite{Appels:2016uha}), having a well-defined temperature and entropy.
%\footnote{The confusing aspect of thermodynamics in C-metrics is that the system also has an acceleration horizon -- due to the cosmic string suspending the black hole off center, analogous to an accelerating observer's Rindler horizon -- which raises the question of thermal equilibrium. We will not need to concern ourselves with this consideration, however, see \cite{Appels:2016uha} for more detail.}

When analyzing the thermodynamics, it is useful to introduce the parameters \cite{Emparan:2020znc}
\beq z\equiv\frac{R_{3}}{r_{i}x_{1}}\;,\quad \nu\equiv\frac{\ell}{R_{3}}\;,\quad \alpha\equiv\frac{ax_{1}}{R_{3}}=\frac{\tilde{a}}{x_{1}}\;,\label{eq:paramsforKdS}\eeq
where $r_{i}$ is a positive real root of the bulk blackening factor $H(r)$, representing each horizon of braneworld black hole. We can express $x_{1}$, $\mu$ and $r_{i}$ solely in terms of these parameters,
\beq 
\begin{split}
&x_{1}^{2}=\frac{1+\nu z^{3}}{z^{2}[1+\nu z+\alpha^{2}z(z+\nu)]}\;,\\
&r_{i}^{2}=R_{3}^{2}\frac{1+\nu z+\alpha^{2}z(z+\nu)}{1+\nu z^{3}}\;,\\
&\mu x_{1}=\frac{(z^{2}-1)(1+\alpha^{2}(1+z^{2}))}{1+\nu z^{3}}\;.
\end{split}
\label{eq:paramssub}\eeq
The first expression is found by solving $H(r_{i})=0$ for $x_{1}^{2}$, from which the other two relations readily follow.\footnote{Via the reassignments $\ell_{3}^{2}\to-R_{3}^{2}$ and $a\to-a$, we recover the relevant parameters of the quantum BTZ via $z^{2}\to -z^{2}$, $\nu^{2}\to-\nu^{2}$ and $\nu z\to \nu z$.} Moreover, the bare and renormalized Newton's constants are
\beq G_{4}=2L_{4}G_{3}=\frac{2G_{3}\ell}{\sqrt{1-\nu^{2}}}\;,\quad \mathcal{G}_{3}=\frac{L_{4}}{\ell}G_{3}=\frac{G_{3}}{\sqrt{1-\nu^{2}}}\;.\label{eq:newtonsconsts}\eeq
The limit of vanishing backreaction now coincides with small $\nu$, and we take $\nu^{2}<1$, which guarantees the bulk is asymptotically $\text{AdS}_{4}$. Using the parameters (\ref{eq:paramssub}), we can recast the mass $M$ (\ref{eq:Massid}) and angular momentum $J$ (\ref{eq:angJ}) 
\beq M=\frac{1}{8G_{3}}\sqrt{1-\nu^{2}}\frac{(z^{2}-1)[1+\alpha^{2}(1+z^{2})][9z^{2}-1+8\nu z^{3}+\alpha^{2}(9z^{4}-1+8\nu z^{3})]}{(3z^{2}-1+2\nu z^{3}+\alpha^{2}(1+4\nu z^{3}+3z^{4}))^{2}}\;,\label{eq:massinnuz}\eeq
\beq J=\frac{\alpha R_{3}}{G_{3}}\sqrt{1-\nu^{2}}\frac{z(z^{2}-1)[1+\alpha^{2}(1+z^{2})]\sqrt{(1+\nu z^{3})(1+\nu z+\alpha^{2}z(z+\nu))}}{(3z^{2}-1+2\nu z^{3}+\alpha^{2}(1+4\nu z^{3}+3z^{4}))^{2}}\;.\label{eq:Jinnuz}\eeq

%In the limit of vanishing rotation ($J=\alpha=0$), we recover the mass of the quantum SdS black hole \cite{Emparan:2022ijy}. 

As described in the previous section, the canonically normalized Killing vector $\bar{\zeta}^{b}=\partial_{\bar{t}}^{b}+\Omega_{i}\partial_{\bar{\phi}}$ (\ref{eq:zetakillvec}) generates rotating horizons at the positive roots $r_{i}$ with rotation $\Omega_{i}$ (\ref{eq:Omi}), now expressed as
\beq
\begin{split}
  \Omega_{i}&=\frac{\alpha}{R_{3}}\frac{(z^{2}-1)\sqrt{(1+\nu z^{3})(1+\nu z+\alpha^{2}z(z+\nu))}}{z(1+\nu z)(1+\alpha^{2}(1+z^{2}))}\;.
\end{split}
\label{eq:Ominuz}\eeq
Additionally, the surface gravity $\kappa_{i}$ (\ref{eq:surfacegravs}) relative to $\bar{\zeta}^{b}$ yields a temperature $T_{i}=\kappa_{i}/2\pi$,
\beq
\begin{split}
%&T_{i}=\frac{1}{2\pi R_{3}}\frac{(z^{2}(1+\nu z)+\alpha^{2}(1+2\nu z^{3}+z^{4}))}{z(1+\nu z)(1+\alpha^{2}(1+z^{2}))(1+\nu z+\alpha^{2}z(z+\nu))(3z^{2}-1+2\nu z^{3}+\alpha^{2}(1+3z^{4}+4\nu z^{3}))}\\
%&\times |(1+\nu z+\alpha^{2}z(z+\nu))(2+3\nu z-\nu z^{3}+\alpha^{2}[4z^{2}+\nu z(z^{4}+3)])|\;.
&\hspace{-2mm} T_{i}=\frac{1}{2\pi R_{3}}\frac{(z^{2}(1+\nu z)+\alpha^{2}(1+2\nu z^{3}+z^{4}))|(2+3\nu z-\nu z^{3}+\alpha^{2}(4z^{2}+\nu z(z^{4}+3)))|}{z(1+\nu z)(1+\alpha^{2}(1+z^{2}))(3z^{2}-1+2\nu z^{3}+\alpha^{2}(1+3z^{4}+4\nu z^{3}))}\;.
\end{split}
\label{eq:Titemps}\eeq
We will deal with absolute value more carefully in the next section.

Lastly, the bulk horizon entropy is given by the Bekenstein-Hawking area formula
\beq
\begin{split}
S^{(4)}_{\text{BH}}&=\frac{\text{Area}(r_{i})}{4G_{4}}=\frac{2}{4G_{4}}\int^{2\pi}_{0}d\bar{\phi}\int^{x_{1}}_{0}dx\frac{r_{i}^{2}\ell^{2}}{(\ell+r_{i}x)^{2}}\eta\left(1+\frac{a^{2}x^{2}_{1}}{r_{i}^{2}}\right)\\
&=\frac{\pi}{G_{4}}\frac{\eta\ell x_{1}(r_{i}^{2}+a^{2}x_{1}^{2})}{(\ell+r_{i}x_{1})}\\
&=\frac{\pi R_{3}}{G_{3}}\frac{\sqrt{1-\nu^{2}}z(1+\alpha^{2}(1+z^{2}))}{(3z^{2}-1+2\nu z^{3}+\alpha^{2}(1+3z^{4}+4\nu z^{3}))}\;.
\end{split}
\label{eq:bulkent}\eeq

Altogether, the mass (\ref{eq:massinnuz}), angular momentum (\ref{eq:Jinnuz}), angular velocity (\ref{eq:Ominuz}), temperature (\ref{eq:Titemps}) and entropy (\ref{eq:bulkent}) constitute the thermodynamics of the rotating $\text{AdS}_{4}$ bulk black hole. In the $\alpha=0$ limit, one recovers the thermodynamics of the static $\text{AdS}_{4}$ bulk \cite{Emparan:2022ijy}. One may derive the bulk thermodynamics using a canonical partition function by evaluating the on-shell bulk gravity action via an appropriate modification of the presentation given in \cite{Kudoh:2004ub}. Additionally, by explicit computation it is straightforward to verify\footnote{Here we replace the absolute value in $T_{i}$ by an overall minus sign for reasons we explain momentarily.}
\beq \partial_{z}M-T_{i}\partial_{z}S^{(4)}_{\text{BH}}-\Omega_{i}\partial_{z}J=0\;,\quad \partial_{\alpha}M-T_{i}\partial_{\alpha}S^{(4)}_{\text{BH}}-\Omega_{i}\partial_{\alpha}J=0\;,\eeq
such that the bulk system obeys the first law
\beq dM=T_{i}dS_{\text{BH}}^{(4)}+\Omega_{i}dJ\;,\label{eq:classfirstlaw}\eeq
for all values of the parameters, including any value of the brane tension, as controlled by $\nu$.

\subsection{Semi-classical thermodynamics on the brane}

From the brane perspective, the thermodynamics of the classical bulk system doubles as the thermodynamics of the quantum de Sitter black hole. It is worth mentioning that, even without accounting for backreaction, de Sitter thermodynamics is more subtle than their flat or AdS space counterparts. Firstly, this is because de Sitter space lacks an asymptotic region to introduce boundary conditions which fix thermoodynamic data to define a thermal ensemble. Moreover, the first law of cosmological horizons \cite{Gibbons:1977mu} comes with a minus sign which begs how the thermodynamics of the dS static patch should be understood. In what follows, we ignore these subtleties, though it would be interesting to return to them in the future, adapting the quasi-local approach developed in \cite{Banihashemi:2022jys,Banihashemi:2022htw} (see also \cite{Svesko:2022txo,Anninos:2022hqo}).

%The question of defining a thermal ensemble and the thermodynamic interpretation of the minus sign can both be addressed using a quasi-local approach where one introduces an artificial timelike boundary to fix thermodynamic data \cite{Banihashemi:2022jys,Banihashemi:2022htw} (see also \cite{Svesko:2022txo,Anninos:2022hqo}). 

\subsection*{Thermodynamics with multiple horizons}

The quantum de Sitter black hole comes with three horizons which are generally at different temperatures. Consequently, each horizon generally has its own thermodynamics, satisfying its own first laws, as we now show. The mass (\ref{eq:massinnuz}), angular momentum (\ref{eq:Jinnuz}), and angular velocity (\ref{eq:Ominuz}) of the quantum black hole all take the same form in terms of parameters (\ref{eq:paramsforKdS}). The temperature (\ref{eq:Titemps}) encodes the temperature of each horizon of the quantum black hole, where we remind the reader the outer and inner black hole horizons correspond to the outer and inner bulk black hole horizons localized on the brane, while the cosmological horizon arises from the bulk acceleration horizon intersecting the brane. To distinguish each horizon, it is useful to slightly modify the notation for $z$ via $z_{c}=R_{3}/r_{c}x_{1}$ and $z_{\pm}=R_{3}/r_{\pm}x_{1}$ to denote the cosmological and black hole horizons, respectively. Then, from the surface gravities (\ref{eq:surfacegravsv2})
\beq T_{c}=T_{i}(z_{c})\;,\quad T_{+}=-T_{i}(z_{+})\;,\quad T_{-}=T_{i}(z_{-})\;,\eeq
 where we used $r_{-}<r_{+}<r_{c}$ such that $z_{-}>z_{+}>z_{c}$. Consequently, the black hole horizon is generally hotter than the cosmological horizon, $T_{c}<T_{+}$, such that the system is not in thermal equilibrium; an observer located between the cosmological and (outer) black hole horizon is in a system characterized by two temperatures. There are three special cases, where the horizons degenerate, when the outer black hole and cosmological horizons are in thermal equilibrium, as we explore below. 

 The most notable difference between the bulk and brane black hole thermodynamics is the interpretation of the entropy (\ref{eq:bulkent}). On the brane, this entropy $S_{\text{BH}}^{(4)}$ is equal to the sum of gravitational entropy and the entanglement entropy of the holographic CFT \cite{Emparan:2020znc}. Thus, the bulk entropy is identified with the generalized entropy on the brane $S_{\text{gen}}^{(3)}$, 
 \beq S_{\text{BH}}^{(4)}=S_{\text{gen}}^{(3)}=S_{\text{grav}}^{(3)}+S_{\text{CFT}}^{(3)}\;.\eeq
 This relation is exact to all orders in semi-classical backreaction codified by $\nu$. The gravitational  entropy is computed using Wald's entropy functional \cite{Wald:1993nt}, 
 \beq S_{\text{Wald}}=-2\pi\int_{\mathcal{H}}dA\frac{\partial\mathcal{L}}{\partial R^{abcd}}\epsilon_{ab}\epsilon_{cd}\;,\eeq
 where $dA=d^{d-2}x\sqrt{q}$ is the codimension-2 area element of the bifurcate horizon $\mathcal{H}$, with $q_{ab}=h_{ab}+n_{a}n_{b}-u_{a}u_{b}$ being the induced metric, for spacelike and timelike unit normals $n_{a}$ and $u_{a}$, respectively. The binormal $\epsilon_{ab}=(n_{a}u_{b}-n_{b}u_{a})$ satisfies $\epsilon^{2}=-1$, and we define  $(d-1)$-dimensional metric in directions orthogonal to the horizon. Moreover, $\mathcal{L}$ refers to the Lagrangian density defining the theory. With respect to the induced theory of gravity on the brane (\ref{eq:indthebrane3D}), the gravitational entropy is
\beq S_{\text{grav}}^{(3)}=\frac{1}{4G_{3}}\int_{\mathcal{H}} dx\sqrt{q}\left[1+\ell^{2}\left(\frac{3}{4}R-g^{ab}_{\perp}R_{ab}\right)+\mathcal{O}(\ell^{4}/R_{3}^{6})\right]\;.\eeq
We see higher-curvature corrections to entropy enter at order $\ell^{2}$, such that the dominant contribution to the entropy at leading order in backreaction is the three-dimensional Bekenstein-Hawking entropy 
\beq S_{\text{BH}}^{(3)}=\frac{1}{4G_{3}}\int_{\mathcal{H}}dx\sqrt{q}=\frac{2\pi r_{i}\eta}{4G_{3}}\left(1+\frac{a^{2}x_{1}^{2}}{r_{i}^{2}}\right)=\frac{1+\nu z}{\sqrt{1-\nu^{2}}}S_{\text{gen}}^{(3)}\;.\eeq
Therefore, the Bekenstein-Hawking entropy includes semi-classical backreaction effects. 

Formally, the matter entropy $S_{\text{CFT}}^{(3)}$ is given by the difference
\beq S_{\text{CFT}}^{(3)}=S_{\text{gen}}^{(3)}-S_{\text{grav}}^{(3)}\;.\eeq
Notably, the matter entropy enters at linear order in $\nu$, 
\beq S_{\text{CFT}}^{(3)}\approx S_{\text{gen}}^{(3)}-S_{\text{BH}}^{(3)}=-\nu zS_{\text{BH}}^{(3)}\;,\eeq
in contrast with the higher-curvature contributions to the gravitational entropy which enter at order $\nu^{2}$. Recall that the central charge $c=L_{4}^{2}/G_{4}\approx \nu R_{3}/2G_{3}$, such that $S_{\text{CFT}}^{(3)}$ is proportional to $c$. As in the quantum BTZ case \cite{Emparan:2020znc}, generally the matter entropy will be dominated by entanglement across the horizon(s) in CFT states with large Casimir effects.

Interpreting $S_{\text{BH}}^{(4)}$ as the generalized entropy of the quantum black hole, the bulk first law (\ref{eq:classfirstlaw}) leads to a semi-classical first law for each horizon\footnote{It is possible to assign a dynamical thermodynamic pressure $P_{3}\propto-\Lambda_{3}$ to the quantum black hole, whose variations are induced by variations in the tension of the brane \cite{Frassino:2022zaz}. The first law then acquires a $VdP_{3}$ term, where $V$ is the `thermodynamic volume' conjugate to $P_{3}$ \cite{Kastor:2009wy,Dolan:2010ha}.}
\beq dM=T_{+}dS^{(3)}_{\text{gen},+}+\Omega_{+}dJ\;,\eeq
\beq dM=-T_{c}dS^{(3)}_{\text{gen},c}+\Omega_{c}dJ\;,\eeq
\beq dM=-T_{-}dS^{(3)}_{\text{gen},-}+\Omega_{-}dJ\;,\eeq
 where $\Omega_{c}=\Omega_{i}(z_{c})$ and $\Omega_{\pm}=\Omega_{i}(z_{\pm})$ are the angular speeds of the cosmological and black hole horizons. Combining the first two
 first laws yields
 \beq 0=T_{+}dS_{\text{gen}}^{(3)}+T_{c}dS_{\text{gen}}^{(3)}+(\Omega_{+}-\Omega_{-})dJ\;.\eeq
Our first law is consistent with the semi-classical first laws for static two-dimensional (A)dS black holes in \cite{Svesko:2022txo,Pedraza:2021cvx}. Notice the minus sign in the first law of the cosmological horizon remains even in the quantum-backreacted geometry. Consequently, adding positive energy into the static patch reduces the total entropy of the system, with the entropy of pure dS being maximal, such that de Sitter black holes behave as instantons constraining the states of the original de Sitter degrees of freedom (cf. \cite{Morvan:2022ybp,Draper:2022xzl,Morvan:2022aon}). 
%It would be interesting to compute the nucleation rate of quantum Kerr-dS black holes, generalizing  \cite{Emparan:2022ijy}. 

At this stage, there are two limits of interest. The first is the quantum de Sitter limit, at $z=1$ or $\mu=0$, and, consequently, 
\beq M=J=\Omega_{i}=0\;,\eeq
\beq S_{\text{gen}}^{(3)}=\frac{2\pi R_{3}}{4G_{3}}\frac{\sqrt{1-\nu^{2}}}{1+\nu}\;,\quad T_{c}=\frac{1}{2\pi R_{3}}\;,\eeq
where we see the temperature of the quantum $\text{dS}_{3}$ cosmological horizon is the same as classical $\text{dS}_{3}$. Second, when backreaction vanishes $\nu\to0$, then $z_{\pm}\to\infty$ since $r_{\pm}\to0$ and we have
\beq 
\begin{split}
&M=\frac{1}{8G_{3}}\frac{(z^{2}-1)(1+\alpha^{2}(1+z^{2}))(9z^{2}-1+\alpha^{2}(9z^{4}-1))}{(3z^{2}-1+\alpha^{2}(1+3z^{4}))^{2}}\;,\\
&J=\frac{\alpha R_{3}}{G_{3}}\frac{z(z^{2}-1)(1+\alpha^{2}(1+z^{2}))\sqrt{1+\alpha^{2}z^{2}}}{(3z^{2}-1+\alpha^{2}(1+3z^{4}))^{2}}\;,\\
&\Omega_{c}=\frac{\alpha(z^{2}-1)\sqrt{1+\alpha^{2}z^{2}}}{R_{3}z(1+\alpha^{2}(1+z^{2}))}\;,\\
&T_{c}=\frac{1}{2\pi R_{3}}\frac{2(1+2\alpha^{2}z^{2})(z^{2}+\alpha^{2}(1+z^{4}))}{z(3z^{2}-1+\alpha^{2}(1+3z^{4}))(1+\alpha^{2}(1+z^{2}))}\;,\\
&S_{c}=\frac{\pi R_{3}}{G_{3}}\frac{z(1+\alpha^{2}(1+z^{2}))}{(3z^{2}-1+\alpha^{2}(1+3z^{4}))}\;,
\end{split}
\eeq
where it is understood that here $z=z_{c}$. It is straightforward to show the resulting thermodynamics reproduces that of the classical Kerr-$\text{dS}_{3}$ (see Appendix \ref{app:KerrdS3}), namely, 
\beq S_{\text{gen}}^{(3)}|_{\nu=0}=\frac{\pi R_{3}}{4G_{3}}\left(\sqrt{(1-8G_{3}M)+i\frac{8G_{3}J}{R_{3}}}+\sqrt{(1-8G_{3}M)-i\frac{8G_{3}J}{R_{3}}}\right)=S_{\text{KdS}_{3}}\;,\eeq
where we used the 
relation $\sqrt{x+iy}+\sqrt{x-iy}=2\sqrt{x+\sqrt{x^{2}+y^{2}}}/\sqrt{2}$.

%, with the Nariai black hole, as the largest black hole to fit inside the static patch, the state of minimum entropy (

\subsection*{Thermodynamics of degenerate horizons}

As described in Section \ref{sec:GeomKdS3}, the quantum Kerr black hole has special limits where two or more horizons become degenerate. Of interest are the extremal ($r_{+}=r_{-}$), Nariai ($r_{c}=r_{+}$), and lukewarm ($T_{c}=T_{+}$) geometries. The extremal black hole is one with a vanishing temperature, $T_{\text{ext}}=0$. Naively, the Nariai black hole will have a vanishing temperature, however, in its near horizon geometry, the temperature of the black hole and cosmological horizon will be in thermal equilibrium at a non-zero temperature $T_{\text{N}}$. The precise form of the temperature can be found, for example, by removing the conical singularity in the Euclideanized section of the (naive) Nariai geometry (\ref{eq:Nariaimet}), given via the Wick rotation $\hat{\tau}\to i\hat{\tau}_{E}$ and $a\to ia_{E}$, resulting in $T_{\text{N}}=(2\pi\sqrt{\Gamma})^{-1}$.
%\footnote{Start from the near-horizon Nairiai geometry and co-rotate the $\hat{\tau}$ and $\hat{\phi}$ coordinate into each others to remove the shift vector. Then, expand the metric close to the horizon and work out the deficit angle by comparing the result with the standard metric in polar coordinates. Finally, the temperature of the horizon is given by the inverse of the periodicity of the rescaled Euclidean time $\hat{T}$.} 
To connect to the canonical geometry, we relate the Nariai radii $r_{\text{N}}$ and $\bar{r}_{\text{N}}$ via (\ref{eq:rsdef}). Lastly, the lukewarm limit occurs when the outer black hole and cosmological horizons are in thermal equilibrium at a temperature different from the Nariai temperature. Though the resulting expression is cumbersome and not very illustrative, the precise temperature can be solved for explicitly by setting $T_{+}=T_{c}$ (using the surface gravities (\ref{eq:surfacegravsv2})) and following the method described in Appendix \ref{app:limitsqKdS}. The lukewarm temperature is proportional to $(r_{c}-r_{+})/2\pi R_{3}^{2}$, with $r_{c}\neq r_{+}$.

%%%%%%%%%%%%%%%%%%%%%%%%%%%%%%%%%%%%%%%%%%%%%%%%%%%%%%%%%%%%%%%%%
\section{Discussion} \label{sec:disc}

In this article we used braneworld holography to construct a three-dimensional quantum-corrected Kerr-de Sitter black hole exactly accounting for backreaction effects due to a conformal field theory. By stark contrast, there are no de Sitter black holes in three-dimensions, only conical defect geometries with a single cosmological horizon. Thus, semi-classical backreaction alters the defect geometry so as to induce inner and outer black hole horizons, which hide a ring singularity, sharing many qualitative features with the classical four-dimensional Kerr-de Sitter solution. With three horizons, we uncovered the extremal, Nariai, and `ultracold' limits of the semi-classical black hole, which appear as fibered products of a circle and $\text{AdS}_{2}$, $\text{dS}_{2}$, or two-dimensional Minkowski space, respectively. 

Moreover, the thermodynamics of the classical bulk black hole, described by the rotating $\text{AdS}_{4}$ C-metric, has a dual interpretation on the brane as thermodynamics of the semi-classical Kerr-$\text{dS}_{3}$ black hole. Specifically, the standard first law of thermodynamics in the bulk becomes a semi-classical first law, where the four-dimensional Bekenstein-Hawking area-entropy is identified with the three-dimensional generalized entropy, given by the sum of the Wald entropy due to higher curvature corrections, and the matter entropy of the CFT. In essence, we have derived the semi-classical generalization of the first law of cosmological horizons of Gibbons and Hawking \cite{Gibbons:1977mu}. As in the classical four-dimensional Kerr-dS solution, the limiting geometries of the quantum Kerr-dS black hole give rise to scenarios of thermal equilibrium, including the Nariai and lukewarm limits where the temperatures of the cosmological and outer black hole horizons coincide.
%The limiting geometries, where at least two of the three horizons degenerate, lead to situations where the the thermodynamics of the aforementioned limiting geometries, as well as the `lukewarm' limit, where the temperatures of the cosmological and outer black hole horizons coincide at a temperature different from the Nariai black hole. 
Therefore, quantum-corrections greatly enrich the thermodynamic structure of three-dimensional de Sitter solutions.

There are multiple future directions to take this work, some of which we list below.
% expanding on ideas we have only briefly touched on thus far.

\vspace{4mm}

\noindent \textbf{Other three-dimensional quantum black holes:} Here we focused on neutral rotating quantum de Sitter black holes. It is natural to ask whether other types of three-dimensional quantum black holes are possible using a similar braneworld construction. Firstly, one may consider charged quantum black holes simply by starting from the charged $\text{AdS}_{4}$ C-metric. Although there is no need for counterterms for the Maxwell field in AdS$_4$ \cite{Emparan:1999pm,Chamblin:1999tk},  a Maxwell action is nevertheless generated on a brane at finite distance in the bulk, further modifying the geometry of the quantum black hole \cite{inprep}. Similarly, starting from the accelerating Taub-NUT $\text{AdS}_{4}$ black hole, one would conceivably find a quantum Taub-NUT black hole on the brane. Altogether, 
%in principle one could combine each of these constructions
via suitable modifications to the $\text{AdS}_{4}$ C-metric, one could develop a catalog of charged, rotating, Taub-NUT quantum (A)dS black holes in three-dimensions.

A further generalization would be to consider quantum black holes with scalar hair. One way to do this is to consider bulk Einstein gravity in addition to a conformally coupled scalar field. Black hole solutions to this theory have a rich history, dating back to Bekenstein \cite{Bekenstein:1974sf,Bekenstein:1975ts}, including exact generalizations of the charged C-metric \cite{Charmousis:2009cm} and Plebanski-Demianski family of metrics \cite{Anabalon:2009qt}.
%, where the bulk spacetime geometry takes same form as when there is no scalar hair, and for non-zero bulk cosmological constant, the divergence of the scalar field may be hidden behind the black hole event horizon. 
In holographic renormalization, adding scalar fields in the bulk requires additional counterterms thereby modifying the induced theory on the brane, resulting in, presumably, an exact quantum black hole with scalar hair.

\noindent \textbf{Higher dimensional quantum black holes:} The quantum Kerr-$\text{dS}_{3}$ black hole is another example of an exact description of a localized three-dimensional black hole in a Randall-Sundrum braneworld, belonging to the class of the solutions uncovered in \cite{Emparan:1999fd,Emparan:1999wa} (see also \cite{Anber:2008qu}, where the brane tension was detuned from the bulk acceleration). It is natural to wonder whether one can construct higher-dimensional quantum black holes in a similar fashion. Extrapolating from the four-dimensional bulk models, holographic considerations predict backreaction due to conformal fields is expected to similarly induce quantum corrections to the geometry. For example, a semi-classical four-dimensional brane black hole would include a $\ell \mu/r^{2}$ correction to the standard $1/r$ gravitational potential, a behavior inherited from its parent $\text{AdS}_{5}$ black hole.

Thus far, however, there are no known exact quantum black holes in higher dimensions. This is because finding static brane localized black holes in higher dimensions has proven challenging, both analytically and numerically (for a review, see \cite{Tanahashi:2011xx}). The essential feature of the four-dimensional C-metric which is exploited is that there is a natural location to place the brane, at $x=0$, where the Israel-junction conditions are automatically satisfied. A higher-dimensional analog of the C-metric exuding this feature is not known to exist \cite{Podolsky:2006du}, making the construction of exact quantum black holes difficult. Perhaps numerical techniques together with the large-$D$ approximation of bulk Einstein gravity, as was recently accomplished to describe evaporating brane black holes \cite{Emparan:2023dxm}, can be adapted to construct exact quantum black holes in higher-dimensions.

\noindent \textbf{Dimensional reduction and deviations away from extremality:} Spherical reduction of a $d\geq4$ dimensional classical de Sitter black hole near its extremal, Nariai, or ultracold limits results in different two-dimensional effective theories of dilaton-gravity (see, e.g., \cite{Maldacena:2019cbz,Castro:2022cuo}). Alternatively, the low-energy dynamics of spherical reduced three-dimensional empty dS is characterized by de Sitter Jackiw-Teitelboim gravity, where the dilaton only ranges over positive values. Spherical reduction of three-dimensional quantum de Sitter black holes in their nearly extremal limits will be described by modified dilaton theories of gravity, including $R^{2}$ corrections owed to reduction of the induced theory on the brane. It would be interesting to study these effective two-dimensional theories to better understand deviations away from extremal quantum dS black holes, along the lines of \cite{Castro:2022cuo}.

%\noindent \textbf{Quantum black holes as constrained instantons:}

%\noindent \textbf{Quasinormal modes?}

\noindent \textbf{Double holography and quantum Kerr-dS/CFT:} Black holes localized on Karch-Randall braneworlds enjoy a `doubly-holographic' perspective, where the $d$-dimensional induced gravity theory on the brane has a dual description in terms of a defect $\text{CFT}_{d-1}$, which interacts with the $\text{CFT}_{d}$ on the $\text{AdS}_{d+1}$ boundary \cite{Karch:2000gx,Karch:2000ct,Takayanagi:2011zk,Fujita:2011fp}. As recognized in \cite{Emparan:2022ijy}, the de Sitter braneworld provides a means to explore dS/CFT holography, where gravity on the brane may be characterized by defect Euclidean CFTs at $\mathcal{I}^{\pm}$ of the dS hyperboloid. The hope would be to use this type of double holography to study dS/CFT in a controlled way.

%however, bulk quantum corrections are expected to introduce big bang and big crunch-like singularities at $\mathcal{I}^{\pm}$, eliminating the Lorentzian AdS cylinders to the past and future. 

The quantum Kerr-black hole presents an opportunity to study another kind of holographic duality, namely, the Kerr-CFT correspondence \cite{Guica:2008mu} (see also \cite{Hartman:2008pb,Lu:2008jk}). In particular, a four-dimensional extremal Kerr-(A)dS black hole is dual to (half of a chiral) two-dimensional CFT, a consequence that there exist boundary conditions of the near horizon extremal Kerr solution which enhance the $U(1)$ isometry of $SL(2,\mathbb{R})\times U(1)$ to a Virasoro algebra with non-trivial central charge. Moreover, in the case of Kerr-dS, the rotating-Nariai black hole has a dual description in terms of a two-dimensional (Euclidean) CFT \cite{Anninos:2009yc}, akin to the dS/CFT correspondence. Since the limiting geometries of the quantum Kerr-$\text{dS}_{3}$ have the same isometry subgroup as their four-dimensional classical counterparts, it is plausible these quantum black holes have an additional dual description in terms of a two-dimensional chiral CFT, however, with a different central charge. It would be interesting to pursue this duality further and better understand the dual CFT using the classical bulk.

\noindent \textbf{Holographic information of quantum de Sitter black holes:} Exact semi-classical black holes allow one to explore proposals of information theoretic descriptions of gravity including backreaction effects. For example, in the context of the (static) quantum BTZ black hole, both `complexity=volume' and `complexity=action' conjectures were analyzed \cite{Emparan:2021hyr}, where complexity=volume was found to have a consistent semi-classical expansion, while complexity=action was unable to produce the classical limit. Likewise, quantum de Sitter black holes offer the chance to study holographic complexity in de Sitter space, which thus far has largely been unexplored (see, however, \cite{Reynolds:2017lwq,Chapman:2021eyy,Jorstad:2022mls}). There are at least two ways forward with the de Sitter braneworld set-ups. One could attempt to exploit the aforementioned doubly holographic interpretation to understand the complexity of bulk quantum fields, however, this would require deeper insight into the relation between the $\text{CFT}_{3}$ and the Euclidean defect $\text{CFT}_{2}$. Alternately, complexity in terms of holographic state preparation, as in \cite{Pedraza:2021mkh,Pedraza:2021fgp,Pedraza:2022dqi}, can be adapted into models of de Sitter braneworlds, where the de Sitter hyperboloid represents time evolution of the dual boundary CFT state prepared by a Euclidean path integral. 

Specific to rotating qBTZ and quantum Kerr-dS black holes, it would be worthwhile to see how complexity of formation -- the volume-complexity of a black hole minus the analogous volume of the relevant vacuum spacetime -- is modified due to backreaction. In fact, complexity of formation for classical rotating black holes was found to be controlled by the thermodynamic volume familiar to the framework of extended black hole thermodynamics \cite{AlBalushi:2020rqe}. Likewise, it would be interesting to see whether the thermodynamic volume of rotating quantum black holes yields a similar description of their complexity of formation.  

Lastly, doubly holography of AdS braneworlds have provided a means to study the black hole information paradox by computing the fine grained entropy of Hawking radiation using the `island rule' \cite{Chen:2020uac,Chen:2020hmv,Almheiri:2019psy}.  It would be worth exploring the formation of islands in quantum de Sitter black hole backgrounds which incorporate all orders of semi-classical backreaction. Progress in this regard has already been made in the static case \cite{inprep2}.  

%Analogous information paradoxes arise in cosmological and de Sitter backgrounds \cite{Hartman:2020khs,Aalsma:2021bit,Kames-King:2021etp}, where quantum extremal surfaces and islands play a key role. Since the qSdS solution  uncovered here naturally incorporates the effect of backreaction, we have analytic control over the extremization of the generalized entropy on the brane to study detailed aspects of quantum extremal islands in dS. 

\noindent \noindent\section*{Acknowledgments}
We are grateful to Roberto Emparan, Juan Pedraza, and Manus Visser for discussions and useful correspondence. EP is supported by the Cosmoparticle Initiative at University College London. AS is supported by the Simons Foundation via \emph{It from Qubit: Simons Collaboration on quantum fields, gravity, and information}, and EPSRC.

%%%%%%%%%%%%%%%%%%%%%%%%%%%%%%%%%%%%%%%%%%%%%%%%%%%%%%%%%%%%%%%%%%%%
\appendix

\section{Geometry of classical Kerr-dS$_3$}\label{app:KerrdS3}

In this appendix we review the geometry of the classical Kerr-dS$_3$ black hole. To appreciate its properties, we first briefly recall facts about conical geometries.

\subsection*{Conical singularities}

In Riemannian geometry, a conical singularity is a singular point of the $n$-dimensional manifold in the proximity of which the metric locally looks like the spherical quotient $\text{S}^{n-1}/G$ by a finite subgroup $G$. For example, for the cyclic group
%is some finite group and S$^{n-1}$ is the $(n-1)$-dimensional sphere. 
%The standard conical geometry we all know and love is included in this class of singularities, and it is obtained by choosing a cyclic group as subspace, i.e. 
$G=\mathbb{Z}_k$, the resulting geometry is isomorphic to $\mathbb{R}^n$ with an angular deficit $\delta$, i.e., a wedge of $\delta$-radians cut out from the $n$-dimensional plane with the edges identified, where $\delta=2\pi(1-\gamma)$,
%The order $G$ is related to $\delta$ translates to the deficit angle in the following way:
%\beq  \delta%= 2\pi\left(1- \frac{1}{k}\right)  = 2\pi\left(1- \gamma\right)\ ,
%\end{equation}
with $\gamma\equiv1/k$.\footnote{When $0<\gamma<1$, $\delta$ is referred to as an angular deficit, while $\delta$ is an angular excess when $\gamma<0$.} 
%When $0<\gamma<1$, then $\delta$ is referred to as an angular \textit{deficit}, while for $\gamma < 0$ it is an angular \textit{excess}. %(in GR, these are sourced by negative energy). 
Unsurprisingly, the geometry of a cone is an example of a manifold with a conical singularity. Explicitly, 
%Here, we focus on 2-dimensional Euclidean manifolds with angular deficits. The metric for a conical singularity looks naively just like the flat metric. Indeed, 
the metric on the (flat) 2-cone in polar coordinates $(r,\varphi)$ is
\begin{equation}
    ds^2 = dr^2+r^2 d\varphi^2 \ .
\end{equation}
where, due to conical geometry, the angular variable is not $2\pi$-periodic but rather,
%However, there is a subtlety related to the coordinate ranges. In particular, recall that we have obtained the conical geometry by cutting away a wedge of angle $\delta$ and then glued the two resulting sides together. This means that, really, $\varphi$ is not 2$\pi$ periodic, 
%\begin{equation}
  $0\leq \varphi <2\pi-\delta$.
 % \ .
%\end{equation}
To make the difference between the cone and the flat space metric (where $\varphi$ is $2\pi$-periodic) explicit, rescale the angular coordinates to have standard $2\pi$-periodicity. Namely, define $\phi$ such that $\varphi=\gamma\phi$, where $\phi$ is $2\pi$-periodic.\footnote{More generally, near a conical singularity in spaces of constant (Gaussian) curvature $K$, the two-dimensional geometry has line element $ds^2 = dr^2 + \gamma^2 F_K(r)^2 d\phi^2$, with $F_{0}=r$, $F_{K}=K^{-1/2}\sin(\sqrt{K}r)$ when $K>0$ (elliptic), or $F_{K}=|K|^{-1/2}\sinh(\sqrt{|K|}r)$ when $K<0$ (hyperbolic). Moreover, the range of the $r$ coordinate changes with $K$; being $0<r<\infty$ for $K\leq 0$ and $0<r<\frac{\pi}{2\sqrt{K}}$ for $K>0$.}
%\begin{equation}
%    \varphi =\left(1-\frac{\delta}{2\pi}\right) \phi = \gamma \phi \ ,
%\end{equation}
%we obtain an angular coordinate with the right periodicity. 
%However, this comes at the cost of having a different effective radius along lines of constant $\varphi$:
%\begin{equation}
%    ds^2 = dr^2 + \gamma^2 r^2 d\phi^2 \ .
%\end{equation}
%More generally, near a conical singularity in spaces of constant (Gaussian) curvature $K$, the two-dimensional geometry is
%\begin{equation}
%    ds^2 = dr^2 + \gamma^2 F_K(r)^2 d\phi^2 \ ,
%\end{equation}
%with $F_{0}=r$, $F_{K}=K^{-1/2}\sin(\sqrt{K}r)$ when $K>0$ (elliptic), or $F_{K}=|K|^{-1/2}\sinh(\sqrt{|K|}r)$ when $K<0$ (hyperbolic).
%\beq
%F_K(r) =  \begin{cases} \frac{1}{\sqrt{K}} \sin\left(\sqrt{K}r\right) &  K>0 \\ r & K=0 \\ \frac{1}{\sqrt{|K|}} \sinh \left(\sqrt{|K|}r\right) & K<0 \end{cases} \ .
%\eeq
%This gives the general form of a conical singularity in two spatial dimension for elliptical, flat and hyperbolic space respectively. 
%Moreover, the range of the $r$ coordinate changes with $K$; being $0<r<\infty$ for $K\leq 0$ and $0<r<\frac{\pi}{2\sqrt{K}}$ for $K>0$. 

Of interest are $(2+1)$-dimensional spacetimes with conical singularities. 
%Consider for example a (2+1)-dimensional spacetime of constant curvature, such that near the tip of the cone the line element takes the form
%Without loss of generality, spacetimes of constant curvature near the tip of a cone take the form
%\begin{equation}
%   ds^2 = -dt^2 + dr^2 + \gamma^2 F_K(r)^2 d\phi^2\;.
%\end{equation}
%The Ricci scalar is $R=-2F_{K}^{-1}\partial_{r}^{2}F_{K}$ where $F_{K}=r$, the geometry appears flat. Note, however,  
Generally the spacetime is singular at the tip of the cone due to geodesic incompleteness and because there is a curvature singularity at the tip.
%can be understood from two different perspectives. The first one is geodesic incompleteness: there is no unique way to continue a geodesic that reaches the tip of the cone. On the other hand, we have a divergent curvature at the origin. This is intuitive and can be rigorously proven by regularising the geometry, i.e. smoothing out the tip, and evaluating the curvature tensors in the non-singular spacetime.
The stress-energy tensor sourcing such geometries is that of a point particle of mass $M$ (and spin $J$ if the spacetime is stationary), whose angular deficit $\delta$ is generally a function of the mass and spin.
%where, for small $M$, $\delta=8\pi G_{3}M$.
%the one of a point particle of mass $M$ (with the presence of an appropriate cosmological constant for $K\neq0$). The mass of the particle is related to the angular deficit in a different manner for each of the three possible geometries, but for small $M$ they all reduce to:
%\begin{equation}
 %   \delta = 8\pi G M \ .
%\end{equation}
%Hence, in $(2+1)$ dimensions, point particles correspond to conical singularities whose angular defect is a function of the mass. 
This also means that, classically, there are no three-dimensional flat or de Sitter black holes as we now briefly review.

\subsection*{(2+1)-dimensional Kerr-de Sitter}

The general solution for (2+1)-dimensional Einstein's equations of a point particle of mass $m$ and spin $J$ in a background with positive cosmological constant $\Lambda=\frac{1}{R_3^2}$ is
%As a second example, add rotation and consider the Kerr-$\text{dS}_{3}$ geometry, In static patch coordinates the line element is 
%(cf. \cite{Klemm:2002ir})
\beq ds^{2}=-N(r)dt^{2}+N^{-1}(r)dr^{2}+r^{2}(d\phi+N_{\phi}dt)^{2}\;,\label{eq:KdS3metapp}\eeq
with lapse and shift metric functions
\beq N(r)=m-\frac{r^{2}}{R_{3}^{2}}+\frac{16 G_{3}^{2}J^{2}}{r^{2}}\;,\quad N_{\phi}=\frac{4G_{3}J}{r^{2}}\;.\eeq
%When the spin $J$ vanishes, the solution reduces to the Schwarzschild-de Sitter background, where $m=1-8G_{3}M$.
The mass and spin are respectively related to the conserved charges associated with the time-translation and rotational Killing symmetries \cite{Klemm:2002ir}.
%\beq Q_{\partial_{t}}=-\frac{m}{8G_{3}}\;,\quad Q_{\partial_{\phi}}=J\;.\eeq
%The lapse $N$ has a single positive root $r_{c}$
%\beq r_{c}^{2}=\frac{1}{2}\left(m R_{3}^{2}\pm |m|R_{3}^{2}\sqrt{1+4\left(\frac{4G_{3}J}{mR_{3}}\right)^{2}}\right)\;,\label{eq:rcKerrdS3}\eeq
%where we take either `$+$' or `$-`$ depending on the sign of $m$. 
The lapse $N$ has roots $r_{\pm}$
\beq r^{2}_{\pm}=\frac{R_{3}^{2}}{2}\left(m\pm\sqrt{m^{2}-\left(\frac{8G_{3}J}{R_{3}}\right)^{2}}\right)\;,\label{eq:rcKerrdS3}\eeq
with only a single positive root $r_{+}$
\beq r_{+}=\frac{R_{3}}{2}\left(\sqrt{m+i\frac{8G_{3}J}{R_{3}}}+\sqrt{m-i\frac{8G_{3}J}{R_{3}}}\right)\;.\eeq
In the main text we set $m=1-8G_{3}M$.
%however, it is common to express $m=-8G_{3}\tilde{M}$, where $8G_{3}\tilde{M}=8G_{3}M-1$. 
Notice $R_{3}^{2}m=r_{+}^{2}+r_{-}^{2}$ while $J=-ir_{+}r_{-}/4G_{3}R_{3}$. When $J=0$, we recover Schwarzschild-$\text{dS}_{3}$, where $r_{-}=0$.

%This root is identified as the cosmological horizon. Notice $mR_{3}^{2}=r_{c}^{2}-\alpha^{2}$, where $\alpha\equiv-4G_{3}J/r_{c}$.

%, in line with the choice adopted in the main text and in the previous section.

The spacetime (\ref{eq:KdS3metapp}) is simply Kerr-$\text{dS}_{3}$, however, it is not a black hole. Here %spacetime may be understood as a point particle with mass parameter $m$ and spin $J$ where 
$r_{+}$ is identified as the cosmological horizon $r_{c}$, there being no black hole horizon for reasons given in the introduction. Rather, Kerr-$\text{dS}_{3}$ is a conical defect geometry. To see this, introduce dimensionless parameters $\gamma\equiv r_{c}/R_{3}$ and $\alpha\equiv-4G_{3}J/\gamma R_{3}=ir_{-}/R_{3}$, satisfying $m=\gamma^{2}-\alpha^{2}$, and consider the following coordinate transformation \cite{Deser:1983nh,Bousso:2001mw}
\beq 
\begin{split}
 \tilde{t}= \gamma t+\alpha R_{3}\phi\;,\quad \tilde{\phi}=\gamma\phi-\alpha t/R_{3}\;,\quad \tilde{r}/R_{3}=\sqrt{\frac{(r/R_{3})^{2}+\alpha^{2}}{\gamma^{2}+\alpha^{2}}}\;.  
\end{split}
\label{eq:coordtranstocon}\eeq
This brings the Kerr-$\text{dS}_{3}$ geometry (\ref{eq:KdS3metapp}) into an empty $\text{dS}_{3}$ form
\beq ds^{2}=-\left(1-\frac{\tilde{r}^{2}}{R_{3}^{2}}\right)d\tilde{t}^{2}+\left(1-\frac{\tilde{r}^{2}}{R_{3}^{2}}\right)^{-1}d\tilde{r}^{2}+\tilde{r}^{2}d\tilde{\phi}^{2}\;,\label{eq:conicaldS3}\eeq
however, the coordinates $\tilde{t}$ and $\tilde{\phi}$ do not have the same periodicity as genuine $\text{dS}_{3}$, where $(t,r,\phi)\sim(t,r,\phi+2\pi)$. Now 
\beq (\tilde{t},\tilde{\phi})\sim(\tilde{t}+2\pi R_{3}\alpha,\tilde{\phi}+2\pi\gamma)\;.\label{eq:periodcondsKdS3}\eeq
 This reveals Kerr$-\text{dS}_{3}$ is a conical defect geometry with angular deficit $\delta=2\pi\gamma$, and is a quotient of $\text{dS}_{3}$. Moreover, the Schwarzschild-$\text{dS}_{3}$ geometry (where $J=0=\alpha$) is also a conical defect with deficit $\delta=\sqrt{1-8G_{3}M}$, or a particle of mass $m$ whose stress-energy tensor $T_{ab}=m\delta(r)\delta^{0}_{a}\delta^{0}_{b}$ sources the geometry. 
 %Hence, there are no black solutions to three-dimensional Einstein's equations with positive cosmological constant.

 The thermodynamics of the cosmological horizon is straightforward to work out (cf. \cite{Park:1998qk})
 \beq
 \begin{split}
 &m=\gamma^{2}-\alpha^{2}\;,\quad J=-\frac{\alpha\gamma R_{3}}{4G_{3}}\;,\quad \Omega_{c}=-\frac{\alpha}{\gamma R_{3}}\;,\\
&T_{c}=\frac{\gamma^{2}+\alpha^{2}}{2\pi\gamma R_{3}}\;,\quad S_{\text{BH}}=\frac{2\pi r_{c}}{4G_{3}}=\frac{2\pi R_{3}\gamma}{4G_{3}}\;,
\end{split}
\eeq
satisfying the first law 
\beq dM=-T_{c}dS_{\text{BH}}+\Omega_{c}dJ\;,\eeq
where $m=1-8G_{3}M$. Moreover, the system obeys the following Smarr relation
\beq 0=-T_{c}S_{\text{BH}}+\Omega_{c}J-2P V_{c}\;,\eeq
where $P=-\frac{\Lambda}{8\pi G_{3}}=-\frac{1}{8\pi G_{3}R_{3}^{2}}$ is a thermodynamic pressure and $V_{c}=\pi r_{c}^{2}$ the conjugate thermodynamic volume. If we allow for variations to the dynamical pressure, the above first law is extended to include a $+V_{c}dP$ term.  

%In contrast, there are three-dimensional black hole solutions to Einstein's equations with a cosmological constant. Specifically, the Banados-Teitelboim-Zanelli (BTZ) spacetime \cite{Banados:1992gq,Banados:1992wn}. The form of the line element follows from the Kerr-$\text{dS}_{3}$ geometry via the Wick rotation\footnote{Moreover, $r_{\pm}\to -ir_{\pm}^{\text{BTZ}}$ such that $\gamma\to\gamma_{\text{BTZ}}$ and $\alpha\to\alpha_{\text{BTZ}}=ir_{-}^{\text{BTZ}}/\ell_{3}=-i4G_{3}J/\gamma_{\text{BTZ}}\ell_{3}$.} $R_{3}\to -i\ell_{3}$, $J\to-J$, and where typically one writes $m=-8G_{3}\tilde{M}$. When $\tilde{M}>0$ and $0\leq \tilde{M}^{2}\ell_{3}^{2}-J^{2}<\infty$, the blackening factor now has two positive roots leading to outer and inner black hole horizons $r_{\pm}$, which follow from a Wick rotation of the parameters in (\ref{eq:rcKerrdS3}), with $r_{+}\geq r_{-}$. The reason a black hole forms in this case is because there is an innate gravitational attraction due to the natural tendency to collapse in AdS. However, when $\tilde{M}<0$ and $0\leq \tilde{M}^{2}\ell_{3}^{2}-J^{2}<1$, the geometry describes spinning particles, i.e., conical $\text{AdS}_{3}$. \AS{Probably exclude final paragraph.}

%%%%%%%%%%%%%%%%%%%%%%%%%%%%%%%%%%%%%%%%%%%%%%%%%%%%%%%%%%%%%%%%%%%%%%%
\section{Perturbative backreaction in Kerr-$\text{dS}_{3}$} \label{app:pertbackKdS}

Here we study perturbative backreaction in the Kerr-$\text{dS}_{3}$ conical defect geometry due to a free conformally coupled scalar field $\Phi$. The complete theory is characterized by
\beq I=\frac{1}{16\pi G_{3}}\int d^{3}x\sqrt{-g}(R-2\Lambda)-\frac{1}{2}\int d^{3}x\sqrt{-g}\left(\frac{1}{8}R\Phi^{2}+(\nabla\Phi)^{2}\right)\;.\eeq
 The energy-momentum tensor for $\Phi$ is found by varying the matter action
\beq T_{\mu\nu}\equiv-\frac{2}{\sqrt{-g}}\frac{\delta I_{\Phi}}{\delta g^{\mu\nu}}=\frac{3}{4}\nabla_{\mu}\Phi\nabla_{\nu}\phi-\frac{1}{4}g_{\mu\nu}(\nabla\Phi)^{2}-\frac{1}{4}\Phi\nabla_{\mu}\nabla_{\nu}\Phi+\frac{1}{4}g_{\mu\nu}\Phi\Box\Phi+\frac{1}{8}G_{\mu\nu}\Phi^{2}\;.\label{eq:stresstensor}\eeq
We will be interested in the case when  $G_{\mu\nu}=-g_{\mu\nu}\Lambda$, where $\Lambda=\frac{1}{R_{3}^{2}}$ and $R= 6/R_{3}^{2}$. Upon invoking the scalar equation of motion,
\beq \left(\Box-\frac{1}{8} R\right)\Phi=0\;,\label{eq:scaleom}\eeq
 it follows that the stress-energy tensor is both traceless and conserved.

Below we compute the renormalized quantum stress-energy tensor $\langle T_{\mu\nu}\rangle$ of the free scalar field. We do this in two steps, primarily following the techniques developed for the BTZ black hole and conical $\text{AdS}_{3}$ \cite{Steif:1993zv,Casals:2016ioo,Casals:2019jfo}, extending the analysis in \cite{Emparan:2022ijy}. First we determine the Green function of the conical $\text{dS}_{3}$ defect geometry (\ref{eq:conicaldS3}) related to Kerr-$\text{dS}_{3}$ using the method of images. We then use point-splitting to compute the renormalized $\langle T_{\mu\nu}\rangle$.

\subsubsection*{Green function in conical $\text{dS}_{3}$}

We begin with the Green function $G(x,x')$ of pure $\text{dS}_{3}$ which solves the scalar field equations of motion (\ref{eq:scaleom}). Imposing transparent boundary conditions,\footnote{We choose transparent boundary conditions because the holographic computation naturally selects these boundary conditions. More generically, the Green function solving the scalar field equation of motion is $4\pi G(x,x')=|x-x'|^{-1}+\lambda|x+x'|^{-1}$, where $\lambda$ is a parameter related to the boundary conditions one imposes; transparent ($\lambda=0$), Dirichlet ($\lambda=-1$) and Neumann ($\lambda=1$).} the Green function is \cite{Avis:1977yn,Lifschytz:1993eb}
\beq G(x,x')=\frac{1}{4\pi}\frac{1}{|x-x'|}\;,\eeq
where %$\lambda$ is a real parameter we will discuss momentarily, and
$|x-x'|\equiv\sqrt{(x-x')^{a}(x-x')_{a}}$ is the chordal or geodesic distance between $x$ and $x'$ in the four-dimensional embedding space $\mathbb{R}^{1,3}$. The embedding coordinates $x^{a}=(X_{1},X_{2},T_{1},T_{2})^{T}$ for empty $\text{dS}_{3}$ are
\beq \hspace{-2mm} T_{1}=\sqrt{\tilde{r}^{2}-R_{3}^{2}}\cosh(\tilde{t}/R_{3})\,,\;\; T_{2}=\sqrt{\tilde{r}^2-R_{3}^{2}}\sinh(\tilde{t}/R_{3})\,,\;\; X_{1}=\tilde{r}\cos\tilde{\phi}\,,\;\; X_{2}=\tilde{r}\sin\tilde{\phi}\,,\label{eq:embeddingcoorddS3}\eeq
obeying $-T_{1}^{2}+T_{2}^{2}+X_{1}^{2}+X_{2}^{2}=R_{3}^{2}$, and where the metric $ds^{2}=-dT_{1}^{2}+dT_{2}^{2}+dX_{1}^{2}+dX_{2}^{2}$ yields empty $\text{dS}_{3}$ in static patch coordinates. Moreover, it is easy to verify
\beq \left(\Box-\frac{3}{4R_{3}^{2}}\right)G_{\text{dS}_{3}}(x,x')=0\;\eeq
for $x\neq x'$, with the chordal distance being
\beq
\begin{split}
|x-x'|&=[-(T_{1}-T'_{1})^{2}+(T_{2}-T'_{2})^{2}+(X_{1}-X'_{1})^{2}+(X_{2}-X'_{2})^{2}]^{1/2}\\
&=\left[2R_{3}^{2}+2\sqrt{\tilde{r}^{2}-R_{3}^{2}}\sqrt{\tilde{r}'^{2}-R_{3}^{2}}\cosh\left(\frac{\tilde{t}-\tilde{t}'}{R_{3}}\right)-2\tilde{r}\tilde{r}'\cos(\tilde{\phi}-\tilde{\phi}')\right]^{1/2}\;.
\end{split}
\eeq

To construct the Green function $G_{\text{CdS}_{3}}(x,x')$ for the conical defect spacetime (\ref{eq:conicaldS3}) we use the method of images, exploiting  the fact the conical defect geometry is an orbifold due to discrete identifications of $\text{dS}_{3}$. Namely, the Green function is given by summing over the distinct images under the action respecting the periodicity conditions (\ref{eq:periodcondsKdS3}). In particular,  identified points are related by an element $H\in SO(1,3)$ on the embedding space coordinates  (\ref{eq:embeddingcoorddS3}), except where now $\tilde{\phi}\sim\tilde{\phi}+2\pi\gamma$ and $\tilde{t}\sim\tilde{t}+2\pi\alpha'$, where we defined $\alpha'=R_{3}\alpha$ in (\ref{eq:periodcondsKdS3}) for notational convenience. 
%Further, $\gamma\equiv 1/N_{+}$ $\tilde{\alpha}\equiv1/N_{-}$ for positive integers $N_{\pm}$ obeying $N_{-}>N_{+}$. 
Explicitly,
\beq H(\gamma,\alpha')=\begin{pmatrix} \cos(2\pi\gamma)&\sin(2\pi\gamma)&0&0\\-\sin(2\pi\gamma)&\cos(2\pi \gamma)&0&0\\ 0&0&\cosh(2\pi\alpha')&-\sinh(2\pi\alpha')\\0&0&-\sinh(2\pi\alpha')&\cosh(2\pi\alpha')\end{pmatrix}\;.\label{eq:identificationmatrixH}\eeq
For integer $n$ we observe $H^{n}(\gamma,\alpha')=H(n\gamma,n\alpha')$. When $\alpha=0$, we recover the identification matrix for static $\text{dS}_{3}$ related to the Schwarzschild-$\text{dS}_{3}$ solution \cite{Emparan:2022ijy}.

The Green function $G_{\text{CdS}_{3}}(x,x')$ for the conical defect spacetime (\ref{eq:conicaldS3}) then follows using the method of images, where one sums over all distinct images of a point obtained by the embedding space identification: 
\beq G_{\text{CdS}_{3}}(x,x')=\frac{1}{4\pi}\sum_{n\in I}G_{\text{dS}_{3}}(x,H^{n}x')=\frac{1}{4\pi}\sum_{n\in I}\frac{1}{|x-H^{n}x'|}\;,\label{eq:GreenfuncdS3}\eeq
%\beq G_{\text{CdS}_{3}}(x,x')=\sum_{n=-\infty}^{\infty}G_{\text{dS}_{3}}(x,H^{n}x')=\frac{1}{4\pi}\sum_{n\in\mathbb{Z}}\frac{1}{|x-H^{n}x'|}\;,\label{eq:imagesum}\eeq
with
\beq
\begin{split}
|x-H^{n}x'|&=\biggr[2\sqrt{\tilde{r}^{2}-R_{3}^{2}}\sqrt{\tilde{r}'^{2}-R_{3}^{2}}\cosh\left(\frac{\tilde{t}-\tilde{t}'+2\pi n\alpha}{R_{3}}\right)\\
&-2\tilde{r}\tilde{r}'\cos\left(\tilde{\phi}-\tilde{\phi}'+2\pi n\gamma\right)+2R_{3}^{2}\biggr]^{1/2}\;.
\end{split}
\eeq
%Notably, in the conical defect spacetime, the infinite sum becomes a finite, 
%\beq G_{\text{CdS}_{3}}(x,x')=\frac{1}{4\pi}\sum_{n=0}^{N-1}\frac{1}{|x-H^{n}x'|}\;,\label{eq:GreenfuncdS3}\eeq
%and where the number $N$ of distinct images, corresponding to the finite number of geodesics connecting two points on a cone \cite{Matschull:1998rv}, is given by the least common multiple of $N_{+}$ and $N_{-}$.
The summation range $I\subset\mathbb{Z}$ depends on the number of distinct images, and is related to the nature of the identification matrix $H$. For the case of the Kerr-dS geometry, the identification matrix (\ref{eq:identificationmatrixH}) will act transitively on $\mathbb{R}^{1,3}$ such that there are a countably infinite number of distinct images, i.e., $I=\mathbb{Z}$. By contrast, in the limit of vanishing rotation $\alpha=0$, the identification matrix $H$ becomes cyclic, such that there are a finite number of distinct images, $N-1$, where $N$ is the smallest positive integer such that $H^{N}=1$. This implies $\gamma$ is a rational number, which without loss of generality can be set to $\gamma=1/N$. The cyclic property is broken for the Kerr-$\text{dS}_{3}$ identification matrix due to the lower block matrix. An analogous story carries over for rotating BTZ and (static or rotating) conical $\text{AdS}_{3}$ \cite{Casals:2016ioo,Casals:2019jfo}. Notably, at this stage, upon the Wick rotation $\ell_{3}= iR_{3}$, and $J\to-J$,\footnote{Moreover, $r_{\pm}\to -ir_{\pm}^{\text{BTZ}}$ such that $\gamma\to\gamma_{\text{BTZ}}$ and $\alpha\to\alpha_{\text{BTZ}}=ir_{-}^{\text{BTZ}}/\ell_{3}=-i4G_{3}J/\gamma_{\text{BTZ}}\ell_{3}$.} one recovers the scalar field Green function in conical $\text{AdS}_{3}$ \cite{Casals:2016ioo}, however, Wick rotating  the identification matrix (\ref{eq:identificationmatrixH}) does not yield the appropriate identification matrix for conical $\text{AdS}_{3}$ or the rotating BTZ.

%\textcolor{red}{The main point of the method of images (in analogy to the electrostatic case) is to sum over ALL the images that contribute to the Green's function, which in our case are those who can be reached by successive applications of the matrix $H$. In the case of conical AdS$_3$ and non rotating dS$_3$, $H$ is cyclic, meaning that there exists only a finite number of images $N$. This is not the case for Kerr-dS$_3$, with the lower diagonal block breaking such cyclic property. Therefore, the sum must be performed on infinite images (both positive and negative $n$). A good sanity check is noting that, if the sum didn't go to negative $n$, we would not have the parity under $x \leftrightarrow x'$.} \textcolor{green}{Also, I am questioning whether we should set $\alpha=1/N_-$ here. Indeed, that is a nice trick for the cyclic part of the matrix, but does not really matter in the hyperbolic sector. I believe we should just set $\gamma=1/N$ and leave $\alpha$ as it is.}%When $\alpha=0$, i.e., vanishing rotation $J$, we recover the Green function for conical $\text{dS}_{3}$ associated with the Schwarzschild-$\text{dS}_{3}$ solution \cite{Emparan:2022ijy}. 

\section*{Renormalized quantum stress-tensor}

The renormalized quantum stress tensor $\langle T_{\mu\nu}\rangle$ is obtain from $G(x,x')$ using the point-splitting method \cite{Christensen:1976vb,Steif:1993zv,Souradeep:1992ia,Casals:2016ioo}. Specifically, 
\beq \langle T_{\mu\nu}(x)\rangle=\lim_{x'\to x}\left(\frac{3}{4}\nabla^{x}_{\mu}\nabla^{x'}_{\nu}G-\frac{1}{4}g_{\mu\nu}g^{\alpha\beta}\nabla^{x}_{\alpha}\nabla^{x'}_{\beta}G-\frac{1}{4}\nabla^{x}_{\mu}\nabla^{x}_{\nu}G+\frac{1}{16R_{3}^{2}}g_{\mu\nu}G\right)\;.\label{eq:qustresstensorads3}\eeq
Here $G(x,x')=G_{\text{CdS}_{3}}(x,x')$ is the Green function (\ref{eq:GreenfuncdS3}), the metric $g_{\mu\nu}$ is a function of the spacetime point $x$, $\nabla_{\mu}^{x}$ denotes a covariant derivative with respect to $x$, and $\nabla_{\mu}^{x'}$ denotes a derivative with respect to the point $x'$. Moreover, the coincident limit $x\to x'$ amounts to evaluating the resulting expression at $x'=x$. Note that while normally the renormalization of the stress tensor is difficult, here we simply subtract off the divergent $n=0$ contribution in the image sum in the coincident limit.

To evaluate each component of the renormalized stress tensor in the conical defect background, we use the fact $G(x,x')$ is a symmetric biscalar, while its covariant derivatives are bitensors. Thus, we invoke a generalization of Synge's theorem for bitensors \cite{Christensen:1976vb} (see also Eq. (54) of \cite{Herman:1995hm}):
\beq \lim_{x'\to x}(\nabla^{x'}_{\mu}A_{\alpha_{1}})=\nabla^{x}_{\mu}\lim_{x'\to x}(A_{\alpha_{1}})-\lim_{x'\to x}(\nabla^{x}_{\mu}A_{\alpha_{1}})\;,\label{eq:syngerule}\eeq
where $A_{\alpha_{1}}$ is a bivector with equal weight at both $x$ and $x'$, whose coincidence limit exists. Consequently, applying Synge's rule (\ref{eq:syngerule}) to the quantum stress tensor (\ref{eq:qustresstensorads3}) we have:
\begin{align}
 \langle T_{\mu\nu}(x)\rangle&=\frac{3}{4}\left[\nabla_{\nu}^{x}\lim_{x'\to x}(\nabla^{x}_{\mu}G)-\lim_{x'\to x}(\nabla^{x}_{\nu}\nabla^{x}_{\mu}G)\right]-\frac{1}{4}g_{\mu\nu}g^{\alpha\beta}\left[\nabla^{x}_{\beta}\lim_{x'\to x}(\nabla_{\alpha}^{x}G)-\lim_{x'\to x}(\nabla^{x}_{\beta}\nabla_{\alpha}^{x}G)\right] \nonumber \\
&+\lim_{x'\to x}\left(-\frac{1}{4}\nabla^{x}_{\mu}\nabla^{x}_{\nu}G+\frac{1}{16R_{3}^{2}}g_{\mu\nu}G\right)\;.
\label{eq:qustresstensorsynge}
\end{align}
To clarify, 
\beq
\begin{split}
\nabla^{x}_{\nu}\lim_{x'\to x}(\nabla^{x}_{\mu}G)&=\partial_{\nu}^{x}\left(\lim_{x'\to x}\partial_{\mu}^{x}G\right)-\Gamma^{\rho}_{\;\nu\mu}\left(\lim_{x'\to x}\partial_{\rho}^{x}G\right)\;, 
\end{split}
\eeq
where the coincident limit is taken before evaluating the $\partial^{x}_{\nu}$ derivative. Meanwhile, 
\beq
\begin{split}
 \lim_{x'\to x}(\nabla^{x}_{\nu}\nabla^{x}_{\mu}G)&=\lim_{x'\to x}(\partial_{\mu}^{x}\partial_{\nu}^{x}G-\Gamma^{\rho}_{\;\mu\nu}\partial_{\rho}G)\;, 
\end{split}
\eeq
where the limit $x\to x'$ is taken at the end.

Evaluating the quantum stress-tensor (\ref{eq:qustresstensorsynge}) in the defect geometry (\ref{eq:conicaldS3}) and performing the inverse coordinate transformation of (\ref{eq:coordtranstocon}) to return to $(t,r,\phi)$ coordinates yields,
\beq
\begin{split}
&\langle T^{t}_{\;t}\rangle=\frac{1}{8\pi}\sum_{n=1}^{\infty}\frac{(4r^{2}[(\beta_{+}^{2}+\beta_{-}^{2})b_{n}-3a_{n}]+2R_{3}^{2}g_{n})c_{n}-3\beta_{+}\beta_{-}e_{n}[8r^{2}+(\beta_{-}^{2}-\beta_{+}^{2})R_{3}^{2}]}{(\beta_{+}^{2}+\beta_{-}^{2})^{2}d_{n}(r)^{5/2}}\;,\\
&\langle T^{r}_{\;r}\rangle=\frac{1}{16\pi}\sum_{n=1}^{\infty}\frac{c_{n}}{d_{n}(r)^{3/2}}\;,\\
&\langle T^{\phi}_{\;\phi}\rangle=-\frac{1}{8\pi}\sum_{n=1}^{\infty}\frac{(4r^{2}[3\bar{a}_{n}-(\beta_{+}^{2}+\beta_{-}^{2})b_{n}]+2R_{3}^{2}\bar{g}_{n})c_{n}-3\beta_{+}\beta_{-}e_{n}[8r^{2}+(\beta_{-}^{2}-\beta_{+}^{2})R_{3}^{2}]}{(\beta_{+}^{2}+\beta_{-}^{2})^{2}d_{n}(r)^{5/2}}\;,\\
&\langle T^{t}_{\;\phi}\rangle=-\frac{3R_{3}}{8\pi}\sum_{n=1}^{\infty}\frac{\beta_{+}\beta_{-}c_{n}[4r^{2}(c_{n}-4)-R_{3}^{2}a_{n}]+e_{n}[4r^{2}(\beta_{+}^{2}-\beta_{-}^{2})+2\beta_{-}^{2}\beta_{+}^{2}R_{3}^{2}]}{(\beta_{+}^{2}+\beta_{-}^{2})^{2}d_{n}(r)^{5/2}}\;,\\
&\langle T^{\phi}_{\;t}\rangle=-\frac{3}{8\pi R_{3}}\sum_{n=1}^{\infty}\frac{\beta_{+}\beta_{-}c_{n}[4r^{2}(c_{n}-4)-R_{3}^{2}a_{n}]+e_{n}[4r^{2}(\beta_{+}^{2}-\beta_{-}^{2})-R_{3}^{2}(\beta_{+}^{4}+\beta_{-}^{4})]}{(\beta_{+}^{2}+\beta_{-}^{2})^{2}d_{n}(r)^{5/2}}\;.
\end{split}
\label{eq:compsofTapp}\eeq
Here we introduced parameters $\beta_{+}\equiv 2\gamma$ and $\beta_{-}\equiv2\alpha'/R_{3}$ and defined
\beq a_{n}=2\left[\beta_{-}^{2}\sin^{2}\left(\frac{n\pi\beta_{+}}{2}\right)+\beta_{+}^{2}\sinh^{2}\left(\frac{n\pi\beta_{-}}{2}\right)\right]\;,\eeq
\beq \bar{a}_{n}=2\left[\beta_{+}^{2}\sin^{2}\left(\frac{n\pi\beta_{+}}{2}\right)+\beta_{-}^{2}\sinh^{2}\left(\frac{n\pi\beta_{-}}{2}\right)\right]\;,\eeq
\beq b_{n}\equiv2\left[\sinh^{2}\left(\frac{n\pi\beta_{-}}{2}\right)+\sin^{2}\left(\frac{\pi n\beta_{+}}{2}\right)\right]\;.\eeq
\beq c_{n}\equiv2+\cos(\pi n\beta_{+})+\cosh(\pi n\beta_{-})\;,\eeq
\beq e_{n}\equiv 2\sin(n\pi\beta_{+})\sinh(n\pi\beta_{-})\;,\eeq
\beq g_{n}\equiv \beta_{-}^{2}(\beta_{+}^{2}-2\beta_{-}^{2})\sin^{2}\left(\frac{n\pi\beta_{+}}{2}\right)-\beta_{+}^{2}(\beta_{-}^{2}-2\beta_{+}^{2})\sinh^{2}\left(\frac{n\pi\beta_{-}}{2}\right)\;,\eeq
\beq \bar{g}_{n}\equiv\beta_{+}^{2}(\beta_{+}^{2}-2\beta_{-}^{2})\sinh^{2}\left(\frac{n\pi\beta_{-}}{2}\right)-\beta_{-}^{2}(\beta_{-}^{2}-2\beta_{+}^{2})\sin^{2}\left(\frac{n\pi\beta_{+}}{2}\right)\;,\eeq
and with denominator
\beq d_{n}(r)\equiv\frac{4R_{3}^{2}}{\beta_{+}^{2}+\beta_{-}^{2}}\biggr[\beta_{-}^{2}\sin^{2}\left(\frac{n\pi\beta_{+}}{2}\right)-\beta_{+}^{2}\sinh^{2}\left(\frac{\pi n\beta_{-}}{2}\right)+2R_{3}^{-2}r^{2}b_{n}\biggr]\;,\eeq
We have already removed the divergent $n=0$ contribution. Note that we recover components of the renormalized stress-tensor for a free conformally coupled scalar field in conical $\text{AdS}_{3}$ \cite{Casals:2019jfo} upon the Wick rotations $R_{3}\to -iL$ and $\beta_{-}\to i\beta_{-}$. To arrive at these expressions we used the summation symmetry over negative and positive integers $n\in\mathbb{Z}$ such that $\sum_{n\in \mathbb{Z}/\{0\}}f_{n}=\frac{1}{2}\sum_{n=1}^{\infty}(f_{n}+f_{-n})$, where $f_{-n}=\pm f_{n}$ depending on each summand $f_{n}$. For example, $\langle T^{\tilde{r}}_{\;\tilde{t}}\rangle$ has a numerator $\sin^{2}(n\pi\gamma)\sinh(2\pi n\alpha'/R_{3})$ which is eliminated under the sum symmetry between positive and negative integers.\footnote{In the non-rotating case a similar symmetry argument is made using $H^{N}=1$ (see Eq. (3.42) in \cite{Casals:2019jfo}).}
%$$\sum_{n=1}^{N-1}f_{n}=\frac{1}{2}\sum_{n=1}^{N-1}(f_{n}+f_{n-N})=\frac{1}{2}\sum_{n=1}^{N-1}(f_{n}+f_{-n})\;,$$
%a consequence of $H^{N}=1$ \cite{Casals:2019jfo}.
This symmetry eliminates all mixed components with $\tilde{r}$. 
%\footnote{Explicitly, 
%$$t=\frac{R_{3}^{2}\left(\gamma\tilde{t}-\alpha'\phi\right)}{(R_{3}\gamma)^{2}+(\alpha')^{2}}\;,\quad \phi=\frac{(\alpha'\tilde{t}+R_{3}^{2}\gamma\tilde{\phi})}{(R_{3}\gamma)^{2}+(\alpha')^{2}}\;,\quad \tilde{r}^{2}=R_{3}^{2}\left(\frac{r^{2}+(\alpha')^{2}}{(R_{3}\gamma)^{2}+(\alpha')^{2}}\right)\;.$$}

To compare to the holographic stress-tensor it is useful to define
\beq r_{n}\equiv d_{n}^{1/2}\;,\quad d_{n}=r^{2}d_{n}^{(1)}+R_{3}^{2}d_{n}^{(2)}\;,\eeq
with
\beq d_{n}^{(1)}=\frac{8b_{n}}{(\beta_{+}^{2}+\beta_{-}^{2})}\;,\quad d^{(2)}_{n}=\frac{4}{\beta_{+}^{2}+\beta_{-}^{2}}\biggr[\beta_{-}^{2}\sin^{2}\left(\frac{n\pi\beta_{+}}{2}\right)-\beta_{+}^{2}\sinh^{2}\left(\frac{\pi n\beta_{-}}{2}\right)\biggr]\;,\eeq
so as to suggestively write the stress-tensor components as in (\ref{eq:compsstresstenscds3}) with coeffcients
%\beq
%\begin{split}
%&\langle T^{t}_{\;t}\rangle=\frac{1}{8\pi}\sum_{n=1}^{\infty}\frac{1}{r_{n}^{3}}\left(A_{n}+\frac{\tilde{A}_{n}}{r_{n}^{2}}\right),\quad \langle T^{r}_{\;r}\rangle=\frac{1}{16\pi}\sum_{n=1}^{\infty}\frac{c_{n}}{r_{n}^{3}},\quad \langle T^{\phi}_{\;\phi}\rangle=-\frac{1}{8\pi}\sum_{n=1}^{\infty}\frac{1}{r_{n}^{3}}\left(B_{n}+\frac{\tilde{A}_{n}}{r_{n}^{2}}\right),\\
% &\langle T^{t}_{\;\phi}\rangle=-\frac{3R_{3}}{8\pi}\sum_{n=1}^{\infty}\frac{1}{r_{n}^{3}}\left(E_{n}+\frac{\tilde{E}_{n}}{r_{n}^{2}}\right),\quad \langle T^{\phi}_{\;t}\rangle=-\frac{3}{8\pi R_{3}}\sum_{n=1}^{\infty}\frac{1}{r_{n}^{3}}\left(E_{n}+\frac{F_{n}}{r_{n}^{2}}\right),
%\end{split}
%\eeq
%where 
\beq 
\begin{split}
&A_{n}\equiv\frac{A'_{n}}{r_{n}^{2}}\;,\quad A'_{n}=\frac{(4r^{2}[(\beta_{+}^{2}+\beta_{-}^{2})b_{n}-3a_{n}]+2R_{3}^{2}g_{n})c_{n}}{(\beta_{+}^{2}+\beta_{-}^{2})^{2}}\;,\\
&B_{n}\equiv\frac{B'_{n}}{r_{n}^{2}}\;,\quad B'_{n}=\frac{(4r^{2}[3\bar{a}_{n}-(\beta_{+}^{2}+\beta_{-}^{2})b_{n}]+2R_{3}^{2}\bar{g}_{n})c_{n}}{(\beta_{+}^{2}+\beta_{-}^{2})^{2}}\;,\\
&E_{n}\equiv\frac{E'_{n}}{r_{n}^{2}}\;,\quad E'_{n}=\frac{\beta_{+}\beta_{-}c_{n}[4r^{2}(c_{n}-4)-R_{3}^{2}a_{n}]}{(\beta_{+}^{2}+\beta_{-}^{2})^{2}}\;,\\
&\tilde{A}_{n}=-\frac{3\beta_{+}\beta_{-}e_{n}[8r^{2}+(\beta_{-}^{2}-\beta_{+}^{2})R_{3}^{2}]}{(\beta_{+}^{2}+\beta_{-}^{2})^{2}}\;,\\
&\tilde{E}_{n}=\frac{e_{n}[4r^{2}(\beta_{+}^{2}-\beta_{-}^{2})+2\beta_{-}^{2}\beta_{+}^{2}R_{3}^{2}]}{(\beta_{+}^{2}+\beta_{-}^{2})^{2}}\;,\quad F_{n}=\frac{e_{n}[4r^{2}(\beta_{+}^{2}-\beta_{-}^{2})-R_{3}^{2}(\beta_{+}^{4}+\beta_{-}^{4})]}{(\beta_{+}^{2}+\beta_{-}^{2})^{2}}\;.
\end{split}
\eeq

\section*{Quantum-corrected geometry}

With the renormalized quantum stress-tensor at hand, let us proceed and compute the quantum corrections to Kerr-$\text{dS}_{3}$ by solving the three-dimensional semi-classical Einstein equations 
\beq 
\label{eq:semiclass}
G_{\mu\nu}+\frac{1}{R_{3}^{2}}g_{\mu\nu}=8\pi G_{3}\langle T_{\mu\nu}\rangle\eeq
perturbatively in $L_{\text{P}}$. Our strategy closely follows the $\text{AdS}_{3}$ analysis presented in \cite{Casals:2019jfo}. Our starting point is the general form of the stationary (rotating) balck hole:
\beq
ds^2 = N(r)^2 f(r) dt^2+ \frac{dr^2}{f(r)}+r^2(d\theta+k(r) dt)^2 \ ,
\eeq
where $k$, $f$ and $N$ are the functions to be determined. Expanding to linear order in $L_p$
\beq
\begin{split}
N(r) = N_0(r) + L_P N_1(r) + \mathcal{O}(L_P^2) \ , \\ 
f(r) = f_0(r) + + L_P f_1(r)+ \mathcal{O}(L_P^2) \ , \\ 
k(r) = k_0(r)+ L_P k_1(r)+ \mathcal{O}(L_P^2) \ .
\end{split}
\eeq
We substitute these expressions back into the left hand side of (\ref{eq:semiclass}) and match order by order in $L_P$ with the expectation value for the stress energy tensor. At $\mathcal{O}(L_P^0)$ we obtain 
%(where S is linear in $L_P$):
\beq N_{0}=1\;,\quad f_{0}=m-\frac{r^{2}}{R_{3}^{2}}+\frac{J^{2}}{4r^{2}}\;,\quad k_{0}=\frac{J}{2r^{2}}\;,\eeq
recovering the Kerr-$\text{dS}_{3}$ metric (\ref{eq:KdS3metapp}), to a minor redefinition of the constants of integration $J$. Note that we have also made the choice of a vanishing shift vector at infinity, which sets a third integration constant to zero.

At linear order in $L_P$, we solve the differential equations for the linear corrections to the Kerr metric using that we have computed the stress-tensor to linear order in $L_{P}$. It is helpful to use the traceless property of the perturbative stress energy tensor,
\beq
\langle T_{tt}\rangle = f_0^2\langle T_{rr} \rangle - \left(\frac{1}{R_3^2}-\frac{m}{r^2}\right) \langle T_{\phi\phi} \rangle + \frac{J}{r^2} \langle T_{t\phi} \rangle \ . 
\eeq
Going through the algebra and requiring the corrections to go to zero for vanishing stress energy tensor, we eventually find
\beq
\begin{split}
\hspace{-2mm}&N_1(r) = \frac{8\pi G_3}{L_P} \int dr\left(2r \langle T_{rr}\rangle+ \frac{\langle T_{\phi\phi} \rangle}{r f_0}\right),\\
&f_1(r)= \int \hspace{-1mm} dr\left[ -2 f_0 N_1'+ \left(\frac{2m}{r}+\frac{J^2}{r^3}\right)N_{1}  +  \frac{2}{r^3} \int  \hspace{-1mm} dr\left (-2mrN_1+\frac{8\pi G_3}{L_P}r^3 f_0 \langle T_{rr}\rangle \right) \right],\\
& Jk_1 = f_1+2f_0 N_1 + 2 \int r dr \left(\frac{2N_1(r)}{R_3^2}-f_0 \frac{8\pi G_3}{L_P}\langle T_{rr} \rangle \right).
\end{split}
\eeq
Substituting in (\ref{eq:compsofTapp}) and integrating, we recover the expressions for $f_{1},N_{1}$ and $k_{1}$ presented in the main text (\ref{eq:N1func}) -- (\ref{eq:k1func}), with
\beq
h_n \equiv (4r^2-R_3 \beta_+^2)(4r^2+R_3^2 \beta_-^2) b_n + (\beta_+^2+ \beta_-^2) \left(4r^2-\frac{(\beta_-^2-\beta_-^2)R_3^2}{2}\right)d_n \;.
\eeq

%\beq
%N_1(r) = \frac{R_3^2}{2(\beta_+^2 + \beta_-^2)}\sum^\infty_{n=1} \frac{a_n c_n-2\beta_+ \beta_- e_n}{b_n d_n^{3/2}} \ , 
%\eeq
%\beq
%f_1(r) = \sum^\infty_{n=1} \frac{4h_n (a_n c_n - 2\beta_+\beta_- e_n)-c_n d_n^2 (\beta_+^2 + \beta_-^2)^{3}}{64r^2(\beta_+^2 + \beta_-^2)b_n^2 d_n^{3/2}} \ ,
%\eeq
%\beq
%k_1(r) = -\frac{R_3}{8r^2}\sum^\infty_{n=1} \frac{(\beta_+^2-\beta_-^2)e_n+\beta_+ \beta_- c_n(c_n-4)}{b_n^2 d_n^{1/2}} \ ,
%\eeq
%where we have defined

%%%%%%%%%%%%%%%%%%%%%%%%%%%%%%%%%%%%%%%%%%%%%%%%%%%%%%%%%%%%%%%%%%%%
\section{Elements of the AdS C-metric} \label{app:AdSCmetprops}

The $\text{AdS}_{4}$ C-metric is a solution to Einstein's equation with negative cosmological constant, arising from a particular rescaling of the Plebanski-Demianski solution \cite{Plebanski:1976gy}:\footnote{Here we follow the conventions of \cite{Emparan:1999fd}. To recover the form of the C-metric used in \cite{Emparan:2020znc} one identifies
$$\lambda =\frac{\ell^{2}}{\ell_{3}^{2}}\;,\quad A=\frac{1}{\ell}\;,\quad k=-\kappa\;,\quad 2mA=\mu\;,\quad y=\frac{-\ell}{r}$$
and further rescale $t\to t/\ell$. To connect to the conventions of \cite{Emparan:2022ijy}, we make the same replacement except restrict to $\kappa=+1$ and replace $\ell_{3}^{2}=-R_{3}^{2}$, which amounts to a Wick rotation of $\ell_{3}$.\label{fn:convsapp}}
\beq ds^{2}=\frac{1}{A^{2}(x-y)^{2}}\left[H(y)dt^{2}-\frac{dy^{2}}{H(y)}+\frac{dx^{2}}{G(x)}+G(x)d\phi^{2}\right]\;,\label{eq:genAdSC}\eeq
with
\beq H(y)=-\lambda+ky^{2}-2mAy^{3}\;,\quad G(x)=1+kx^{2}-2mAx^{3}\;,\eeq
and $k=+1,0,-1$ which will determine the topology of the horizon of the black hole solutions when they exist. The parameters $A$ and $m$ can be thought of as acceleration and mass, respectively, while $\lambda$ is related to the cosmological constant. Indeed, the bulk Ricci tensor satisfies $\hat{R}_{AB}=-(3/L_{4}^{2})\hat{g}_{AB}$ where $L_{4}\equiv (A\sqrt{\lambda+1})^{-1}$ sets the scale for the bulk cosmological constant. Maintaining a negative cosmological constant in the bulk requires $\lambda>-1$, however, as summarized below, various ranges of $\lambda$ describe different asymptotic brane geometries.

The overall factor $(x-y)^{-2}$ in (\ref{eq:genAdSC}) implies the point $y=x$ is infinitely far away from points $y\neq x$ (the point $y=x$ corresponds to the asymptotic $\text{AdS}_{4}$ geometry). A curvature singularity is located at $y=-\infty$, which is hidden behind one of the horizons. To maintain a `mostly plus' Lorentzian signature requires $G(x)\geq0$, restricting the range of $x$.
%Also one typically takes $-\infty<y<x$. 
%To get a better handle on the metric, consider the coordinate transformation $y=-1/Ar$, $t\to At$ and $x=\cos\theta$. 
%for which the C-metric becomes 
%\beq ds^{2}=\frac{1}{(1+Ar\cos\theta)^{2}}\biggr\{H(r)dt^{2}-H^{-1}(r)dr^{2}+\frac{r^{2}\sin^{2}\theta}{G(\theta)}d\theta^{2}+r^{2}G(\theta)d\phi^{2}\biggr\}\;,\eeq
%with
%\beq H(r)=-A^{2}r^{2}\lambda+k+\frac{2m}{r}\;,\quad G(\theta)=1+k\cos^{2}\theta-2mA\cos\theta^{3}\;.\eeq
% Upon taking the limit $A\to0$, $\lambda\to0$ and $k\to-1$, the geometry reduces to the four-dimensional (flat) Schwarzschild black hole. Thus, it is natural to interpret $t$ as a time, $y$ as an inverse radial coordinate, and $\phi$ and $x$ as angular coordinates.

 Each zero of $H(y)$ corresponds to a Killing horizon associated with the time translation Killing vector $\partial_{t}$. Meanwhile, the zeros of $G(x)$ correspond to an axis for the rotation symmetry $\partial_{\phi}$, i.e., for $\xi^{a}=\partial_{\phi}^{a}$, then $\xi^{2}\sim G(x)$, vanishing at a zero of $G(x)$. For a range of values of $mA$ and $k$, there will be three distinct real zeros\footnote{Explicitly, the cubic $G(x)=-2mA x^{3}+kx^{2}+1=0$ can be solved by introducing $x=z-\frac{k}{3}$ and express in depressed form, $z^{3}+pz+q=0$, with $p=-\frac{k^{2}}{12(mA)^{2}}$ and $q=-\frac{[2k^{3}+27(4mA)^{2}]}{27(2mA)^{3}}$, such that the discriminant is $\Delta\equiv-(4p^{3}+27q^{2})=-\frac{k^{3}+27(mA)^{2}}{4(mA)^{4}}$. When $\Delta>0$, then $G(x)$ will have three distinct real roots, while if $\Delta<0$ then $G(x)$ will have one real root and two complex roots. In the flat space case \cite{Emparan:1999wa}, where $k=-1$, then $G(x)$ will have three distinct roots $x_{0}<x_{2}<0<x_{1}$ provided $0<mA<\frac{1}{3\sqrt{3}}$.}  to $G(x)$, $\{x_{0},x_{1},x_{2}\}$, with each zero leading to a distinct conical singularity. One singularity can be removed via\footnote{To see this, introduce a coordinate $\tilde{x}^{2}=4(x-x_{i})/G'(x_{i})$. Then expand the $(x,\phi)$ sector of the line element (\ref{eq:genAdSC}) about a zero of $G(x)$, where $G^{-1}(x)dx^{2}+G(x)d\phi^{2}\approx[G'(x_{i})(x-x_{i})]^{-1}dx^{2}+G'(x_{i})(x-x_{i})d\phi^{2}=\tilde{x}^{2}(G'(x_{i})/2)^{2}d\phi^{2}+d\tilde{x}^{2}$. Periodicity (\ref{eq:periodicityphiapp}) follows from imposing regularity at $\tilde{x}=0$.\label{fn:consingperiod}}
\beq \phi\sim \phi+\Delta\phi\;,\quad \Delta\phi=\frac{4\pi}{|G'(x_{i})|} \;,\label{eq:periodicityphiapp}\eeq
where $x_{i}$ is one of the zeros. Once the period of $\phi$ has been fixed in this way, say at $x=x_{1}$, then $\phi$ cannot be readjusted to eliminate the remaining conical singularities at $x=x_{0},x_{2}$. Thus, in general there will be a conical singularity along the axis $x=x_{i}$ with deficit angle
\beq \delta=\frac{4\pi}{G'(x_{i})}-\Delta\phi\;,\eeq
which can be interpreted as a cosmic string with tension $\tau_{cs}=\delta/8\pi$. Note that it is this feature which leads one to interpret the C-metric as a single or pair of accelerating black holes. For example, for a single black hole, a cosmic string attaches at one pole in the background and the black hole, suspending it away from the center of the spacetime, thereby inducing its acceleration (for a detailed analysis on the interpretation of the C-metric, see \cite{Griffiths:2006tk}). This acceleration leads to an additional acceleration horizon, analogous to a Rindler horizon, and an equilibrium thermodynamic description \cite{Emparan:1999fd} (see also \cite{Appels:2016uha}).

We are interested in introducing a brane into the $\text{AdS}_{4}$ spacetime. Generally there will be a discontinuity in the extrinsic curvature
%\footnote{The brane $\mathcal{B}$ is a timelike 3-surface with spacelike unit normal $n_{i}$ and extrinsic curvature $K_{ij}=\nabla_{i}n_{j}$.} 
$K_{ij}[h]$ across the brane which, via the Israel junction conditions (equations of motion for the brane), is related to the stress-tensor introduced by the brane. In the four-dimensional case at hand, where the brane action is purely tensional, the junction conditions are
\beq \Delta K_{ij}-h_{ij}\Delta K^{k}_{\;k}=8\pi G_{4}\tau h_{ij}\;,\eeq
where $\Delta K_{ij}=K_{ij}^{+}-K_{ij}^{-}=2K_{ij}$ and $S_{ij}=-\tau h_{ij}$. Therefore, the tension can be seen as a parameter which fixes the location of the brane $\mathcal{B}$. In the case of the C-metric, a natural choice for the location of $\mathcal{B}$ is the surface $x=0$ since it is \emph{umbilic}. To see this, note that the unit normal to the brane at $x=0$ is $n_{x}^{i}=A\epsilon(x-y)\sqrt{G(x)}\partial^{i}_{x}$, where $\epsilon=\pm1$ corresponds to the orientation of the normal; here we take $\epsilon=+1$ since $x=0$ is a timelike hypersurface. The non-vanishing components of $K_{ij}=\nabla_{i}n_{j}$ obey  $K_{ij}=-A h_{ij}^{(x)}$,
%\beq
%\begin{split}
%&K_{tt}|_{x=0}=-\frac{1}{Ay^{2}}H(y)=-A h^{(x)}_{tt}|_{x=0}\;,\\
%&K_{yy}|_{x=0}=\frac{1}{Ay^{2}H(y)}=-A h^{(x)}_{yy}|_{x=0}\;,\\
%&K_{\phi\phi}|_{x=0}=-\frac{1}{Ay^{2}}=-A h^{(x)}_{\phi\phi}|_{x=0}\;,
%\end{split}
%\eeq
with $h^{(x)}_{ij}$ being the induced metric along the $x=0$ brane. Comparing to the Israel junction conditions we identify the brane tension (\ref{eq:branetens}).  
%We then perform surgery on the bulk spacetime by cutting $\text{AdS}_{4}$ at this surface, for which on one of the sides of $\mathcal{B}$ there will be no conical singularities. We then take two copies of the conical singularity free side of the spacetime and glue them together along $\mathcal{B}$ such that the resulting spacetime is free of conical singularities. 
Similarly, the $y=0$ hypersurface is umbilic. Indeed, with unit normal $n_{y}^{i}=A\epsilon(x-y)\sqrt{H(y)}\partial^{i}_{y}$, then $K_{ij}=-A\epsilon\sqrt{-\lambda}h^{(y)}_{ij}$,
%\beq
%\begin{split}
%&K_{tt}|_{y=0}=-\frac{\epsilon}{Ax^{2}}(-\lambda)^{3/2}=-A\epsilon\sqrt{-\lambda}h^{(y)}_{tt}|_{y=0}\;,\\
%&K_{xx}|_{y=0}=-\frac{\epsilon}{A x^{2} G(x)}\sqrt{-\lambda}=-A\epsilon\sqrt{-\lambda}h^{(y)}_{xx}|_{y=0}\;,\\
%&K_{\phi\phi}|_{y=0}=\frac{-\epsilon G(x)}{A x^{2}}\sqrt{-\lambda}=-A\epsilon\sqrt{-\lambda}h^{(y)}_{\phi\phi}|_{y=0}\;,
%\end{split}
%\eeq
where $h^{(y)}_{ij}$ is the induced metric at $y=0$.
%such that $K_{ij}=(-A\epsilon\sqrt{-\lambda})h^{(y)}_{ij}$.

To gain some intuition for the brane construction, it is helpful to consider the simplifying case when $mA=0$. One can then move to a coordinate frame showing the geometry is locally $\text{AdS}_{4}$ where the brane itself has a three-dimensional cosmological constant $\Lambda_{3}=-\lambda$ \cite{Emparan:1999fd}.
%following coordinate transformation \cite{Emparan:1999fd}
%\beq \tilde{r}=\frac{\sqrt{y^{2}+\lambda x^{2}}}{A(x-y)}\;,\quad \rho=\sqrt{\frac{1+kx^{2}}{y^{2}+\lambda x^{2}}}\;,\eeq
%the metric (\ref{eq:genAdSC}) becomes
%\beq ds^{2}=\frac{d\tilde{r}^{2}}{\frac{\tilde{r}^{2}}{\ell_{4}^{2}}-\lambda}+\tilde{r}^{2}\left[-(\lambda\rho^{2}-k)dt^{2}+\frac{d\rho^{2}}{\lambda\rho^{2}-k}+\rho^{2}d\phi^{2}\right]\;.\label{eq:localAdS4m0}\eeq
%Locally the geometry is $\text{AdS}_{4}$, where surfaces of constant $\tilde{r}$ have constant Riemann curvature 
%with a three-dimensional cosmological constant $\Lambda_{3}=-\lambda$.
Thus, the sign of $\lambda$ denotes different constant curvature slicings of $\text{AdS}_{4}$. There are three distinct cases: \textbf{(1)} $\lambda=0$, a flat slicing. In this case one must choose $k=\pm1$, where for $k=-1$ the coordinate $t$ is timelike everywhere; \textbf{(2)} $-1<\lambda<0$, leads to a three-dimensional de Sitter slicing. One must select $k=-1$ to have $\text{dS}_{3}$ in static patch coordinates and cosmological horizons, and \textbf{(3)} $\lambda>0$, an $\text{AdS}_{3}$ slicing where the three different values of $k$ distinguish three distinct slicings of $\text{AdS}_{3}$: global coordinates ($k=-1$), the massive BTZ black hole ($k=+1$), and the massless BTZ black hole ($k=0$). The flat ($\lambda=0$) solution was studied in \cite{Emparan:1999wa} while the $\text{AdS}_{3}$ slicings were analyzed in \cite{Emparan:1999fd}.

\subsubsection*{Adding rotation}

For completion, let us briefly review the rotating $\text{AdS}_{4}$ C-metric, following the conventions of \cite{Emparan:1999fd}.\footnote{To recover the form of the metric used in \cite{Emparan:2020znc} one makes the identifications in Footnote \ref{fn:convsapp}, together with $\sqrt{\lambda}a\to a/\ell_{3}$, while to recover the metric used in the main text (\ref{eq:AdS4Ccoord}), one identifies $\sqrt{\lambda}a\to -a/iR_{3}$.} The line element is
\beq 
\begin{split}
ds^{2}&=\frac{1}{A^{2}(x-y)^{2}}\biggr[\frac{H(y)}{\Sigma(x,y)}(dt+ax^{2}d\phi)^{2}-\frac{\Sigma(x,y)}{H(y)}dy^{2}+\frac{\Sigma(x,y)}{G(x)}dx^{2}+\frac{G(x)}{\Sigma(x,y)}(d\phi-ay^{2}dt)^{2}\biggr]\;,
\end{split}
\label{eq:rotCmetapp}\eeq
with metric functions
\beq
\begin{split}
&H(y)=-\lambda+ky^{2}-2mAy^{3}-a^{2}y^{4}\;,\quad \Sigma(x,y)=1+a^{2}x^{2}y^{2}\\
&G(x)=1+kx^{2}-2mAx^{3}+a^{2}\lambda x^{4}\;.
\end{split}
\eeq
%The quartic order in $y^{4}$ and $x^{4}$ also appears for the charged C-metric, and it is therefore straightforward to include charge, however, we will not do so here. 
As in the static case, this spacetime is obeys $\hat{R}_{AB}=-3/L_{4}^{2}\hat{g}_{AB}$ with the  same scale $L_{4}$. When $m\neq0$, there is a curvature singularity when $1/y^{2}\Sigma(x,y)=0$, i.e., when both $y\to-\infty$ and $x=0$, which may be recognized as the standard ring singularity familiar to Kerr black holes. 

The zeros $x_{i}$ of $G(x)$ now correspond to fixed orbits of the rotational Killing vector 
\beq \xi=\partial_{\phi}-ax_{i}^{2}\partial_{t}\;.\eeq
Indeed, the vector $\partial_{\phi}^{\mu}$ no longer has vanishing norm at $x=x_{i}$, while $\xi$ does. Avoiding a conical defect at, say, $x=x_{1}$ requires one identify points along the integral curves of this Killing vector with an appropriate period, amounting to coordinate transformation $\tilde{t}=t+ax_{1}^{2}\phi$, where angular variable $\phi$ has the same period (\ref{eq:periodicityphiapp}). To see this, expand the relevant portion of the metric (\ref{eq:rotCmetapp}) near a zero of $G(x)$. Without loss of generality, consider the slice $y=0$, where, up to the conformal factor
\beq ds^{2}\approx -\lambda(dt+ax_{i}^{2}d\phi)^{2}+\frac{dx^{2}}{G'(x_{i})(x-x_{i})}+G'(x_{i})(x-x_{i})d\phi^{2}\;.\eeq
Aside from the first term, the $(x,\phi)$ sector takes the same form as in the non-rotating case, from which the periodicity of $\phi$ is (\ref{eq:periodicityphiapp}) (see Footnote \ref{fn:consingperiod}). Including rotation, however, this would not be the correct periodicity for $\phi$. The situation is remedied with the coordinate transformation $\tilde{t}=t+ax_{i}^{2}\phi$, such that, at $x=x_{i}$, then $d\tilde{t}=(dt+ax_{i}^{2}d\phi)$. Similarly, at the roots $y_{i}$ of $H(y)$, the Killing vector $\zeta=\partial_{t}+ay_{i}^{2}\partial_{\phi}$
becomes null, defining horizons with angular velocity $\Omega=ay_{i}^{2}$. 
%An essential new feature with the rotating C-metric is that, for non-zero $a$, the function $H(y)$ acquires a new positive root, which is identified as an inner black hole horizon. 

The brane is again placed at $x=0$ since this surface remains umbilic. Indeed, for spacelike unit normal $n_{x}^{i}=A(x-y)\sqrt{G(x)/\Sigma(x,y)}\partial_{x}^{i}$, the extrinsic curvature again satisfies $K_{ij}=-Ah^{(x)}_{ij}$ at $x=0$. Similarly, the $y=0$ hypersurface, with unit normal $n^{i}_{y}=A\epsilon(x-y)\sqrt{H(y)/\Sigma(x,y)}\partial_{y}^{i}$, obeys $K_{ij}=(-A\epsilon\sqrt{-\lambda})h^{(y)}_{ij}$.

 Due to the periodicity in $\phi$, notice that along the orbit $\xi$ the coordinate $t$ is shifted: via $t=\tilde{t}-ax_{1}^{2}\phi$ and $\tilde{t}\sim\tilde{t}$, then $t\sim t-ax_{1}^{2}\Delta\phi$. Consequently, this introduces a rotation of frames in the asymptotic limit; namely, introducing radial coordinate $\rho=-y^{-1}$  and performing the coordinate transformation $(t,y,\phi)\to(\tilde{t},\rho,\phi)$, the $h_{\tilde{t}\phi}$ component of the brane metric at $x=0$ will grow as $\rho^{2}$ and not as a constant \cite{Emparan:1999fd}. To remove this undesired asymptotic growth, one further shifts $\phi=\tilde{\phi}+C\tilde{t}$ for a judiciously chosen constant\footnote{Specifically, take the brane geometry in coordinates $(\tilde{t},\rho,\phi)$ and then perform the shift $\phi=\tilde{\phi}+C\tilde{t}$. In the large $\rho$ limit, one finds that the $d\tilde{t}d\tilde{\phi}$ component will grow as $\rho^{2}$ unless $C\equiv -\lambda ax_{1}^{2}/(1-\lambda a^{2}x_{1}^{4})$. Note this choice of constant $C$ differs from the one in the main text because here we called $t=\tilde{t}-ax_{1}^{2}\phi$ and then redefined $\phi$, whereas in the main text we defined $t=\tilde{t}-ax_{1}^{2}\tilde{\phi}$.} to remove the $\rho^{2}$ growth in the coordinate frame $(\tilde{t},\rho,\tilde{\phi})$. To place the brane metric in more canonically normalized coordinates, one further rescales coordinates $\tilde{t}$ and $\tilde{\phi}$ and redefines the radial coordinate $\rho$.
 %, as carried out in the main text. 

%Again there are three interesting slices depending on the value of $\lambda$ \cite{Emparan:1999fd} \textbf{(1)} $\lambda=0$, then one sets $k=-1$ such that $\partial_{t}$ remains timlike at infinity on the brane. The metric on the brane is \emph{locally} equal to an equatorial slice of an asymptotically flat Kerr-black hole in four-dimensions; \textbf{(2)} $-1<\lambda<0$, one continues to use $k=-1$ to have $\text{dS}_{3}$ and a cosmological horizon. This system describes the (quantum-corrected) rotating $\text{dS}_{3}$ black hole analyzed in this paper; \textbf{(3)} $\lambda>0$ corresponds to rotating $\text{AdS}_{3}$ black holes, where each of the values $k=\pm1$ and $k=0$ correspond to the aforementioned distinct slicings of $\text{AdS}_{3}$. Due to the quartic behavior in $G(x)$, the metric has a number of subtleties, carefully analyzed in \cite{Emparan:1999fd}, and later interpreted as the rotating quantum BTZ black hole \cite{Emparan:2020znc}. 

%%%%%%%%%%%%%%%%%%%%%%%%%%%%%%%%%%%%%%%%%%%%%%%%%%%%%%%%%%%%%%%%%%%%%%%%
\section{Limits of qKdS} \label{app:limitsqKdS}

Here we provide details leading to the various limiting geometries of the qKdS black hole described in the main text. Our analysis primarily follows \cite{Booth:1998gf}. First let us explore the horizon structure of the naive metric (\ref{eq:KdSnaivemet}), corresponding to the roots of the blackening factor $H(r)$. Introduce a function $Q(r)\equiv r^{2}H(r)$
%\beq Q(r)\equiv r^{2}H(r)=r^{2}-\frac{r^{4}}{R_{3}^{2}}-r\mu \ell+a^{2}\;,\label{eq:Qdef}\eeq
such that the roots of $Q$ coincide with the roots of $H$. Since $Q$ is a quartic, it will have either four, two, or zero real roots. Requiring $Q$ to have four real roots, three of which are positive and correspond to the horizons $r_{c}\geq r_{+}\geq r_{-}$, imposes restrictions on the physical parameters of the black hole solution, namely, $a$ and $\mu$. As described in the main text, we can express the parameters $a^{2},R_{3}^{2}$ and $\mu\ell$ in terms of the three horizons $r_{\pm}$ and $r_{c}$.\footnote{For example, to reexpress $R_{3}^{2}$, first subtract $r_{+}H(r_{+})=0$ from $r_{c}H(r_{c})=0$, and similarly subtract $r_{-}H(r_{-})=0$ from $r_{+}H(r_{+})=0$. Subtracting the first of the resultant expressions from the other and rearranging one recovers (\ref{eq:paramsintermsofri}).} The negative root of $Q$, denoted $r_{n}$ and sometimes called the `negative horizon', lies behind the singularity at $r=0$. 

With these roots, we factorize $Q$ as
\beq Q(r)=-\frac{1}{R_{3}^{2}}(r-r_{c})(r-r_{+})(r-r_{-})(r-r_{n})\;.\eeq
It proves convenient to introduce parameters $d,\delta,e$ and $\epsilon$ to parameterize the roots of $Q$,
\beq 
\begin{split}
&r_{c}=e+\epsilon\;,\quad r_{+}=e-\epsilon\;,\\
&r_{-}=d+\delta\;,\quad r_{n}=d-\delta\;.
\end{split}
\label{eq:rootparameterization}\eeq
Since $r_{c}$ and $r_{\pm}$ are all non-negative, we immediately learn $e$ and $d$ are real while $\delta$ and $\epsilon$ must be non-negative real numbers. Expressing $Q(r)$ in terms of these parameters, one fixes $d=-e$ to eliminate the $r^{3}$ contribution.
%In terms of these parameters, we have
%\beq Q(r)=-\frac{1}{R_{3}^{2}}(r-(e+\epsilon))(r-(e-\epsilon))(r-(d+\delta))(r-(d-\delta))\;.\eeq
%As there is no cubic contribution in $Q(r)$, by comparison one fixes $d=-e$. 
Additionally, the root ordering condition, $r_{c}\geq r_{+}\geq r_{-}$ and $r_{n}=-(r_{+}+r_{-}+r_{c})<0$ further imposes 
\beq 0\leq \epsilon< e\;,\quad e\leq \delta< 2e-\epsilon\;,\label{eq:rootordering}\eeq
allowing us to write
\beq Q=-\frac{1}{R_{3}^{2}}((r-e)^{2}-\epsilon^{2})((r+e)^{2}-\delta^{2})\;.\eeq
Moreover, the parameters $R_{3}^{2}$, $a^{2}$ and $\mu\ell$ (\ref{eq:paramsintermsofri}) become
\beq 
\begin{split}
&R_{3}^{2}=2e^{2}+\epsilon^{2}+\delta^{2}\;,\\
&a^{2}=-\frac{(e-\delta)(e+\delta)(e-\epsilon)(e+\epsilon)}{2e^{2}+\epsilon^{2}+\delta^{2}}\;,\\
&\mu\ell=\frac{2e(\delta-e)(\delta+e)}{2e^{2}+\epsilon^{2}+\delta^{2}}\;.
\end{split}
\label{eq:paramseepd}\eeq
Notice from the root ordering (\ref{eq:rootordering}) that $a^{2}\geq0$ and $\mu\geq0$. 

There are multiple cases when $Q$ has degenerate roots, or degenerate horizons. These include: (i) the extremal or `cold' limit, where $r_{+}=r_{-}$; (ii) the (rotating) Nariai limit, where $r_{c}=r_{+}$, and (iii) the `ultracold' limit, when $r_{c}=r_{+}=r_{-}$ coincide. Each are explored below.
%Let us explore each of these limiting geometries in turn. 

\subsubsection*{Extremal limit}

The extremal limit corresponds to when the inner and outer black hole horizons coincide, $r_{+}=r_{-}$. In this limit, the temperature of the outer black hole horizon vanishes, $T_{+}=0$, and is thus sometimes called the `cold' black hole. Given the parameterization (\ref{eq:rootparameterization}), $r_{+}=r_{-}$ imposes $\delta=2e-\epsilon$. Consequently, the parameters (\ref{eq:paramseepd}) become 
\beq a^{2}=\frac{1}{R_{3}^{2}}(e-\epsilon)^{2}(3e-\epsilon)(e+\epsilon)\;,\quad \mu\ell=\frac{8e^{2}(e-\epsilon)}{R_{3}^{2}}\;,\eeq
with $R_{3}^{2}=2(3e^{2}-2e\epsilon+\epsilon^{2})$, and 
\beq Q(r)=-\frac{1}{R_{3}^{2}}(r-(e+\epsilon))(r-e+\epsilon)^{2}(r+3e-\epsilon)\;.\eeq
 %In this limit there is no need to change the form of the metric, however, the global structure does change. This is because now the (double) black hole recedes to an infinite proper distance from all other portions of the geometry, such that the black hole interior is inaccessible from the rest of the spacetime. 

 It is interesting to consider the near horizon limit of the extremal black hole, where $r\approx r_{+}=e-\epsilon$. The coordinates $(t,r,\phi)$ become singular in this limit and we therefore perform the coordinate transformation
 \beq r=e-\epsilon+\lambda\rho\;,\quad t=\frac{\tau}{\lambda}\;,\quad \phi=\varphi-\frac{a\tau}{(e-\epsilon)^{2}\lambda}\;,\eeq
 where $\lambda$ is a dimensionless parameter which for the time being is non-zero. Then,
 \beq Q(\rho)=-\frac{\lambda^{2}\rho^{2}}{R_{3}^{2}}(\lambda\rho-2\epsilon)(\lambda\rho+4e-2\epsilon)\;.\eeq
 Substituting this, performing the coordinate transformation and then taking the limit $\lambda\to0$, the naive brane geometry (\ref{eq:KdSnaivemet}) becomes
 \beq ds^{2}_{\text{ex}}=-\frac{\rho^{2}}{\Gamma}d\tau^{2}+\Gamma\frac{d\rho^{2}}{\rho^{2}}+(e-\epsilon)^{2}\left(d\varphi-\frac{2\rho a}{(e-\epsilon)^{3}}d\tau\right)^{2}\;,\eeq
 with\footnote{Here we use that in the extremal limit $R_{3}^{2}=r_{c}^{2}+3r_{+}^{2}+2r_{+}r_{c}$, such that $r_{c}=\sqrt{R_{3}^{2}-2r_{+}^{2}}-r_{+}$.}
 \beq  \Gamma\equiv \frac{R_{3}^{2}(e-\epsilon)^{2}}{4\epsilon(2e-\epsilon)}=\frac{R_{3}^{2}r_{+}^{2}}{(r_{c}-r_{+})(r_{c}+3r_{+})}=\frac{r_{+}^{2}}{1-6r_{+}^{2}/R_{3}^{2}}\;.\eeq
 Introducing dimensionless coordinates ($\hat{\tau},\hat{\rho})$ such that $\tau= \sqrt{\Gamma}\hat{\tau}$ and $\rho=\sqrt{\Gamma}\hat{\rho}$, we find 
 \beq ds^{2}_{\text{ex}}=\Gamma\left(-\hat{\rho}^{2}d\hat{\tau}^{2}+\frac{d\hat{\rho}^{2}}{\hat{\rho}^{2}}\right)+r_{+}^{2}\left(d\varphi+k\hat{\rho}d\hat{\tau}\right)^{2}\;,\eeq
 where $k\equiv-2a\Gamma/r_{+}^{3}=-2aR_{3}^{2}/r_{+}(R_{3}^{2}-6r_{+}^{2})$.

 For completeness, the leading order contribution to the holographic CFT stress-energy tensor in the extremal background is
 \beq
 \begin{split}
  &\langle T^{\hat{\tau}}_{\;\hat{\tau}}\rangle_{0}=\langle T^{\hat{\rho}}_{\;\hat{\rho}}\rangle_{0}=-\frac{1}{2}\langle T^{\varphi}_{\;\hat{\tau}}\rangle_{0}=\frac{1}{16\pi G_{3}}\frac{\mu\ell}{r_{+}^{3}}\;,\\
 &\langle T^{\varphi}_{\;\varphi}\rangle_{0}=\frac{3\mu\ell aR_{3}^{2}}{8\pi G_{3}(R_{3}^{2}-6r_{+}^{2})}\frac{\hat{\rho}}{r_{+}^{4}}\;,
 \end{split}
\eeq
which has vanishing trace to this order.

\subsubsection*{Nariai limit}

The Nariai black hole is when the outer black hole horizon and cosmological horizon coincide, $r_{c}=r_{+}=r_{\text{N}}$. In this limit, naively, the temperature of the cosmological and black hole horizons vanish, however, we will see the Nariai black hole has a non-zero temperature. From the parameterization (\ref{eq:rootparameterization}), $r_{c}=r_{+}$ is equivalent to $\epsilon=0$, such that
\beq a^{2}=\frac{e^{2}(\delta^{2}-e^{2})}{R_{3}^{2}}\;,\quad \mu\ell=\frac{2e\delta^{2}}{R_{3}^{2}}\;,\eeq
with $R_{3}^{2}=2e^{2}+\delta^{2}$, and 
\beq Q=-\frac{1}{R_{3}^{2}}(r-r_{\text{N}})^{2}((r+r_{\text{N}})^{2}-\delta^{2})\;,\eeq
which vanishes in the limit $r=r_{\text{N}}$. Therefore, the $(t,r,\phi)$ coordinate system is insufficient to describe the Nariai geometry. 

 To this end, consider the following coordinate transformation
\beq r=e+\epsilon\rho\;,\quad t=\frac{\tau}{\epsilon}\;,\quad \phi=\varphi-\frac{a}{e^{2}\epsilon}\tau\;.\eeq
The function $Q$ becomes
\beq Q(\rho)=\frac{\epsilon^{2}}{R_{3}^{2}}(1-\rho^{2})((\epsilon\rho+2e)^{2}-\delta^{2})\;.\eeq
Taking the $\epsilon\to0$ limit, the brane metric (\ref{eq:KdSnaivemet}) takes the form
\beq ds^{2}_{\text{N}}=\Gamma\left(-(1-\rho^{2})d\hat{\tau}^{2}+\frac{d\rho^{2}}{(1-\rho^{2})}\right)+e^{2}\left(d\varphi+k\rho d\hat{\tau}\right)^{2}\;,\label{eq:Nariaigeomapp}\eeq
where $\tau=\Gamma\hat{\tau}$ and
\beq \Gamma\equiv\frac{R_{3}^{2}e^{2}}{4e^{2}-\delta^{2}}=\frac{R_{3}^{2}r_{\text{N}}^{2}}{6r_{\text{N}}^{2}-R_{3}^{2}}\;,\quad k\equiv-\frac{2a\Gamma}{r_{\text{N}}^{3}}\;,\eeq
where we used $\delta=\sqrt{R_{3}^{2}-2r_{\text{N}}^{2}}$. 

The temperature $T_{\text{N}}$ of the Nariai black hole can be found by Wick rotating the near horizon metric into an appropriate Euclidean section. To see this, rescale coordinates $\hat{\rho}=\rho\sqrt{\Gamma}$ and $\tau=\hat{\tau}\sqrt{\Gamma}$, such that the Nariai metric (\ref{eq:Nariaigeomapp}) becomes
\beq ds^{2}_{\text{N}}=-f(\hat{\rho})d\tau^{2}+f^{-1}(\hat{\rho})d\hat{\rho}^{2}+r_{\text{N}}^{2}(d\phi+(k/\Gamma)\hat{\rho}d\tau)^{2}\;,\eeq 
with $k/\Gamma=-2a/r_{\text{N}}^{2}$ and $f(\hat{\rho})=1-\hat{\rho}^{2}/\Gamma$. Next, Wick rotate $\tau\to i\tau_{E}$ and $a\to ia_{E}$, such that the Euclideanized geometry is
\beq ds^{2}_{\text{N}}=f(\hat{\rho})d\tau_{E}^{2}+f^{-1}(\hat{\rho})d\hat{\rho}^{2}+r_{\text{N}}^{2}\left(d\phi+\frac{2a_{E}\hat{\rho}}{r_{\text{N}}^{2}}d\tau_{E}\right)\;.\label{eq:EucNarbh}\eeq
We now zoom in to the near horizon region where $\hat{\rho}_{i}=\sqrt{\Gamma}$ and express the metric in flat polar coordinates and impose regularity to remove the conical singularity. Thus, introduce the radial coordinate
\beq \rho'^{2}=\frac{4(\hat{\rho}-\hat{\rho}_{i})}{f'(\hat{\rho}_{i})}\quad \Rightarrow \quad d\hat{\rho}^{2}=(f'(\hat{\rho}_{i})\rho'/2)^{2}d\rho'^{2}\;.\eeq
Moreover, to eliminate the $d\tau_{E}d\phi$ cross term in the line element, introduce coordinates
\beq \tau_{E}=\frac{1}{(\gamma^{2}+\alpha^{2})}(\gamma\tau_{E}'-r_{\text{N}}\alpha\phi')\;,\quad \phi=\frac{1}{(\gamma^{2}+\alpha^{2})}\left(\frac{\alpha}{r_{\text{N}}}\tau'_{E}+\gamma\phi'\right)\;,\eeq
with $\gamma=\sqrt{\Gamma}/r_{\text{N}}$ and $\alpha=-2a_{E}\Gamma/r_{\text{N}}^{2}$. Then, expand the metric (\ref{eq:EucNarbh}) about $\hat{\rho}=\sqrt{\Gamma}$,
\beq ds^{2}_{\text{N}}\approx \rho'^{2}d(\beta^{-1}\tau'_{E})^{2}+d\rho'^{2}+\frac{r_{\text{N}}^{2}}{\Gamma}d\phi'^{2}\;,\label{eq:nearhorNari}\eeq
where we used $f(\hat{\rho})\approx f'(\hat{\rho}_{i})(\hat{\rho}-\hat{\rho}_{i})$ and
\beq \beta=\frac{\Gamma^{2}}{r_{\text{N}}^{3}R_{3}^{2}}(6r_{\text{N}}^{2}-R_{3}^{2}+4a_{E}^{2}R_{3}^{2})\;.\eeq
The $(\tau_{E}',\rho')$-sector is that of a cone where we remove the singularity at $\rho'=0$ by demanding $\tau_{E}'$ have period $\Delta\tau_{E}'=2\pi\beta$. Additionally, we impose $\phi'$ to have period $\Delta\phi'=0$, such that the geometry (\ref{eq:nearhorNari}) represents flat polar coordinates. We now solve for the periodicity of $\Delta\tau_{E}$ in the Euclidean geometry (\ref{eq:EucNarbh}) via\footnote{Where it is useful to know the inverse coordinate transformation $\tau_{E}'=\gamma\tau_{E}+r_{\text{N}}\alpha\phi$ and $\phi'=\gamma\phi-\frac{\beta}{r_{\text{N}}}\tau_{E}$.}
\beq \Delta\tau_{E}'=2\pi\beta=\gamma\Delta\tau_{E}+r_{\text{N}}\alpha\Delta\phi\;,\quad \Delta\phi'=0=\gamma\Delta\phi-\frac{\alpha}{r_{\text{N}}}\Delta\tau_{E}\;,\eeq
leading to 
\beq \Delta\tau_{E}=\frac{2\pi\beta\gamma}{(\gamma^{2}+\alpha^{2})}=2\pi\sqrt{\Gamma}\;.\eeq
Hence, the temperature of the Nariai black hole is $T_{\text{N}}=(2\pi\sqrt{\Gamma})^{-1}$.

Lastly, the holographic CFT stress-energy tensor in the rotating Nariai geometry is
 \beq
 \begin{split}
  &\langle T^{\hat{\tau}}_{\;\hat{\tau}}\rangle_{0}=\langle T^{\rho}_{\;\rho}\rangle_{0}=-\frac{1}{2}\langle T^{\varphi}_{\;\hat{\tau}}\rangle_{0}=\frac{1}{16\pi G_{3}}\frac{\mu\ell}{r_{\text{N}}^{3}}\;,\\
 &\langle T^{\varphi}_{\;\varphi}\rangle_{0}=\frac{3\mu\ell aR_{3}^{2}}{8\pi G_{3}(6r_{\text{N}}^{2}-R_{3}^{2})}\frac{\rho}{r_{\text{N}}^{4}}\;.
 \end{split}
\eeq
 When $a=0$, we recover the stress-tensor for the static quantum Nariai black hole \cite{Emparan:2022ijy}.

\subsubsection*{Ultracold limit}

The ultracold limit occurs when all three horizons coincide, $r_{c}=r_{+}=r_{-}\equiv r_{\text{uc}}$. Thus, this is a combination of the Nariai and extremal limits, i.e., simultaneously sending $\delta\to 2e-\epsilon$ and $\epsilon\to0$. Here we will arrive at the ultracold limit directly from the Nariai limit, where we make the following change of variables:
\beq \rho=\sqrt{\frac{2e-\delta}{R_{3}}}X\;,\quad \tau=\sqrt{\frac{R_{3}}{2e-\delta}}\frac{R_{3}e}{4}T\;,\eeq
and subsequently take the limit $\delta\to 2e$ such that the Nariai geometry (\ref{eq:Nariaigeomapp}) becomes that of the ultracold black hole,
\beq ds^{2}_{\text{uc}}=\frac{R_{3}r_{\text{uc}}}{4}(-dT^{2}+dX^{2})+r_{\text{uc}}^{2}\left(d\varphi-\frac{2aX}{r_{\text{uc}}^{3}}dT\right)^{2}\;.\eeq
The physical parameters meanwhile are $a^{2}=3e^{4}/R_{3}^{2}$ and $\mu\ell=8e^{3}/R_{3}^{2}$.

\subsubsection*{Lukewarm black hole}

Lastly, as with all Kerr-de Sitter black holes, the qKdS has a lukewarm limit, where the surface gravities of the cosmological and outer black hole horizons coincide $\kappa_{c}=\kappa_{+}$, apart from the surface gravity of the Nariai black hole. For completeness we carry out the analysis of this limiting geometry with respect to the surface gravities of the naive black hole spacetime, leaving the analysis of the lukewarm limit of the regular black hole for the main text. 

To this end, the surface gravities with respect to the naive metric (\ref{eq:KdSnaivemet}) are simply $\kappa_{i}=\frac{1}{2}|H'(r_{i})|$, following the definition $\zeta^{b}\nabla_{b}\zeta^{c}=\kappa\zeta^{c}$, where there is a surface gravity associated with each root of the blackening factor $H(r)$. The temperature of each horizon is then\footnote{One way to derive this expression of the temperature is to move to the Euclidean section of the rotating geometry (\ref{eq:KdSnaivemet}), via the double Wick rotation $t\to i\tau_{E}$ and $a\to ia_{E}$ and then impose regularity to remove the conical singularity along the Euclidean time direction.}
\beq T_{i}=\frac{\kappa_{i}}{2\pi}=\frac{|H'(r_{i})|}{4\pi}\;.\eeq
Then, using $H'(r_{i})=r_{i}^{-2}Q'(r_{i})$, since $H(r_{i})=0$, we have
\beq T_{i}=\frac{|Q'(r_{i})|}{4\pi r_{i}^{2}}=\frac{1}{2\pi r_{i}^{2}R_{3}^{2}}|[(r_{i}-e)((r_{i}+e)^{2}-\delta^{2})+(r_{i}+e)((r_{i}-e)^{2}-\epsilon^{2})]|\;.\eeq
Then, the temperature of the outer horizon $r_{+}=e-\epsilon$ is
\beq T_{+}=\frac{1}{4\pi R_{3}^{2}}\frac{2\epsilon((2e-\epsilon)^{2}-\delta^{2})}{(e-\epsilon)^{2}}\;.\eeq
Via root ordering, this temperature is positive, and vanishes in the extremal limit. Meanwhile, the temperature of the cosmological horizon $r_{c}=e+\epsilon$ is 
\beq T_{c}=\frac{1}{4\pi R_{3}^{2}}\frac{2\epsilon((2e+\epsilon)^{2}-\delta^{2})}{(e+\epsilon)^{2}}\;.\eeq
Taking their difference, 
\beq T_{c}-T_{+}=0\quad \Longleftrightarrow\quad  e\epsilon^{2}(\epsilon^{2}-2e^{2}+\delta^{2})=0\;.\eeq
Here $e=0$ is forbidden via the root ordering while $\epsilon=0$ corresponds to the Nariai limit. Hence, the lukewarm limit corresponds to when $\delta^{2}=2e^{2}-\epsilon^{2}$, with temperature $T_{\text{luke}}=\frac{\epsilon}{2\pi R_{3}^{2}}$. Moreover, since $H(r)$ is non-zero in this limit, the lukewarm geometry is safely covered by the coordinates $(t,r,\phi)$ with blackening factor 
\beq H(r)=-\frac{1}{R_{3}^{2}r^{2}}((r-e)^{2}-\epsilon^{2})((r+e)^{2}-2e^{2}+\epsilon^{2})\;,\eeq
and where (\ref{eq:paramseepd}) become $R_{3}^{2}a^{2}=(e^{2}-\epsilon^{2})^{2}$ and $R_{3}^{2}\mu\ell=4e(e^{2}-\epsilon^{2})$, with $R_{3}^{2}=4e^{2}$.

\bibliography{qdSrefs}

\end{document}